\documentclass[universe,article,accept,pdftex,moreauthors]{Definitions/mdpi} 
\firstpage{1} 
\pubvolume{8}
\issuenum{11}
\articlenumber{578}
\pubyear{2022}
\copyrightyear{2022}
\externaleditor{\textls[-25]{Academic Editor: {Marina Manganaro}}}
\datereceived{9 September 2022} 
\dateaccepted{26 October 2022} 
\datepublished{2 November 2022} 
\hreflink{https://\linebreak doi.org/10.3390/universe8110578} 

\Title{Study of Intra-Day Flux Distributions of Blazars Using \mbox{XMM-Newton} Satellite}

\TitleCitation{Study of Intra-Day Flux Distributions of Blazars Using \mbox{XMM-Newton} Satellite}

\Author{{Kiran Wani}
 $^{1,2}$\orcidA{}  and Haritma Gaur $^{1,}$*}

\AuthorNames{Kiran Wani and Haritma Gaur}

\AuthorCitation{Wani, K.; Gaur, H.}

\address{%
$^{1}$ \quad Aryabhatta Research Institute of Observational Sciences (ARIES), Manora Peak, Nainital 263002, India\\   
$^{2}$ \quad School of Physical Sciences, SRTM University, Nanded 431606, India}

\corres{Correspondence: {harry.gaur31@gmail.com}}

\abstract{We present a study of the flux distribution of a sample of 15 Intermediate and Low-energy peaked blazars using \mbox{XMM-Newton} observations in a total of 57  epochs on short-term timescales. 
We characterise the X-ray variability of all of the light curves using excess fractional variability amplitude and found that only 24 light curves in 7 sources are significantly variable.   
In order to characterise the origin of X-ray variability in these blazars, we {{fit the flux distributions of all} 
} these light curves using Gaussian and lognormal distributions, as any non-Gaussian perturbation
could indicate the imprints of fluctuations in the accretion disc, which could be Doppler boosted through the relativistic jets in blazars. However, intra-day variability, as seen in our observations, is difficult to reconcile using disc components as the emissions in such sources are mostly dominated by jets.  We used Anderson--Darling (AD) and $\chi^{2}$ tests to fit the histograms.
 In 11 observations of 4~blazars, namely, ON\:231, 3C~273, PKS 0235+164 and PKS 0521-365, both models equally fit the flux distributions. In the rest of the observations, we are unable to model them with any distribution.
In two sources, namely, BL\:Lacertae and S4 0954+650, the lognormal distribution is preferred over the normal distribution, which could arise from non-Gaussian perturbations 
from relativistic jets or linear Gaussian perturbation in the particle time scale leading to such flux distributions. }


\begin{document}

\setcounter{section}{0} 
\section{Introduction}
Blazars are well known for their collimated relativistic jets aligned to the line of sight of the observer \cite{Blandford_Konigl1979}. Blazars constitute BL Lacertae objects (BL Lacs)
 and Flat-Spectrum Radio Quasars (FSRQs). The significant difference between these two classes is their optical emission which is strong for FSRQs but weak/absent for BL Lacs \cite{Urry_padovani_1995}. Emissions from  blazars span across the whole electromagnetic spectrum (EM) from radio to high-energy gamma rays.
The spectral energy distribution of blazars consists of two broad humps, where the low-energy hump peaks in the optical/X-ray band  while the high-energy hump peaks at GeV/TeV energies.
\textls[-15]{Recently, Fan et al. (2016) [\citealp{Fan_2016}] (see also [\citealp{Abdo_2010, Yang_2022}]), classified blazars based on the peak frequency of their low-energy component, namely, low-synchrotron frequency peaked blazars (LSP; $\nu_{s} \le 10^{14}$ Hz),
intermediate-synchrotron frequency peaked blazars (ISP; $10^{14} < \nu_{s}$ $ < 10^{15.3}$ Hz) and high-synchrotron peaked blazars (HSP; \mbox{$\nu_{s} \ge 10^{15.3}$ Hz}).} The low-energy component of a SED is ascribed to
synchrotron emission by relativistic electrons in the jet (e.g., \cite{Romero_2017}); whereas the high-energy component is likely due to Inverse Compton (IC) scattering of seed photons off the relativistic
electrons. The seed photons are the synchrotron photons in the synchrotron self Compton (SSC) model, whereas the external Compton (EC) scenario foresees that seed photons may
come from the accretion disc, broad line region (BLR) or hot corona surrounding the accretion disc (e.g., \cite{Dermer_1992,Sikora_1994,Ghisellini_Tavecchio_2009}). Alternate to this
Leptonic interpretation, hadronic models were put forth to explain the broadband spectral features of blazars \cite{Bottcher_2013,Atoyan_Dermer_2003}. 

Blazars exhibit strong flux and polarisation variability on diverse timescales ranging from a few minutes to months to years across the entire EM spectrum \cite{Sembay_1993, Wagner_1995, Ulrich_1997}.
The variability in such sources probed so far are widely split into three categories: variability within a day (intra-day variability (IDV); \cite{Wagner_1995}), variability on
days to months (short-term variability (STV)) and variability over a few months and longer (long-term variability (LTV); \cite{Gaur_2019}).
The physical processes responsible for the observed behaviour are still unclear, but recent continuous advancements in detector technologies enable the study of
variability at diverse timescales in X-ray bands. Most puzzling variations occur on IDV timescales. The cause of variability is still not well understood, but possible clues can
be obtained by studying the intra-day flux distribution of blazars. 

Long-term flux distributions have been {{obtained} 
} in detail for X-ray energies for Seyfert galaxies,
where the emission of these energies is dominated by the accretion disk or its corona. The flux distribution of such sources exhibits a lognormal distribution~\cite{Gaskell_2004}.
 For a linear stochastic process, the flux distribution is expected to be Gaussian, while a lognormal flux distribution is expected to arise from multiplicative processes that
originate in the accretion disc \cite{Uttley_McHardy_2005,McHardy_2010}. The lognormality of the flux distribution in blazars was detected using {\it{Rossi X-ray Timing Explorer (RXTE)} 
}
observations \cite{Giebels_2009}. The lognormal behaviour of flux distribution is observed in other blazars at various timescales and wavelengths~\mbox{(\cite{Chevalier_2015,Sinha_2016,
Sinha_2017,Sinha_2018,Kushwaha_2016,Kushwaha_2020,Shah_2018}} and references therein). However, the dominance of non-thermal emission hinders our understanding
of the accretion disc--jet connection in jet-dominated sources. The variations on IDV timescales are, however, difficult to originate from the accretion disc and favours  originating within
the blazar jets. {{Biteau and Giebels (2012)}} \cite{Biteau_Giebels_2012} explained the flux distributions using the mini jets-in-a-jet model  \cite{Giannios_2009} {{and showed that the flux from a single
randomly oriented mini jet will follow a Pareto distribution, whereas}} the total flux from {{many
isotropically}} oriented mini jets will lead to an $\alpha$-stable distribution, which will converge to a lognormal one subjected to experimental uncertainties.  Therefore, small Gaussian
perturbations can also propagate to produce non-linear flux distributions, which can explain the lognormal behaviour in long-term time series as well as during blazar flaring states.
{{Sinha et al. (2018)}}~\cite{Sinha_2018} provided an alternative interpretation of the \mbox{non-Gaussian} distributions of blazar light curves through linear fluctuations of the underlying particle acceleration
and/or the diffusive escape rate of the emitting electrons.

In this work, we aim to study the statistical properties of flux distributions of a sample of intermediate and low-energy peaked blazars observed using {{the X-ray data from}} the \mbox{XMM-Newton} satellite on
IDV timescales. { For LSP and ISP blazars, X-ray emission is probably due to inverse Compton emissions from lower energy electrons. The emission mechanism is not well understood, and variability could have imprints of an accretion disc component. Therefore, flux distribution studies of LSPs and ISPs are important, as such studies are useful for testing models of blazars,} which are almost entirely dominated by the emission of relativistic jets, contrary to the majority of AGNs where the
emission is mostly dominated by the accretion disc. Therefore, the non-Gaussian flux distribution of blazars, if found on IDV timescales, can be used to test additive models, such as
mini jets-in-a-jet and other scenarios (mentioned above) in blazar jets.
Data analysis and sample selection are provided in the next section, which is followed by analysis techniques in Section \ref{analysis tech} and results in Section \ref{results}. We discuss our results and their implications in Section \ref{Discussion}.
 

\section{Sample Selection and Data Analysis}
We have selected a sample of ISP and LSP blazars observed by the \mbox{XMM-Newton} satellite since its launch. Our sample is derived from  {{Gaur et al. (2018)}} \cite{Gaur_2018}.
We considered all \mbox{XMM-Newton} observations for this sample of blazars {with an exposure time greater than 10 ks.} We also included LBL blazar 3C 273 in our studies. Due to a large number 
of \mbox{XMM-Newton} observations for blazar 3C 273, we selected 3$\sigma$ variable observations from { Gowtami et al. (2022)} \cite{Gowtami_2022} {including the latest observation of 28 June 2022}. \mbox{XMM-Newton} data are obtained from the HEASARC archive\endnote{(\url{https://heasarc.gsfc.nasa.gov/docs/archive.html}, {accessed on 9 September 2022}
).}.  \textls[-15]{We have used European Photon Imaging
Camera (EPIC) pn instrument data. EPIC-pn is 
most sensitive and less affected by the photon pile-up effects \cite{Struder_2001}. Science Analysis System (SAS-19.1.0) is used to analyse the EPIC-pn data. SAS task {{\emph{epproc}}} is used to generate the calibrated and concatenated event lists. An X-ray energy range of 0.3--10 keV is considered since data below 0.3 keV are considerably dominated by detector noise components, and data above 10 keV are dominant with background flaring events and hot pixels. Background flaring in the X-ray energy range of 10--12 keV in the observations is checked and avoided from the event list files. A circular region of 40 arcsec radius and circular annulus of 50 (inner radius) to 60 (outer radius) arcsec centred on the source is used to extract the source and background light curve, respectively, for the X-ray energy range of 0.3--10 keV with a time binning of 100 s.  Pile-ups in the observations are checked and are not found in any of the observations. SAS task {\emph{epiclccorr}} is used to correct for vignetting, PSF variation, bad pixels and to get a background-subtracted light curve.}

\section{Analysis Techniques}
\label{analysis tech}
Excess variance is the measure of intrinsic variability in the light curve and is determined by subtracting the variance due to measurement errors from the total variance of the observed
 light curve \cite{Vaughan(2003a)}. The fractional rms variability amplitude $F_{var}$ is the square root of the normalised excess  variance \cite{EdelsonKrolikPike(1990), Rodriguez-Pascual(1997)}.
 Histograms are used to visualise the density or frequency distribution of any measured variable. The number of bins affects the shape of the histogram. Therefore, we need an optimal data-based 
binning for a histogram. The optimal binning algorithm, i.e., the Knuth method \cite{KNUTH_2019}, is used to obtain the histogram of a count rate distribution. A histogram of a count rate distribution for individual light curves is obtained and fitted with normal and lognormal probability distribution functions. Normal and lognormal distribution functions are defined as follows:
\begin{equation}\small
    f_{norm}(x)=\frac{1}{\sigma\sqrt{2\pi}}\:exp \Biggl[\frac{-(x - \mu)^{2}}{2 \sigma^{2}}\Biggl]
\end{equation}
\begin{equation}\small
    f_{lognorm}(x)=\frac{1}{x\:\sigma\sqrt{2\pi}}\:exp \Biggl[\frac{-(ln(x) - \mu)^{2}}{2 \sigma^{2}}\Biggl]
\end{equation}
where $\mu$ and $\sigma$ are the mean and standard deviation of the distribution, respectively, in units of counts/sec.
The Anderson--Darling test (AD) \cite{AD_1952,Stephens_1977} is used to quantify the lognormal and normal distribution function fitting to the histogram of the count rate 
distribution. {{When} the $p$-value of the AD test is less than or equal to 0.01,  we reject the hypothesis of normality/lognormality}. The $\chi^2$ goodness-of-fit test is also
 used to quantify the 
normal and lognormal distribution fitting for count rate histograms, and we choose the value of {{reduced}} $\chi^2$ close to unity as the one representing
the data better {{with} null hypothesis probability value lying within the range of 0.75--0.05. If the probability value lies outside this range, then the null hypothesis is rejected. The distribution model is preferred only when both \mbox{AD } and $\chi^2$ tests accept the null hypothesis, where the null hypothesis is that the data follow the specified distribution.}

\section{Results}
\label{results}
We analysed all of the \mbox{XMM-Newton} satellite observations of our sample of ISP and LSP blazars for 57 epochs.  
We estimated the fractional root mean square (rms) variability amplitude F$_{var}$ of all of these observations and found significant (3 sigma level) variations in 7 sources
 during 24 observational epochs. In order to study the statistical properties of X-ray variability of these light curves, we considered only significantly variable 
observations in our studies. 
 Our sample of blazars, classifications and F$_{var}$ are tabulated in Table \ref{tab:results}. The light curves of blazars are represented in Figure \ref{fig:lightcurve}. Histograms of count rates of these light curves are obtained and are fitted with lognormal 
and normal probability distribution functions (PDFs). 
Histograms and their corresponding distribution functions (i.e., normal and lognormal PDFs) are plotted in Figure \ref{fig:histogram}. The statistical analysis of the \mbox{AD test} yields the 
AD statistic, $p$-value, and the $\chi^2$ test yields a reduced $\chi^2$ ($\chi^2_{r}$) value and a {{probability} value,} which are presented in Table \ref{tab:results}. 
{ BL Lac is studied by Giebels and Degrange (2009)~\cite{Giebels_2009} using the RXTE-PCA observations, and it is the first blazar in which lognormal X-ray variability is detected, and a linear correlation 
between the excess rms and the average flux was also found.} {The flux distribution of blazar S5 0716+714 in X-ray with \mbox{XMM-Newton} observations was studied on a short-term timescale by Kushwaha and Pal (2020) \cite{Kushwaha_2020} and Mohorian et al.
 (2022) \cite{Mohorian_2022}.} {3C 273 is studied in X-ray using 16~years of Rossi X-ray Timing Explorer (RXTE) archival data by Khatoon et al. 2020 \cite{Khatoon_2020}.
The flux distribution of FSRQ 3C 273 is lognormal, while its photon index distribution is
Gaussian. This result is consistent with linear Gaussian perturbation in the particle acceleration
timescale, which produces lognormal distribution in flux. 
On the decade-long timescale, BL Lac, 3C273, S5 0716+714 and PKS 0235+164
follow the lognormal flux distribution and display a strong linear rms–flux relationship in
optical \cite{Bhatta_2021} and in weekly binned gamma-ray~\cite{Bhatta_Dhital_2020} observations. The monthly average $\gamma$-ray flux distribution favours the lognormal flux distribution,
 (i.e., \cite{Shah_2018}) and $\gamma$-ray (0.1–300 GeV) observations from
Fermi/LAT telescope using maximum likelihood estimation (MLE) methods suggest that the $\gamma$-ray flux variability can be characterised by log-stable distributions \cite{Duda_Bhatta_2021}}.

The results for individual sources are provided below:

{{BL Lac}}:
 This blazar is observed in 5 \mbox{XMM-Newton} observations and found to be variable in only one light curve {of {16 May 2008}}. 
 \mbox{The AD test} shows that the hypothesis of normality is 
rejected since the $p$-value for a normal distribution is less than 0.01, whereas the hypothesis of lognormality is accepted. The $\chi^2$ test also gives a reduced $\chi^2$ value of 
close to 1, {and the probability value lies in the range of 0.75 to 0.05 for lognormal PDF fitting as compared to normal PDF fitting.} Therefore, a lognormal distribution is a preferred model over a \mbox{normal one}. 

{ON 231:}
For observations taken on dates {26 June 2002, 14 June 2008 and 18 June 2008}, \mbox{the AD test} rejects the hypothesis for normal as well as lognormal model. The $\chi^2$ test also yields a reduced $\chi^2$ value much 
greater than 1 for both PDF fittings. Therefore, in these cases, none of the distributions is the preferred model.
For the observation from {12 June 2008}, the \mbox{AD test} accepts the hypothesis for normal as well as lognormal distributions, and the $\chi^2$ test also yields a reduced $\chi^2$ value close to 1 {with a probability value in the range of 0.75 to 0.05 for both  models.} Therefore, 
we cannot distinguish between these two models. For the observation from \mbox{{16 June 2008}}, the \mbox{AD test} rejects the hypothesis of normality and accepts the lognormality hypothesis. However, it has a reduced $\chi^2$ 
value not close to 1 {with a probability value lying outside the prescribed range}. Therefore, in this observation, we can not prefer any distribution.

{S5 0716+714:} For the observation from {24 September 2007}, the \mbox{AD test} rejects the  normal hypothesis as 
well as the lognormal distribution, and the reduced $\chi^2$ value is also not close to 1 for both  models. Therefore, we cannot prefer any distribution for this observation.
 For the observation from {4 April 2004}, the \mbox{AD test} rejects the hypothesis of normality but accepts the lognormality hypothesis. However, the corresponding reduced $\chi^2$ value for the lognormal
distribution is not close to 1, {with a probability value lying outside the specified range. Therefore, in this case, we can not prefer any model over~another}.

{3C 273:} For the observation from {12 December  2011}, the \mbox{AD test} rejects the hypothesis of normality, whereas it accepts the lognormality hypothesis, but its corresponding reduced $\chi^2$ value is not close to 1, {with a probability value lying outside the specified range. Therefore, no distribution is the preferred model.} For observations taken on the dates {13~June 2001, 12 January  2007, 9 December  2008} and {6 July 2020}, the \mbox{AD test} accepts both the normality and lognormality hypotheses, but their corresponding reduced $\chi^2$ values are not close to 1, {with a probability value lying outside the specified range.}
 Therefore, both distributions are not favourable for these observation cases.
 Similarly, for observations from {14 June 2000, 30 June 2004, 8 December 2007}, \mbox{{20 December 2009}}, {10 December 2010, 13 July 2015, 26 June 2017 and 28 June 2022}, the \mbox{AD test} accepts both the normality and lognormality hypotheses and their corresponding reduced $\chi^2$ value 
is also close to 1, {with a probability value lying in the specified range.} Therefore, both the lognormal and normal distributions are favourable models in these observations.
\startlandscape
\begin{table}[H]
\tablesize{\footnotesize}
\caption{Observation log of ISP and LSP blazars with their corresponding $F_{var}$, results of \mbox{AD test} and $\chi^2$ test for lognormal and normal PDF {{fitting}} and results of the RMS--Flux relationship.} 
\label{tab:results}
\newcolumntype{C}{>{\centering\arraybackslash}X}
\begin{tabularx}{\textwidth}{ccccccCCCCCCC}
\toprule
\multirow{2.6}{*}{$\mathbf{Object}$} & \multirow{2.6}{*}{\shortstack{ \textbf{Source}\\ \textbf{Class}}} &\multirow{2.6}{*}{$\bold{Redshift}$}& \multirow{2.6}{*}{\textbf{Observation Date}}&\multirow{2.6}{*}{\textbf{N}} & \multirow{2.6}{*}{$\mathbf{F_{var}}$}  & \multicolumn{2}{c}{\textbf{Lognormal}} & \multicolumn{2}{c}{\textbf{Normal}}&\multicolumn{2}{c}{\textbf{RMS-Flux Relation}}& \multirow{2.6}{*}{$\mathbf{Preferred\:Model}$}\\
\cmidrule{7-12}
&   && & &&$\mathbf{AD}$ (\boldmath{$p$}\textbf{-Value}) &\boldmath{$\chi^2_{r}$} \textbf{(prob)} & $\mathbf{AD}$ (\boldmath{$p$}\textbf{-Value}) & \boldmath{$\chi^2_{r}$} \textbf{(prob)} & \boldmath{$\rho_{10}$} \boldmath{$(p$}\textbf{-Value)}&\boldmath{$\rho_{25}$} \boldmath{$(p$}\textbf{-Value)}&\\\midrule
BL Lac&ISP&0.069&{16 May 2008}&22&7.26 (0.42)&0.32 (0.52)&0.94 (0.54)&2.93 (<0.01)&1.58 (0.04)&0.22 (0.26)&0.36 (0.18)&Lognormal\\
ON 231&ISP&0.102&{26 June 2002}&6&44.54 (0.44)&5.59 (<0.01)&7.43 (<0.01)&5.04 (<0.01)&6.18 (<0.01)&0.66 (<0.01)&0.74 (<0.01)&-\\
ON 231&ISP&0.102&{12 June 2008}&8&14.15 (1.21)&0.29 (0.60)&0.95 (0.48)&0.46 (0.26)&0.92 (0.50)&$-$0.05 (0.91)&0.75 (0.14)&None\\
ON 231&ISP&0.102&14 June 2008&5&38.42 (0.30)&5.95 (<0.01)&16.68 (<0.01)&4.69 (<0.01)&14.27 (<0.01)&0.83 (<0.01)&0.41 (0.24)&-\\
ON 231&ISP&0.102&16 June 2008&5&15.89 (0.54)&0.59 (0.12)&3.74 (<0.01)&2.82 (<0.01)&4.00 (<0.01)&0.05 (0.93)&$-$0.49 (0.40)&-\\
ON 231&ISP&0.102&18 June 2008&4&11.08 (0.67)&1.05 (0.01)&2.51 (0.04)&0.98 (0.01)&1.98 (0.09)&$-$0.09 (0.89)&$-$0.94 (0.22)&-\\
S5 0716+714&ISP&0.3&4 April 2004&14&22.00 (0.25)&0.26 (0.70)&2.98 (<0.01)&9.68 (<0.01)&4.09 (<0.01)&0.83 (<0.01)&0.87 (<0.01)&-\\
S5 0716+714&ISP&0.3&24 September 2007&14&24.09 (0.24)&3.63 (<0.01)&7.88 (<0.01)&3.49 (<0.01)&5.33 (<0.01)&0.52 (<0.01)&0.52 (0.02)&-\\
3C 273 *&LSP&0.1575&14 June 2000&16&0.64 (0.17)&0.22 (0.84)&1.10 (0.35)&0.27 (0.68)&1.16 (0.29)&0.09 (0.67)&0.07 (0.82)&None\\
3C 273 &LSP&0.1575&13 June 2001&15&0.63 (0.12)&0.77 (0.04)&1.93 (0.02)&0.72 (0.06)&1.97 (0.01)&$-$0.06 (0.78)&$-$0.21 (0.54)&-\\
3C 273&LSP&0.1575&30 June 2004&10&0.91 (0.24)&0.17 (0.93)&0.74 (0.68)&0.16 (0.96)&0.71 (0.71)&0.23 (0.66)&1.00 (1.00)&None\\
3C 273&LSP&0.1575&12 January 2007&18&1.29 (0.09)&0.39 (0.38)&0.50 (0.96)&0.35 (0.48)&0.51 (0.95)&$-$0.13 (0.64)&0.59 (0.06)&-\\
3C 273&LSP&0.1575&8 December 2007&11&0.49 (0.16)&0.29 (0.60)&0.94 (0.50)&0.32 (0.54)&0.98 (0.46)&0.09 (0.76)&0.20 (0.67)&None\\
3C 273&LSP&0.1575&9 December 2008&12&1.05 (0.12)&0.13 (0.98)&0.61 (0.83)&0.14 (0.98)&0.63 (0.82)&0.07 (0.80)&$-$0.73 (0.06)&-\\
3C 273&LSP&0.1575&20 December 2009&12&0.89(0.14)&0.20 (0.87)&1.09 (0.36)&0.25 (0.75)&1.07 (0.38)&0.10 (0.77)&0.66 (0.15)&None\\
3C 273&LSP&0.1575&10 December 2010&14&0.73 (0.19)&0.28 (0.63)&1.30 (0.20)&0.55 (0.16)&1.36 (0.16)&0.43 (0.13)&0.82 (0.05)&None\\
3C 273&LSP&0.1575&12 December 2011&12&2.11 (0.11)&0.62 (0.11)&1.91 (0.03)&1.54 (<0.01)&2.11 (0.01)&0.19 (0.51)&0.04 (0.92)&-\\
3C 273&LSP&0.1575&13 July 2015&12&1.06 (0.14)&0.46 (0.27)&0.98 (0.46)&0.43 (0.31)&0.95 (0.49)&$-$0.18 (0.40)&$-$0.45 (0.11)&None\\
3C 273&LSP&0.1575&26 June 2017&21&0.73 (0.20)&0.33 (0.52)&1.27 (0.18)&0.28 (0.63)&1.20 0.24)&0.64 (<0.01)&0.44 (0.38)&None\\
3C 273&LSP&0.1575&6 July 2020&16&1.79 (0.13)&0.17 (0.93)&0.41 (0.98)&0.38 (0.40)&0.47 (0.96)&$-$0.02 (0.92)&$-$0.41 (0.21)&-\\
3C 273&LSP&0.1575&28 June 2022&18&1.03 (0.22)&0.24 (0.78)&1.33 (0.16)&0.25 (0.73)&1.32 (0.16)&0.25 (0.24)&0.30 (0.34)&None\\
S4 0954+650&LSP&0.367&30 September 2007&13&12.57 (0.75)&0.74 (0.05)&1.55 (0.09)&3.92 (<0.01)&3.35 (<0.01)&0.39 (0.19)&0.61 (0.20)&Lognormal\\
PKS 0235+164&LSP&0.94&10 February 2002&8&11.60 (0.50)&0.23 (0.80)&1.50 (0.15)&0.60 (0.12)&1.75 (0.08)&0.36 (0.22)&0.32 (0.49)&None\\
PKS 0521-365&LSP&0.055&10 October 2002&9&2.06 (0.62)&0.36 (0.45)&1.30 (0.23)&0.33 (0.51)&1.15 (0.33)&0.42 (0.14)&0.32 (0.44)&None
\\\bottomrule
\end{tabularx}
\begin{adjustwidth}{+\extralength}{0cm}
\noindent{\footnotesize{* FSRQ; {{N: degree of freedom}; $\chi^2_{r}\sim$$\frac{\chi^2}{N}$; prob: probability value; $\rho_{10}$ and $\rho_{25}$ are the  linear correlation coefficients for 10 and 25 data points per bin, respectively; dash `-' represents no distribution fits well; and `None' represents that flux distributions equally fit with both distributions and
we cannot prefer any model.}}}
\end{adjustwidth}
\end{table}
\finishlandscape

{S4 0954+650:} the \mbox{AD test} rejects the normality hypothesis and accepts the lognormality hypothesis. The reduced $\chi^2$ value for the lognormal distribution is close to unity, {with a probability value lying in the specified range. Therefore, the 
lognormal distribution is the preferred model over the normal one.}

{PKS 0235+164:} \mbox{the AD test} accepts both  normality and lognormality hypotheses. Their corresponding reduced $\chi^2$ values  are close to unity 
for both models, {with a probability value lying in the specified range. Therefore, we cannot prefer any model from this source.}

{PKS 0521-365:} the \mbox{AD test} accepts both normality and lognormality hypotheses. The $\chi^2$ test yields a reduced $\chi^2$ value close to 1, {with a probability value lying in the specified range for normal as well as lognormal models.
 Therefore, in this observation, we cannot prefer any  distribution.}

In order to quantify the degree of intrinsic variability, we estimated the rms--flux relationship for two different binnings of 15 and 25 data points per bin.
We tabulated the results of the linear Pearson correlation coefficient and their corresponding $p$-values for all  sources in Table \ref{tab:results}.

\begin{figure}[H]
\begin{adjustwidth}{-\extralength}{0cm}

   \centering
    \includegraphics[width=5.8cm]{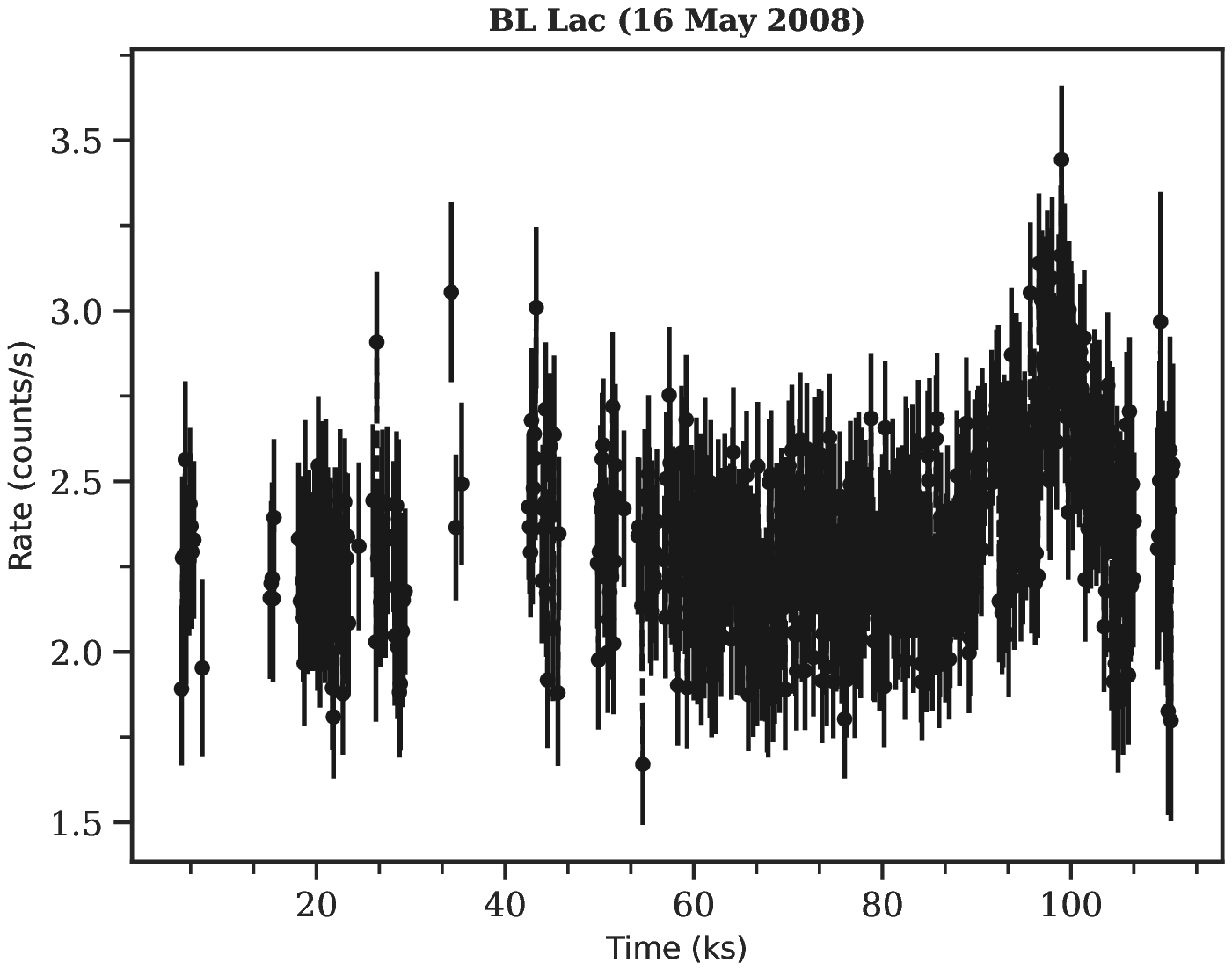}
    \includegraphics[width=5.8cm]{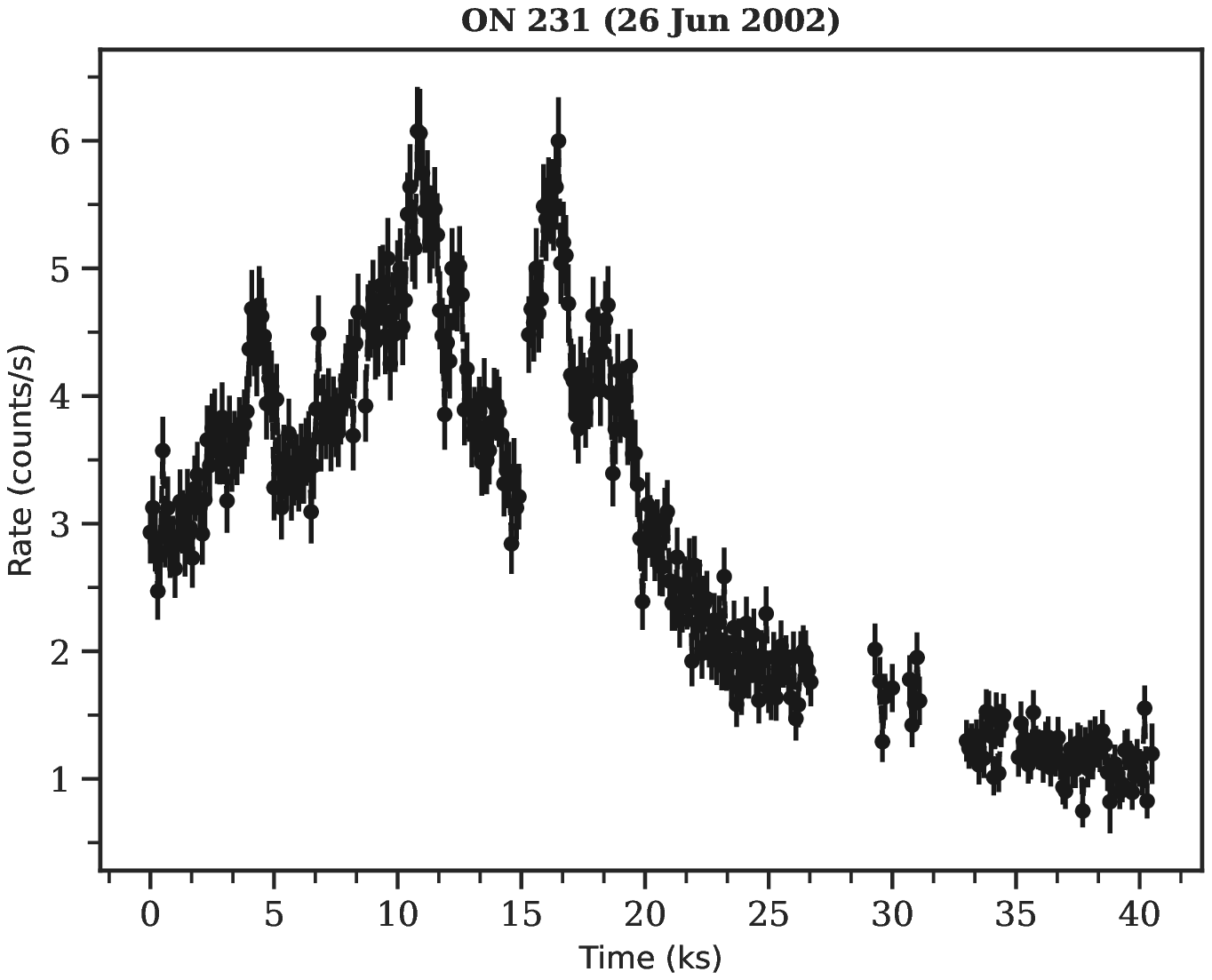}
    \includegraphics[width=5.8cm]{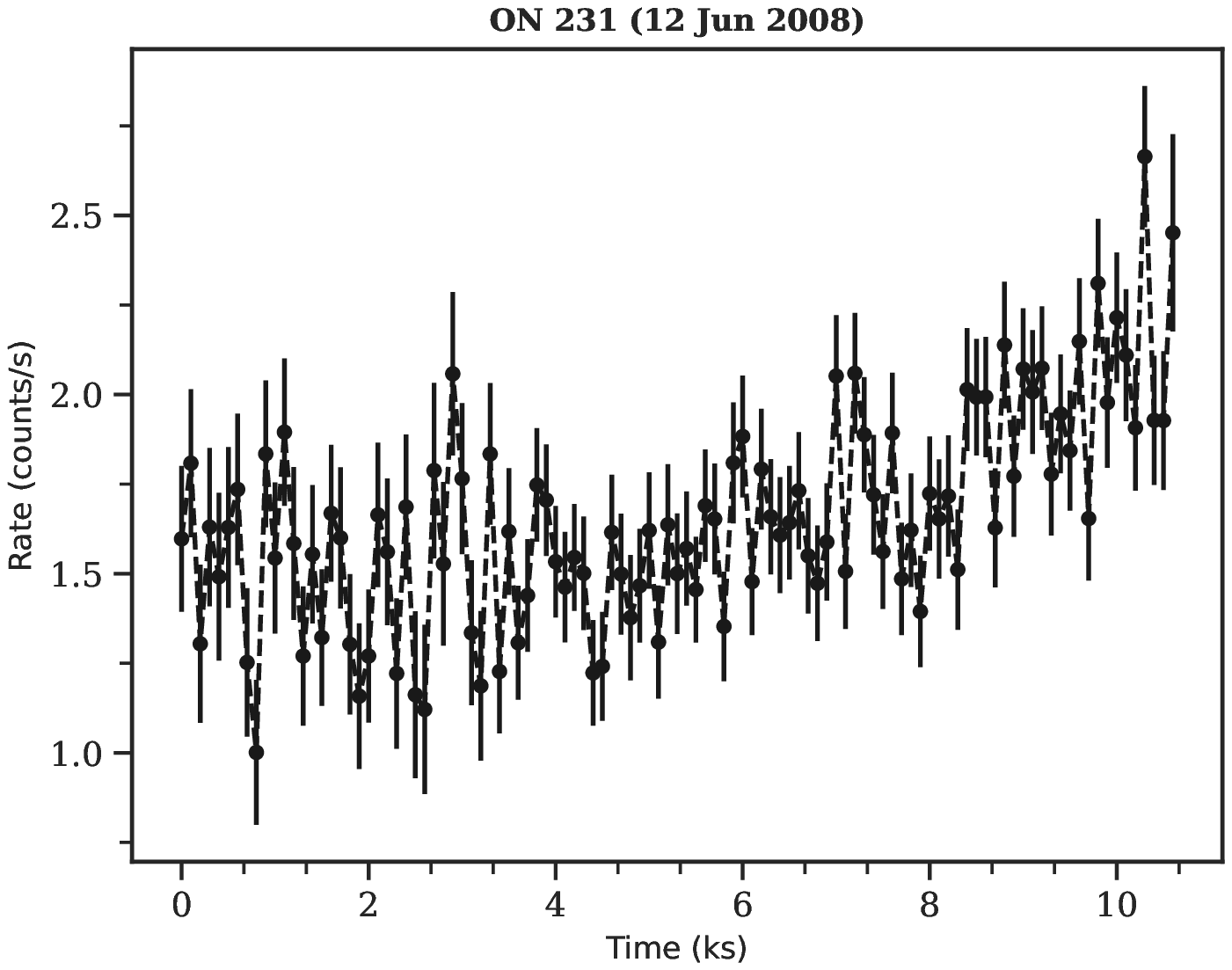}
    \includegraphics[width=5.8cm]{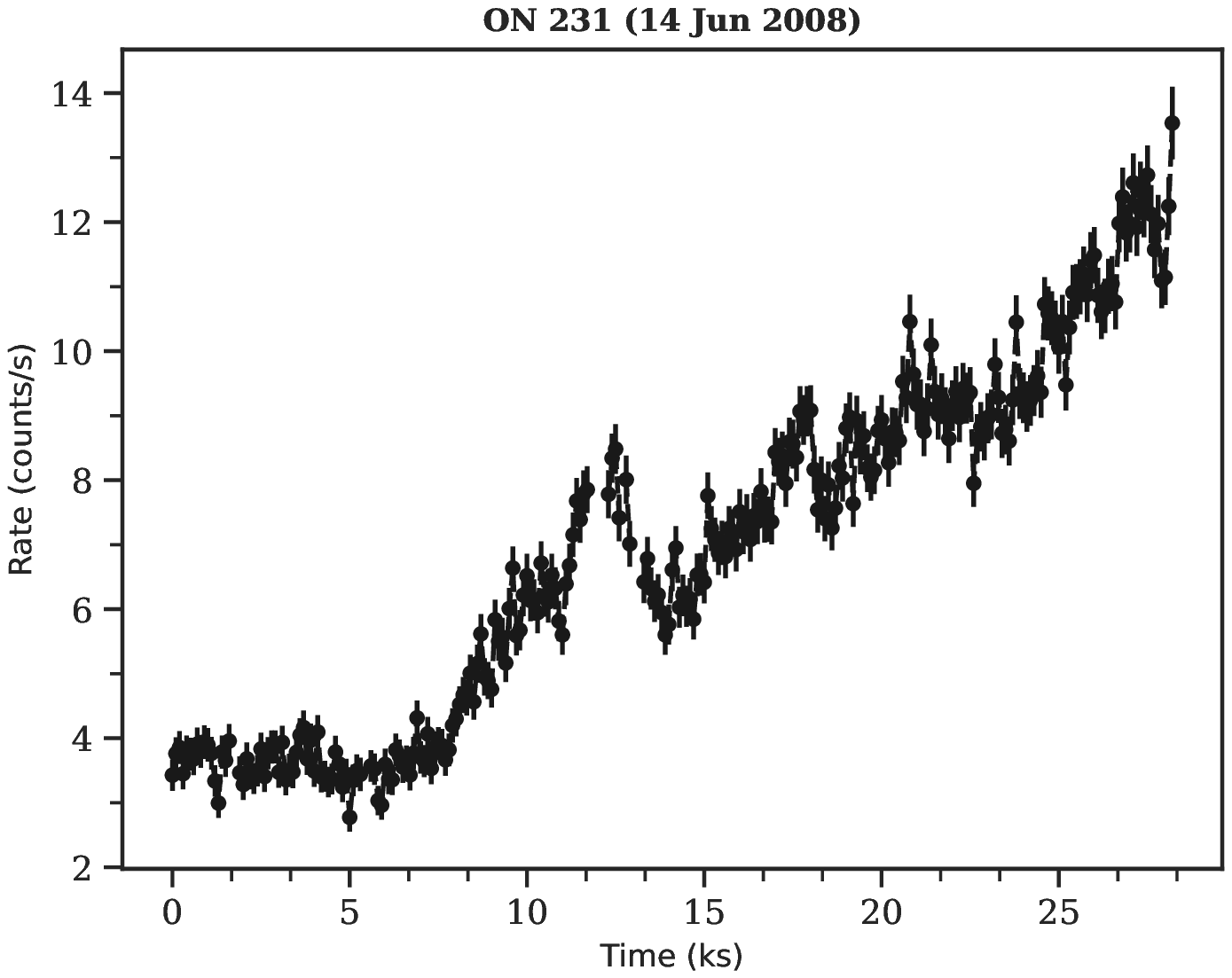}
    \includegraphics[width=5.8cm]{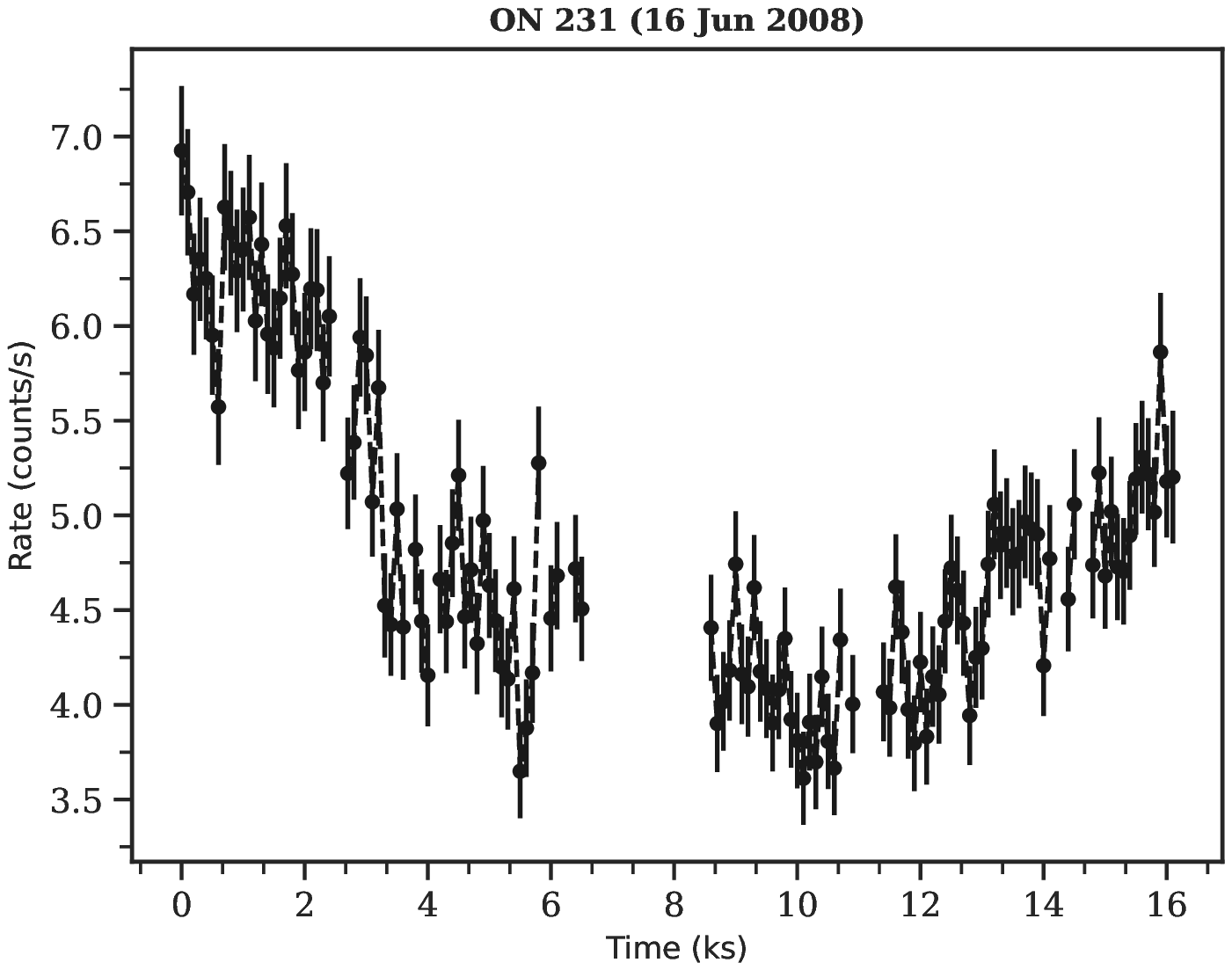}
    \includegraphics[width=5.8cm]{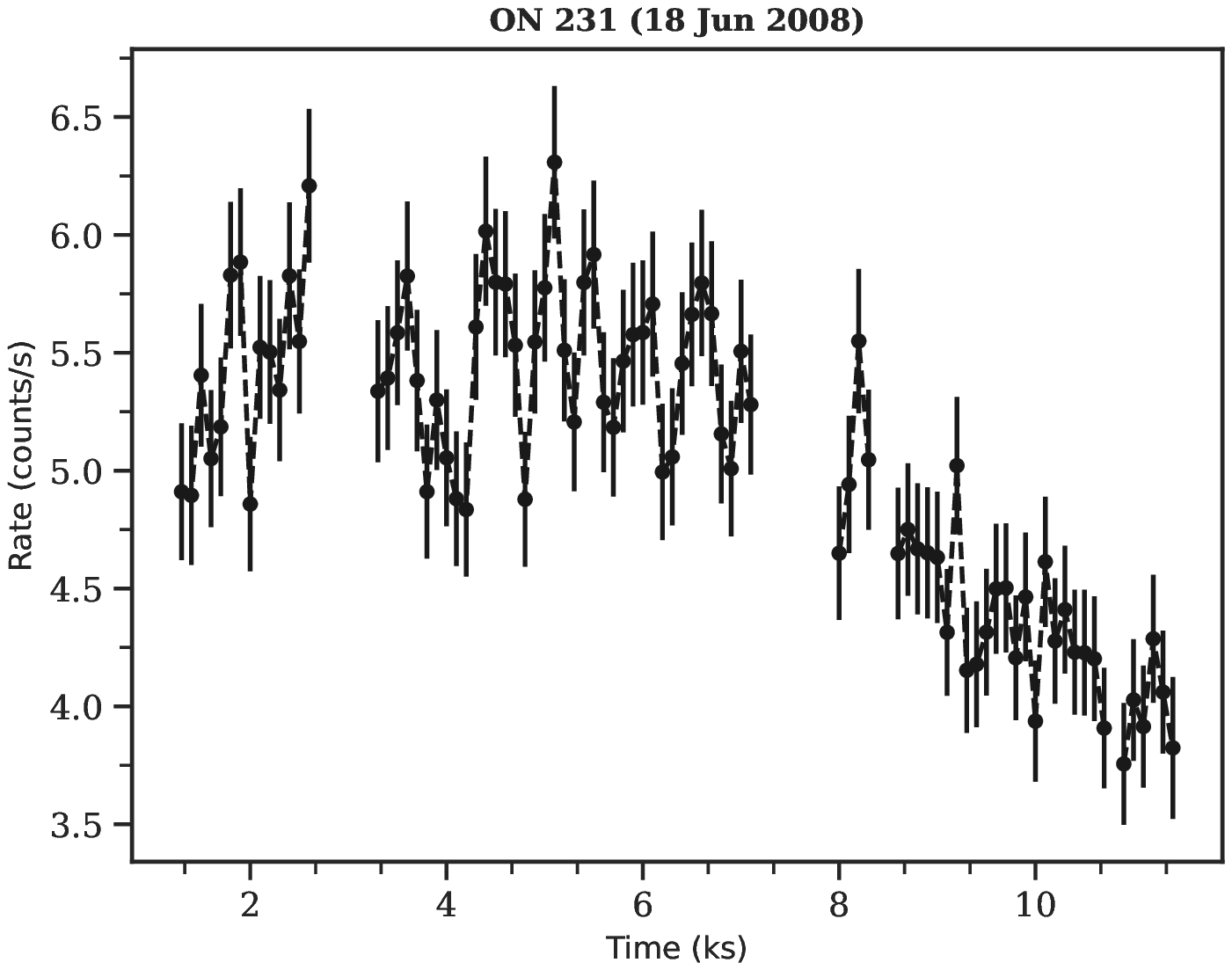}
    \includegraphics[width=5.8cm]{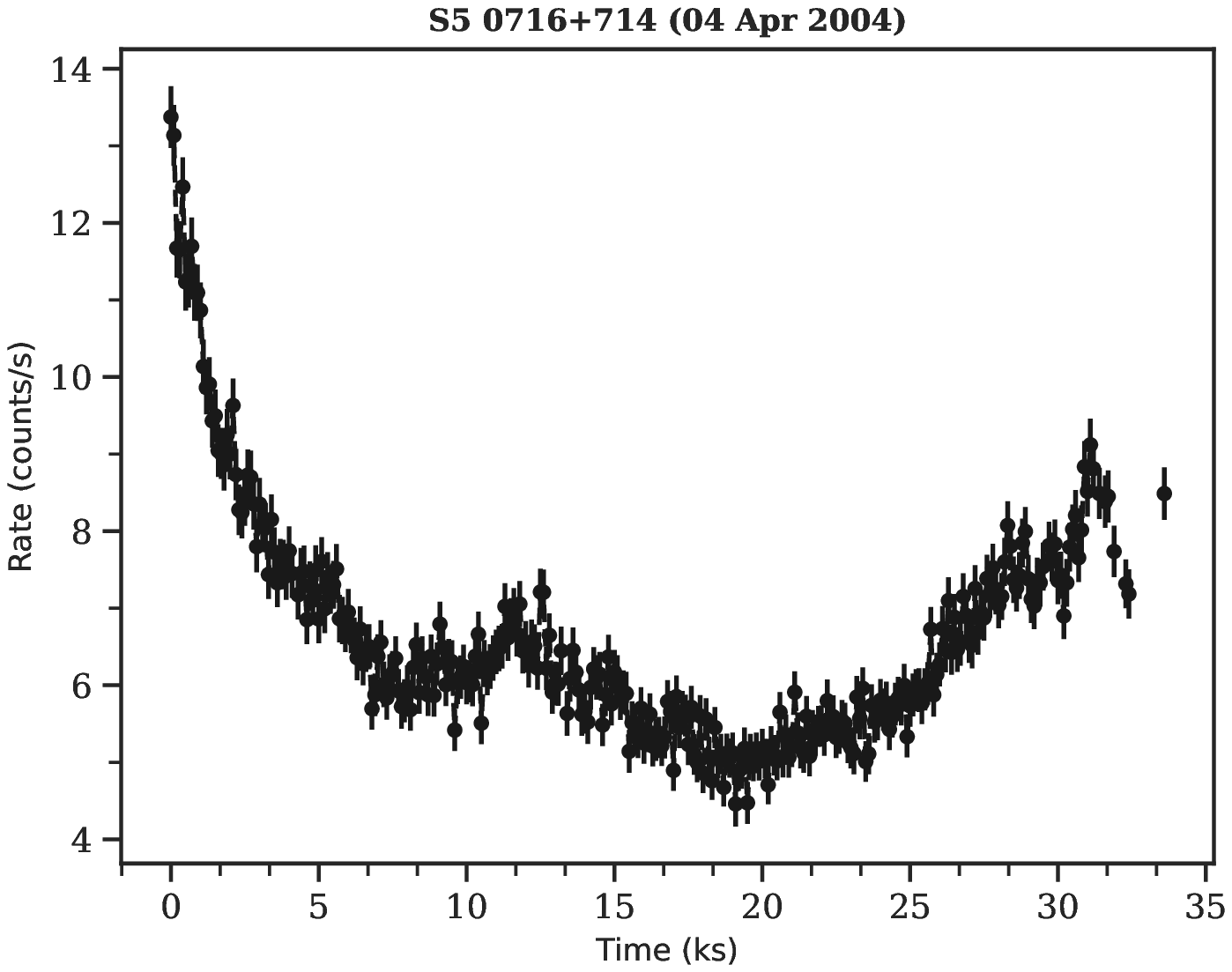}
    \includegraphics[width=5.8cm]{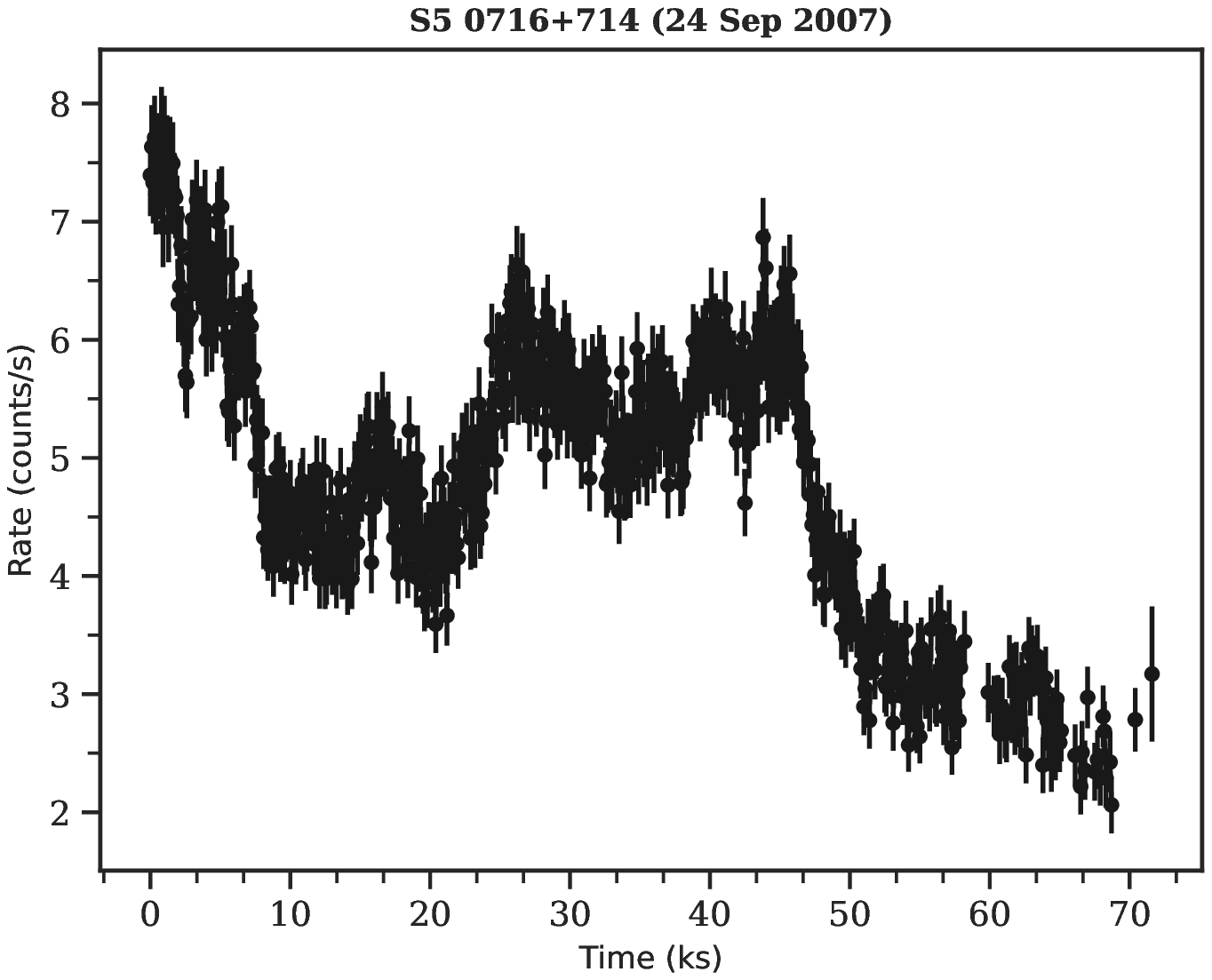}
    \includegraphics[width=5.8cm]{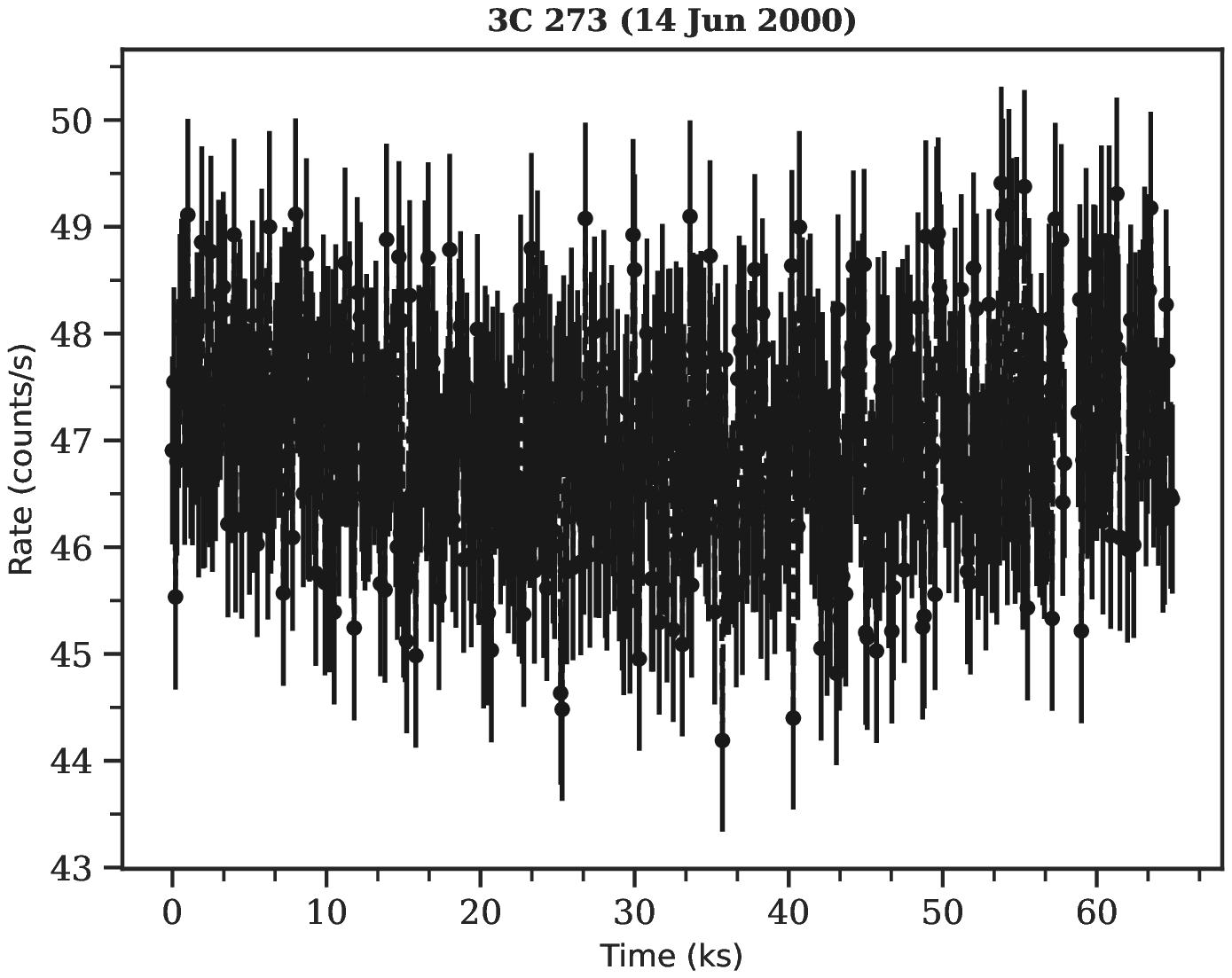}

    \end{adjustwidth}
  \caption{\textit{Cont.}}
\end{figure}
\begin{figure}[H]\ContinuedFloat
\begin{adjustwidth}{-\extralength}{0cm}\centering

    \includegraphics[width=5.8cm]{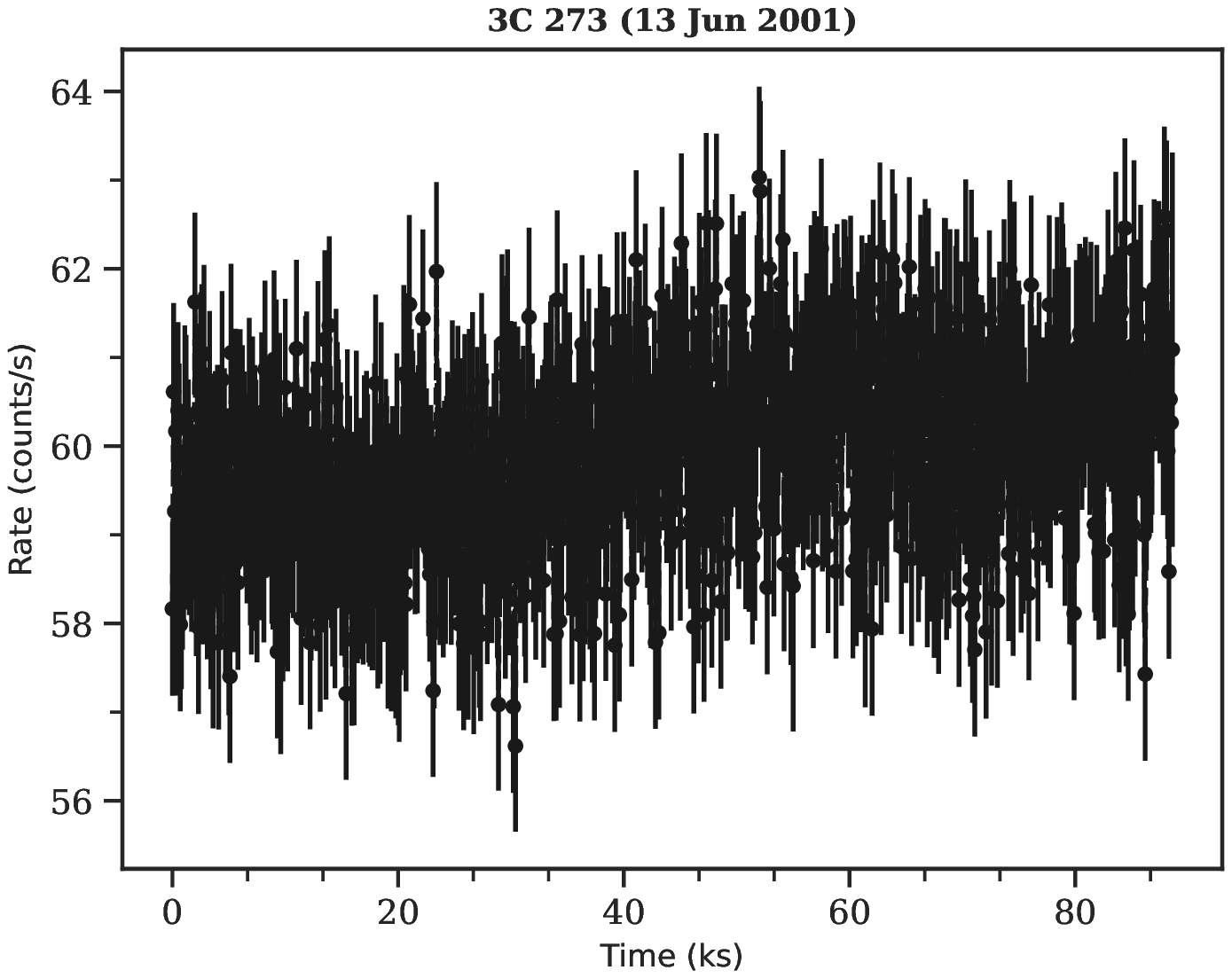}
    \includegraphics[width=5.8cm]{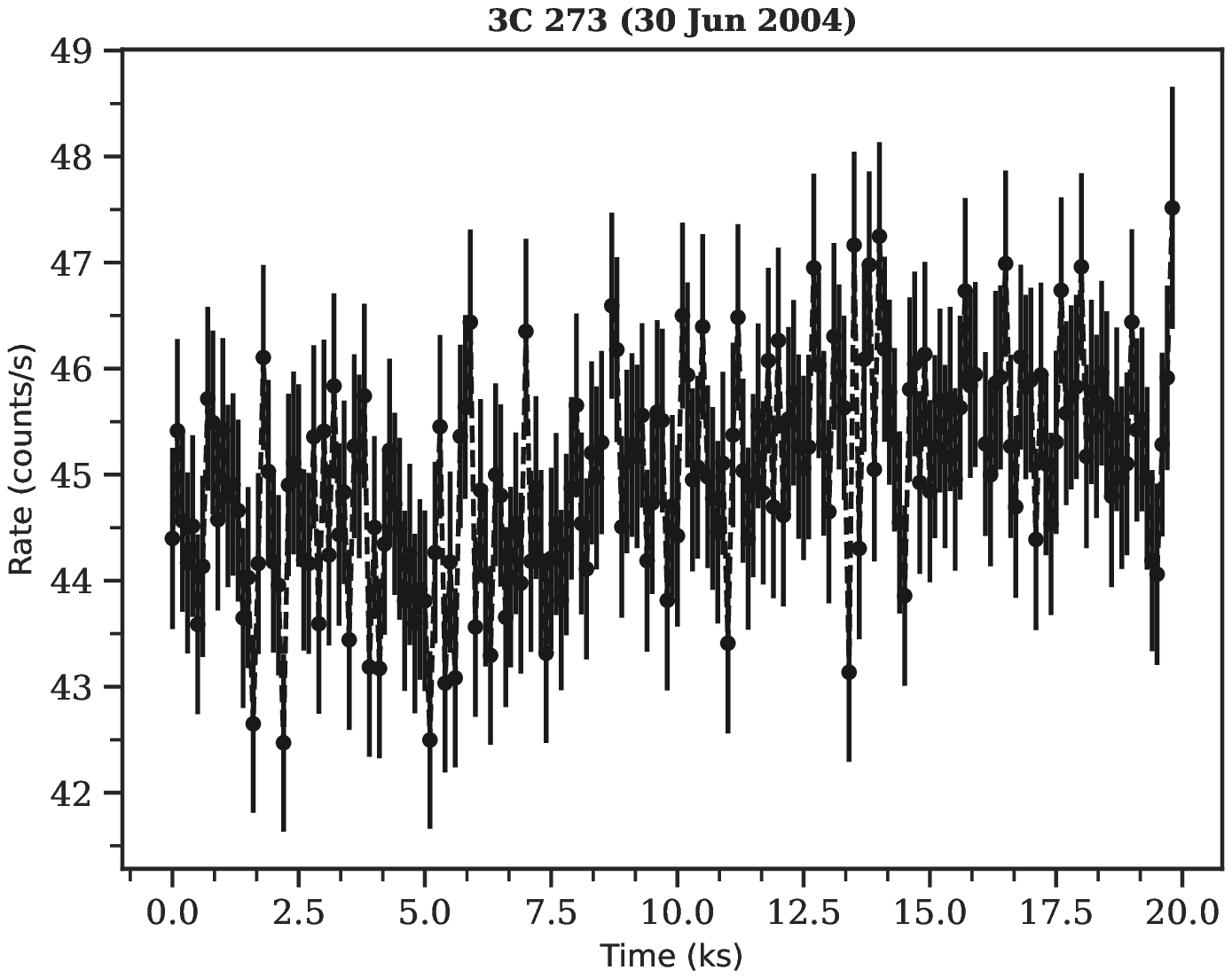}
    \includegraphics[width=5.8cm]{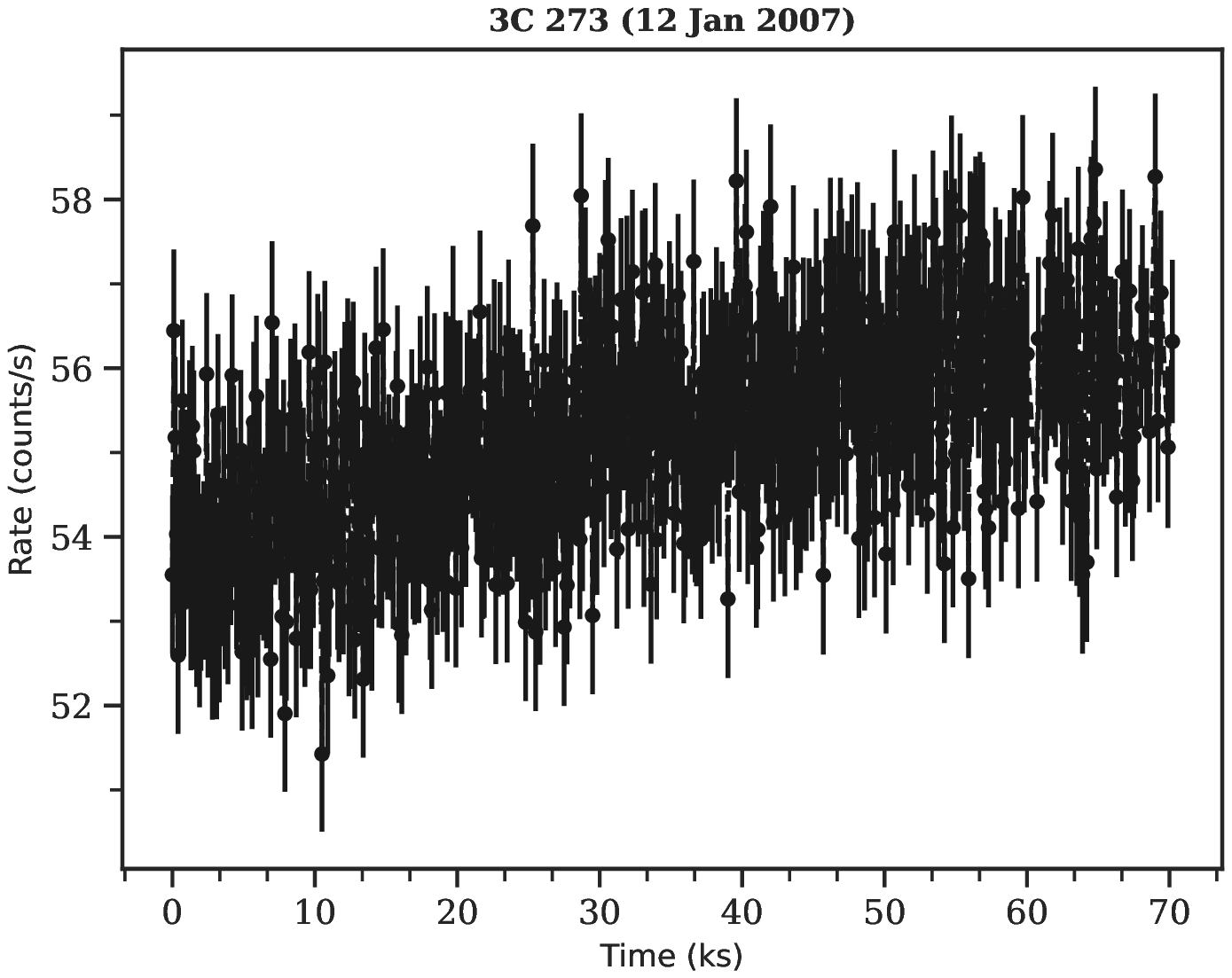}
    \includegraphics[width=5.8cm]{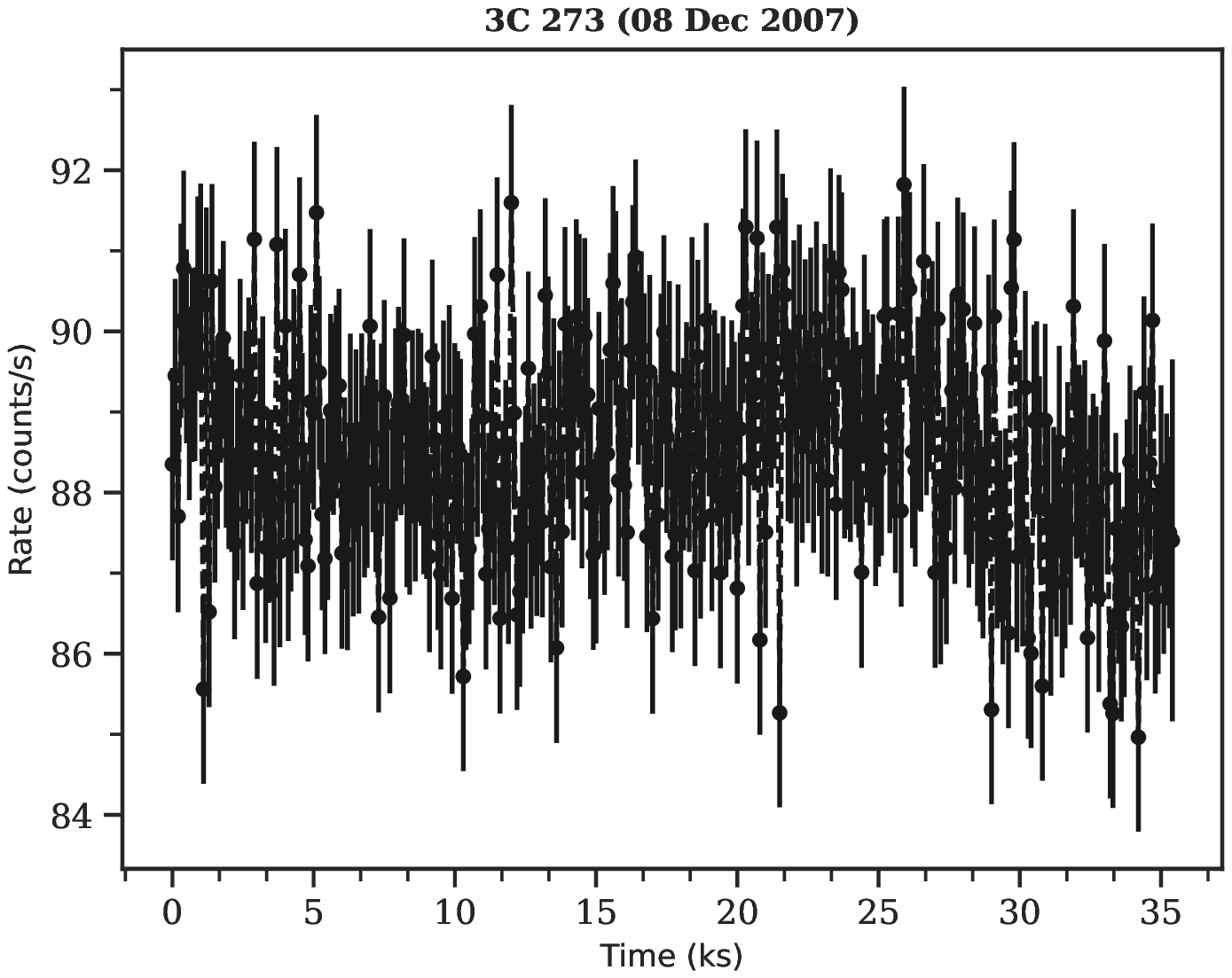}
    \includegraphics[width=5.8cm]{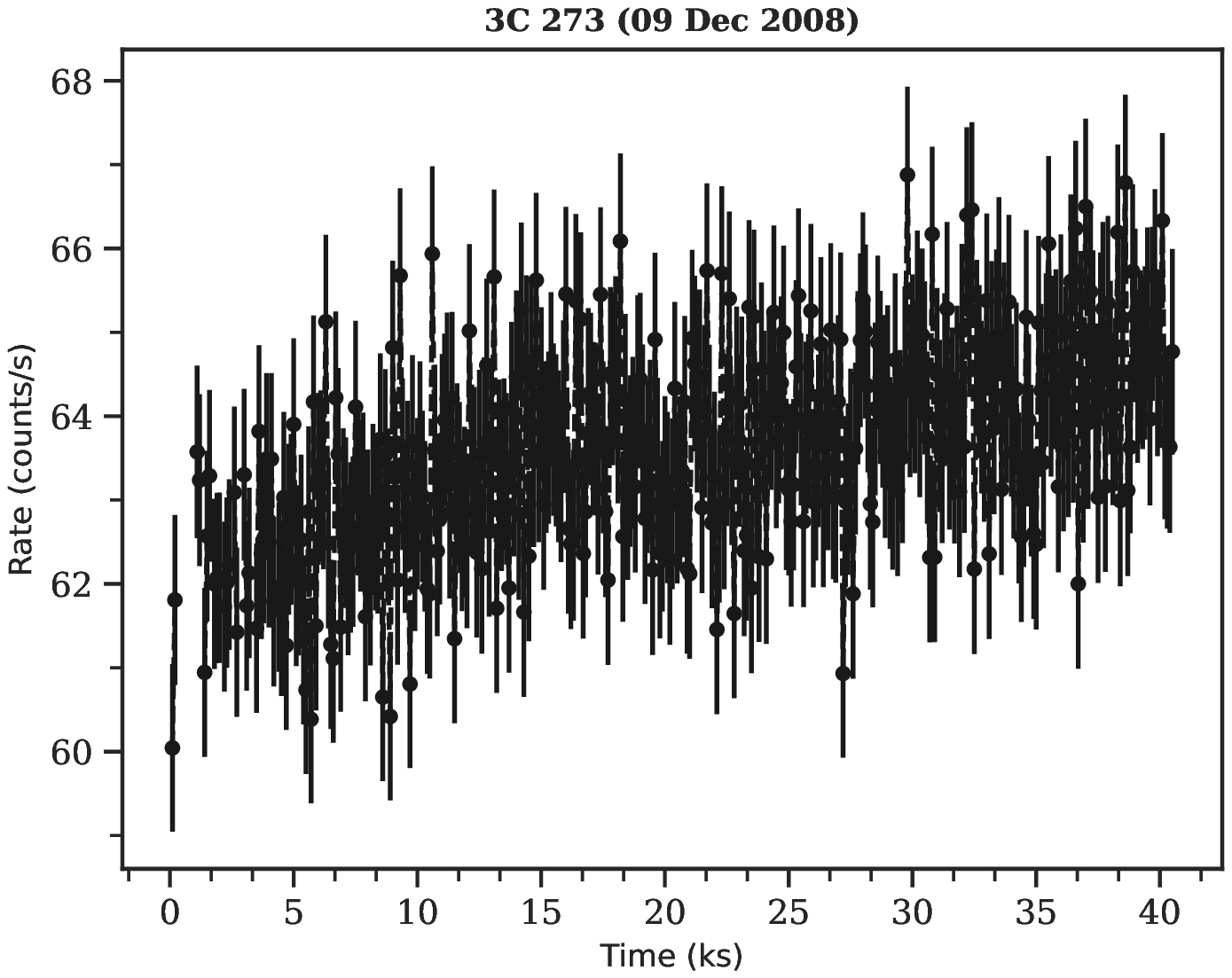}
    \includegraphics[width=5.8cm]{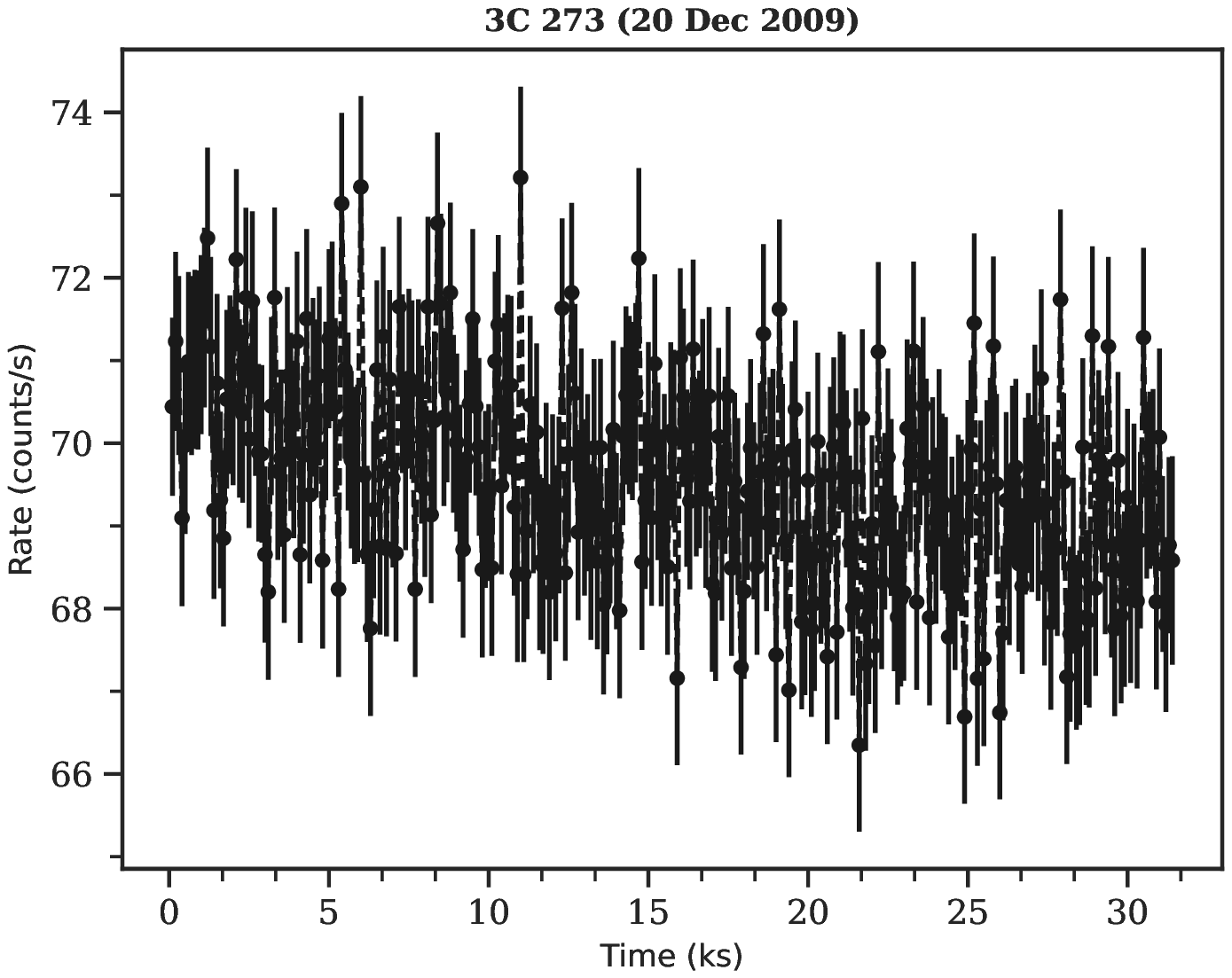}
        \includegraphics[width=5.8cm]{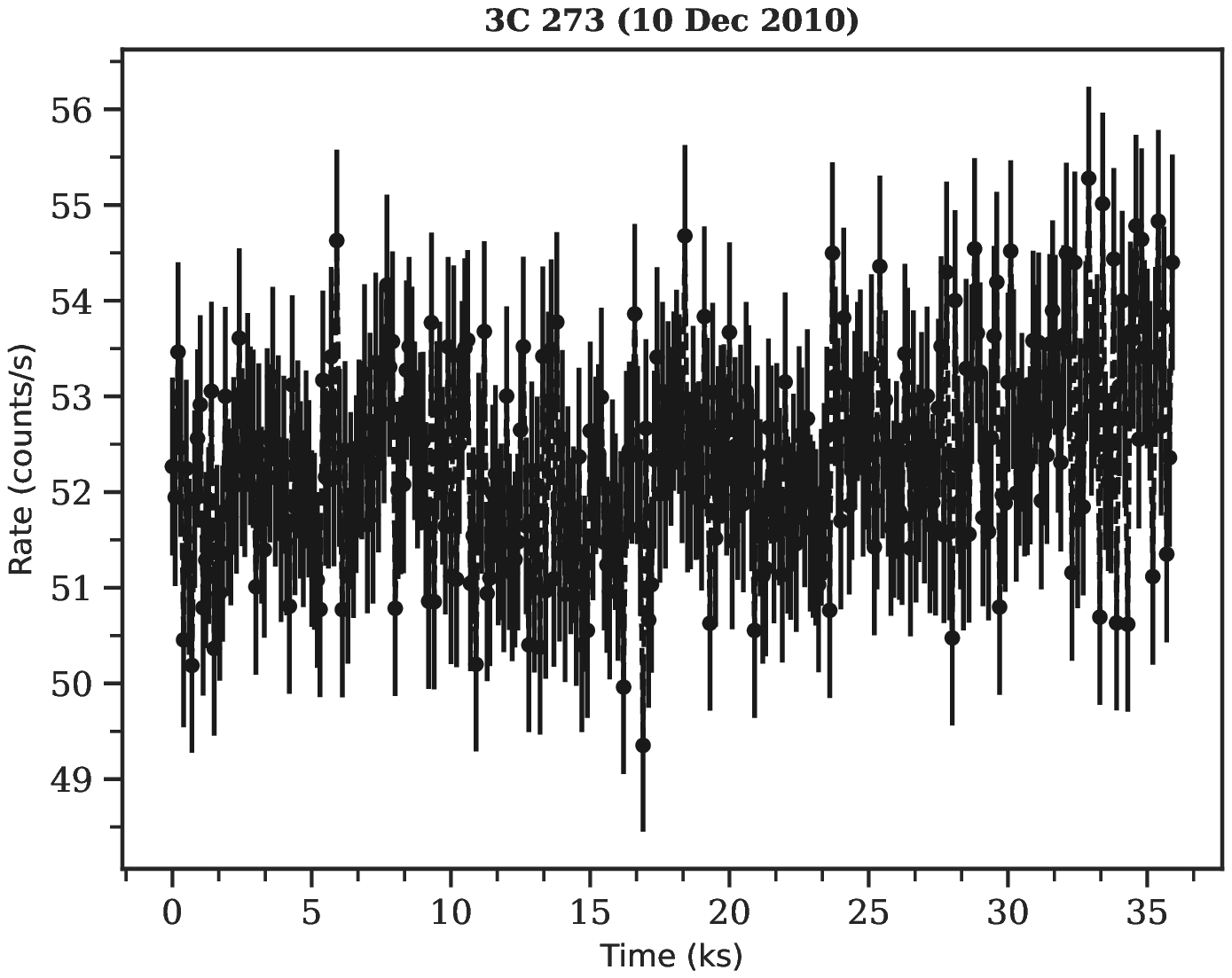}
    \includegraphics[width=5.8cm]{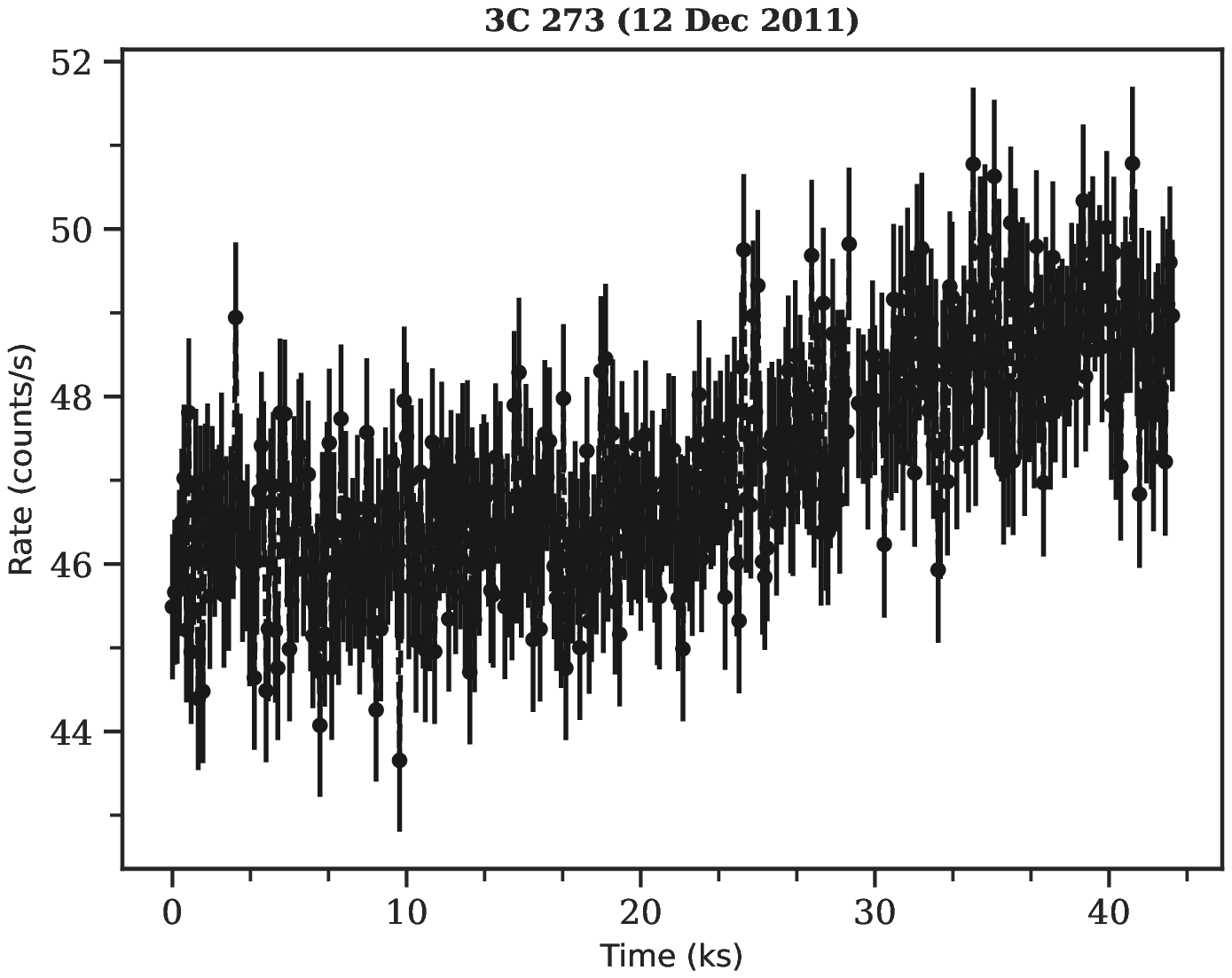}
    \includegraphics[width=5.8cm]{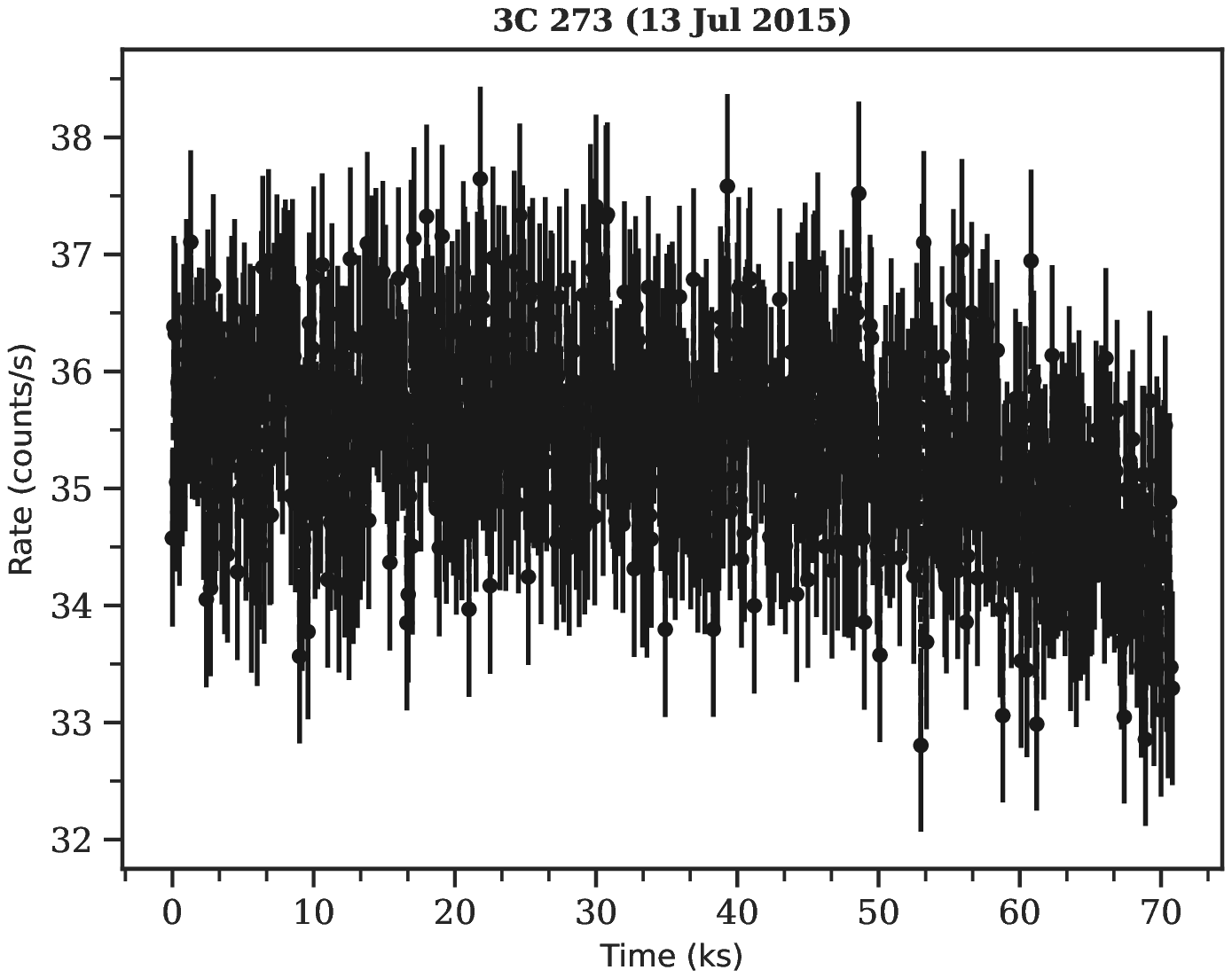}
    
    \includegraphics[width=5.8cm]{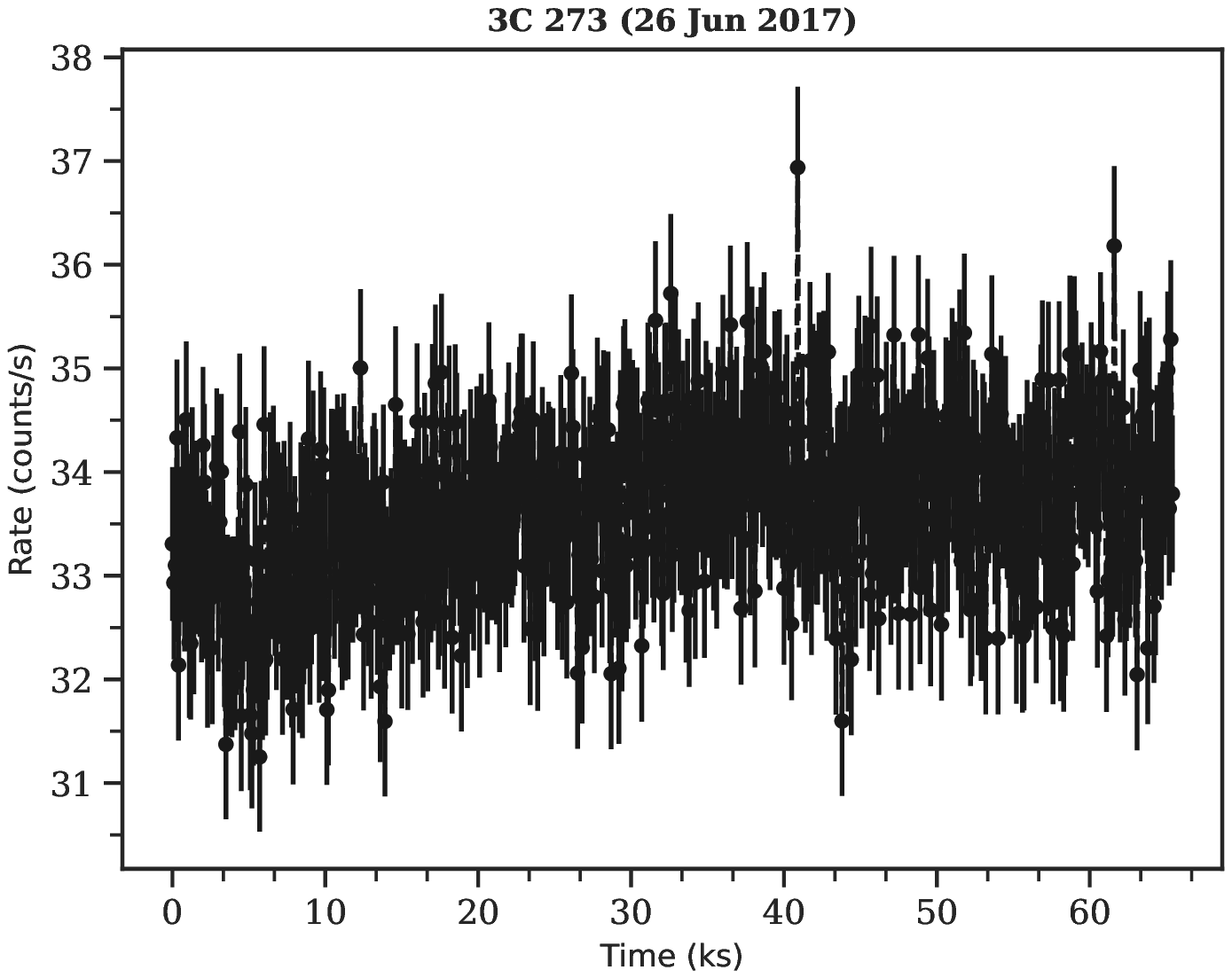}
    \includegraphics[width=5.8cm]{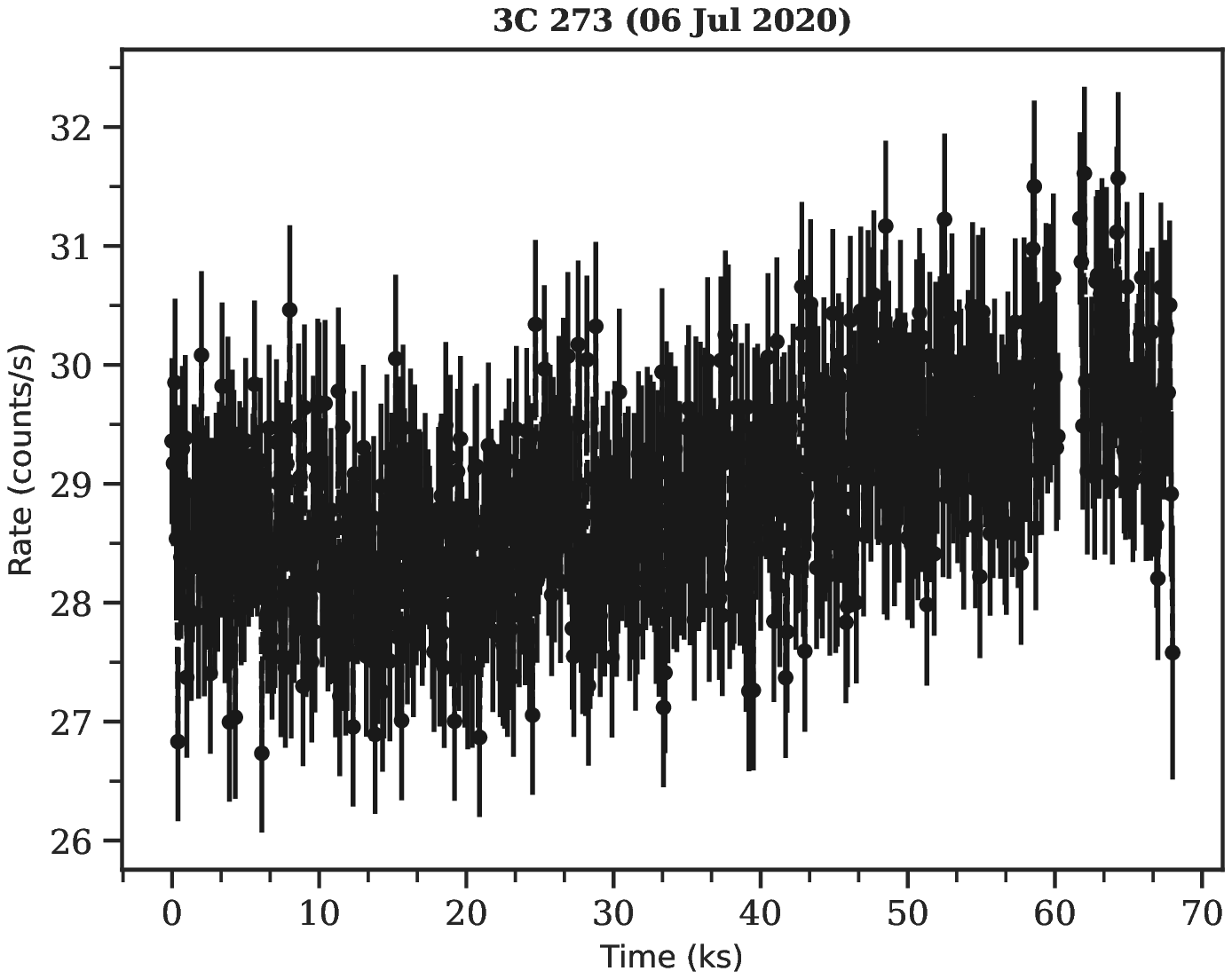}
    \includegraphics[width=5.8cm]{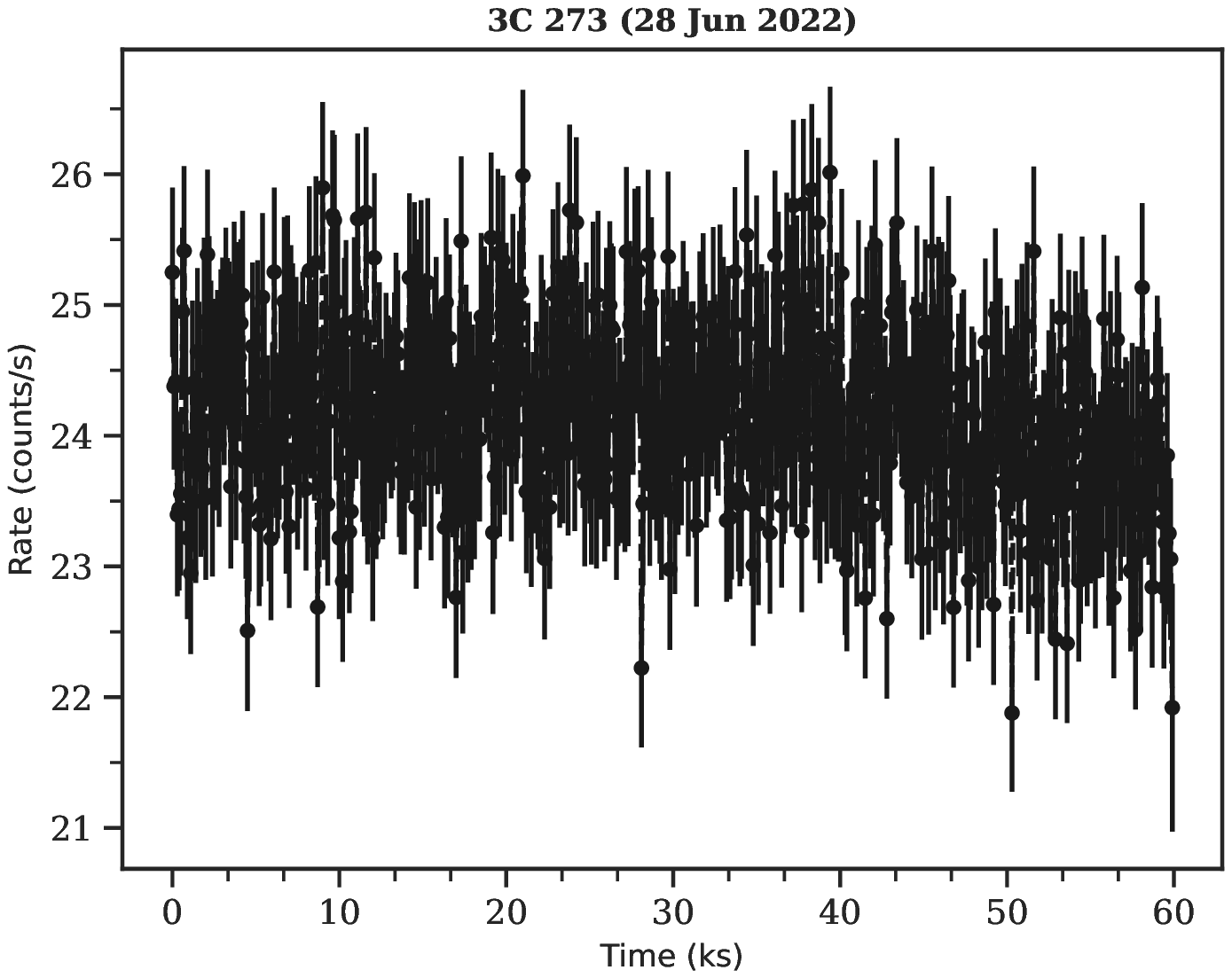}
    \includegraphics[width=5.8cm]{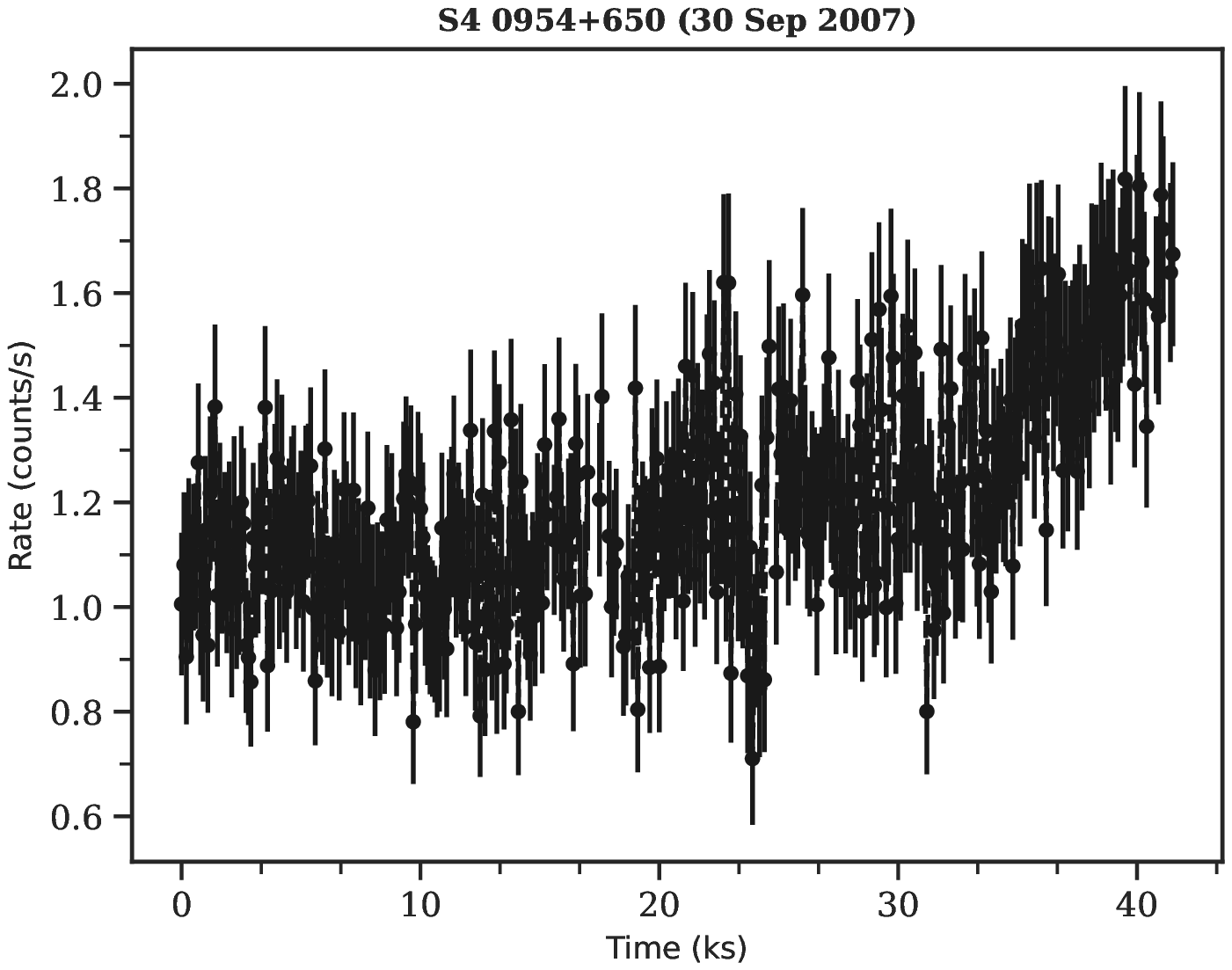}
    \includegraphics[width=5.8cm]{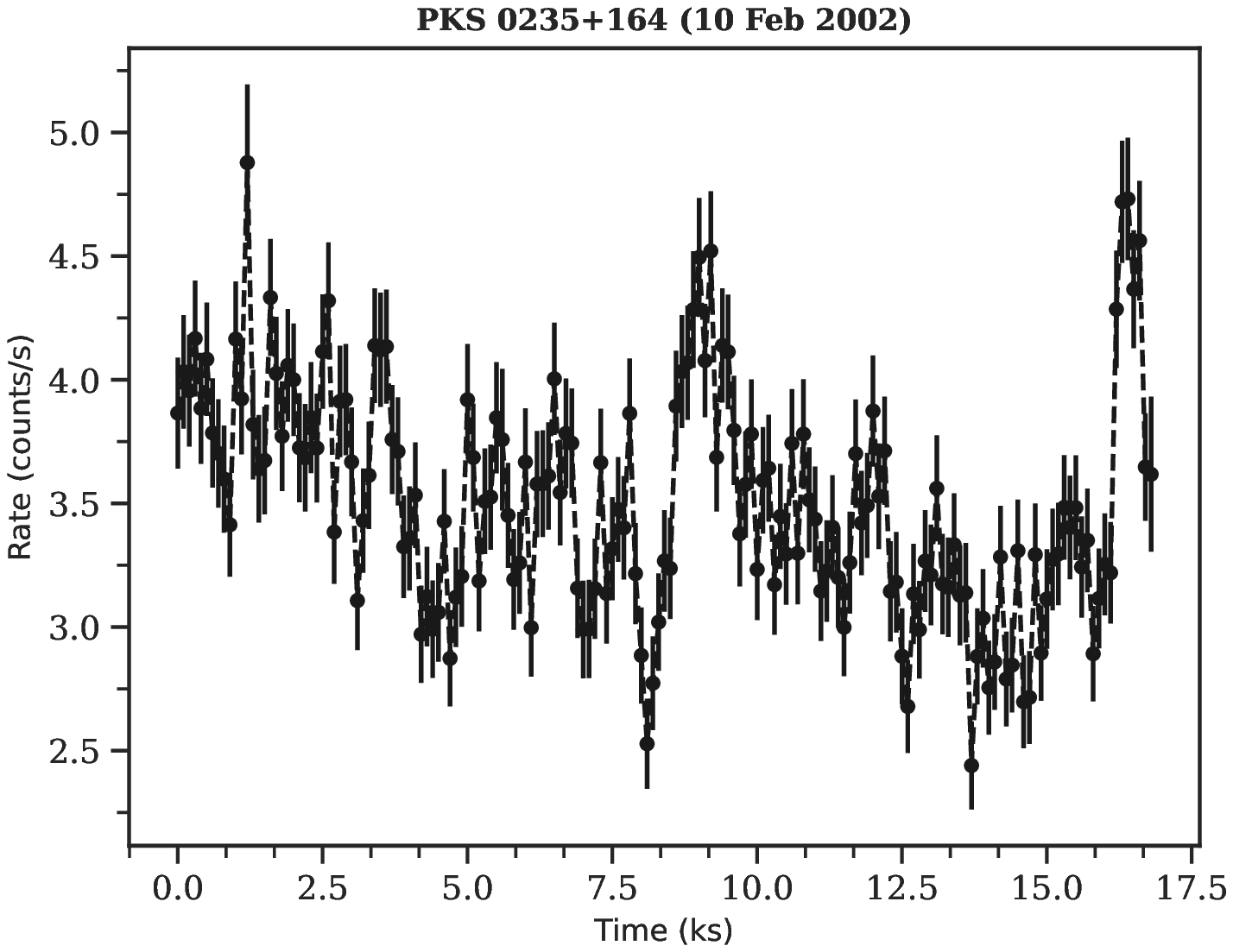}
    \includegraphics[width=5.8cm]{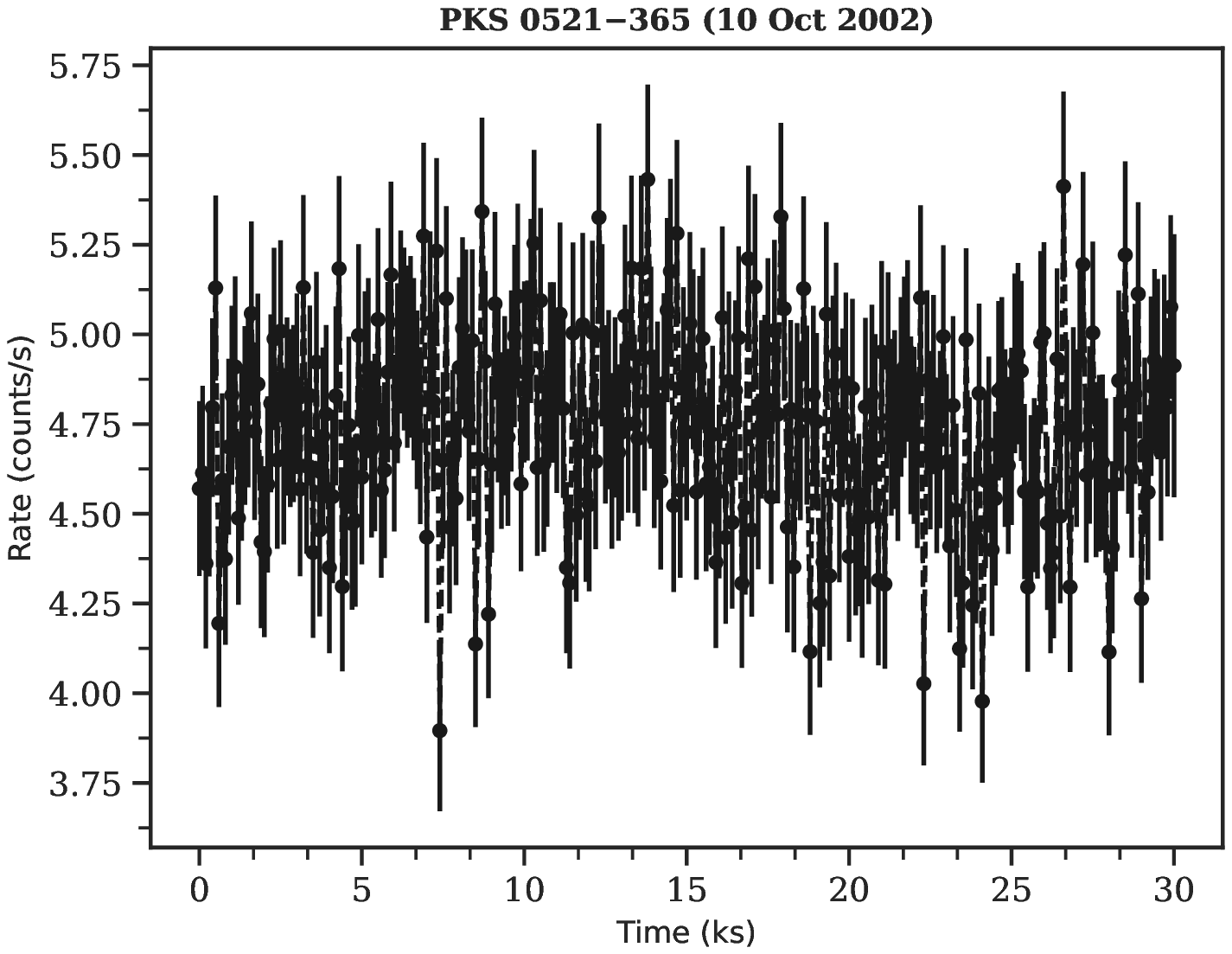}
\end{adjustwidth}
  \caption{{Light curves} of \mbox{XMM-Newton} observations of blazars with a time binning of 100 s.}
\label{fig:lightcurve}  
\end{figure}

\begin{figure}[H]
\begin{adjustwidth}{-\extralength}{0cm}

    \centering
    \includegraphics[width=6cm]{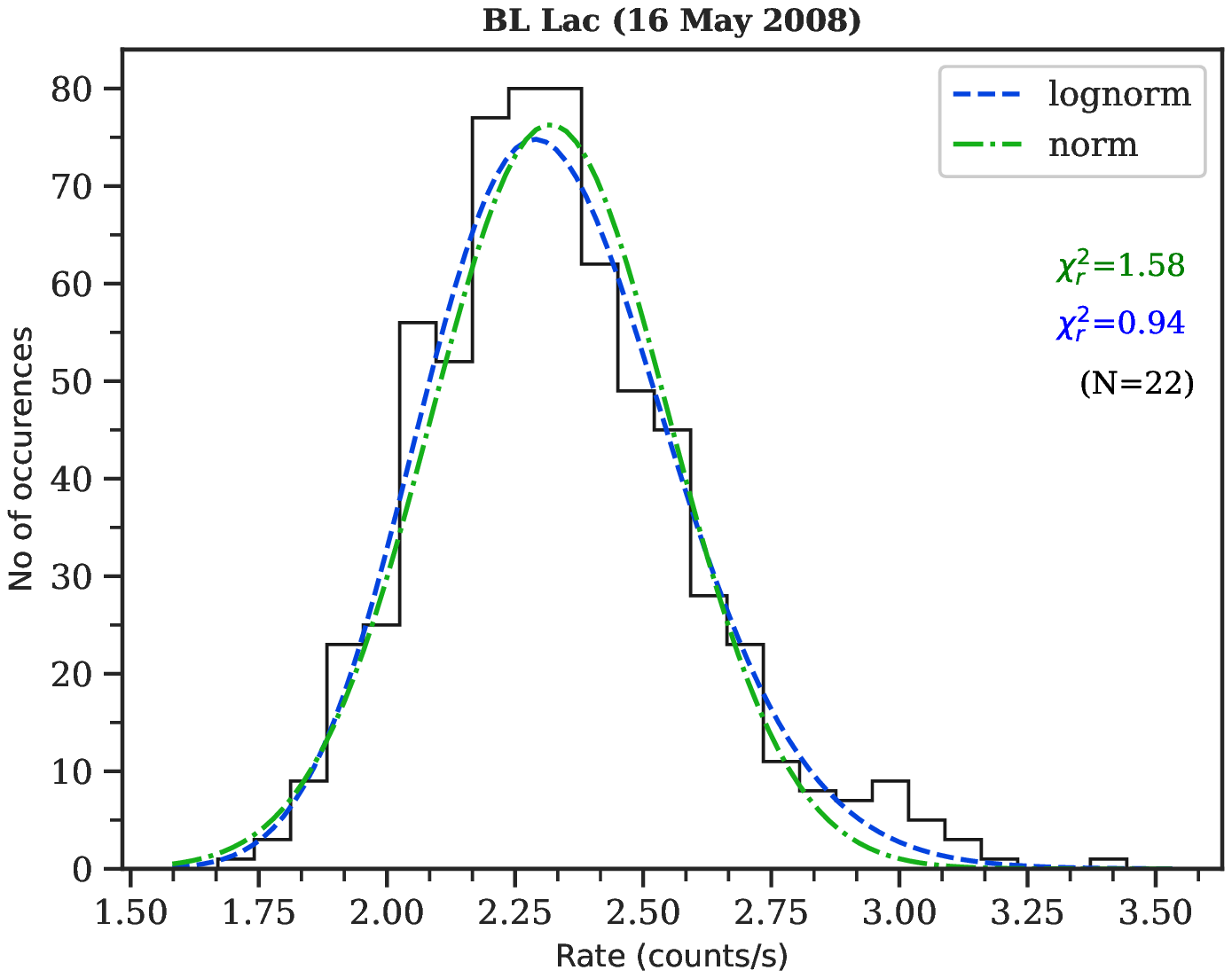}
    \includegraphics[width=6cm]{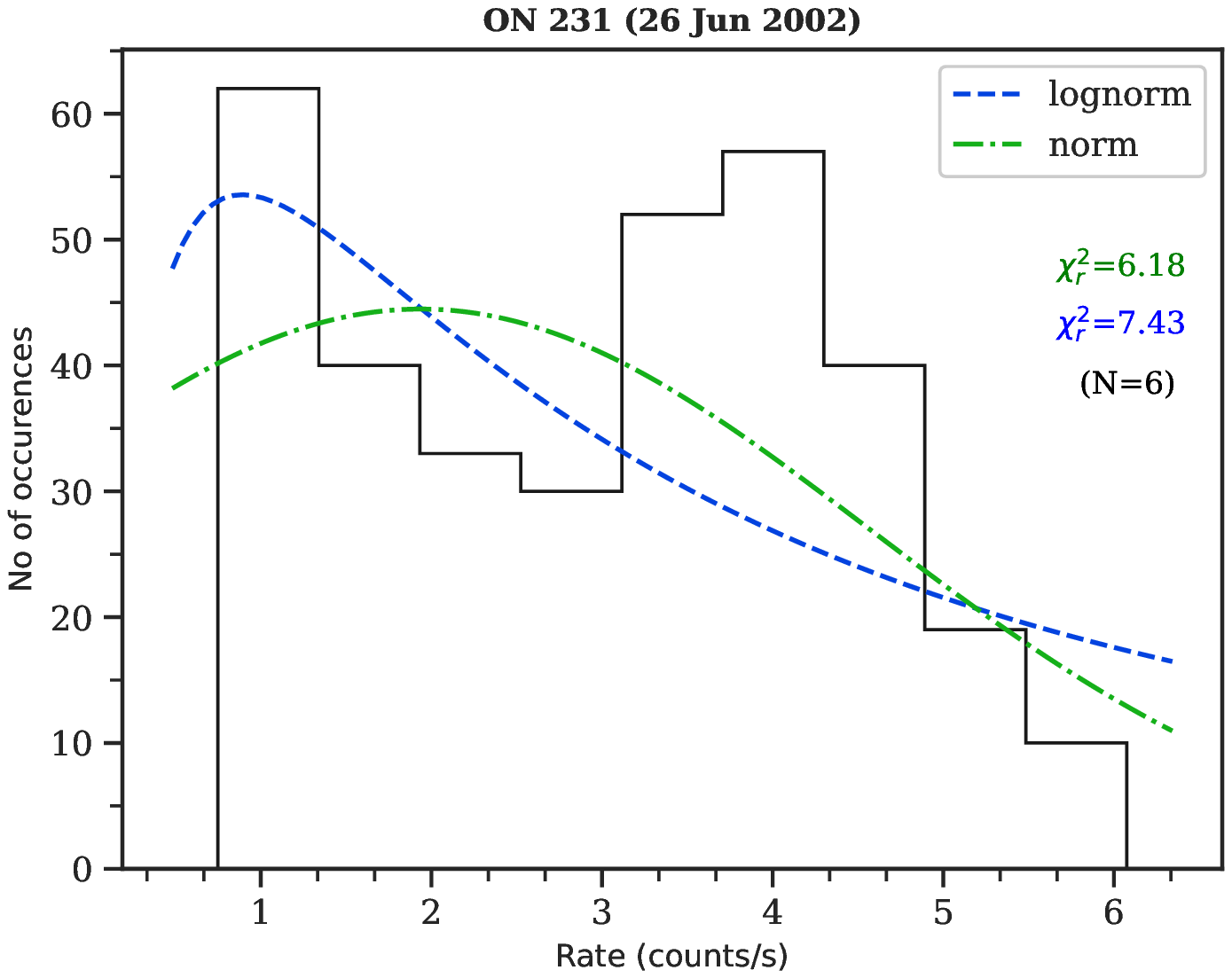}
    \includegraphics[width=6cm]{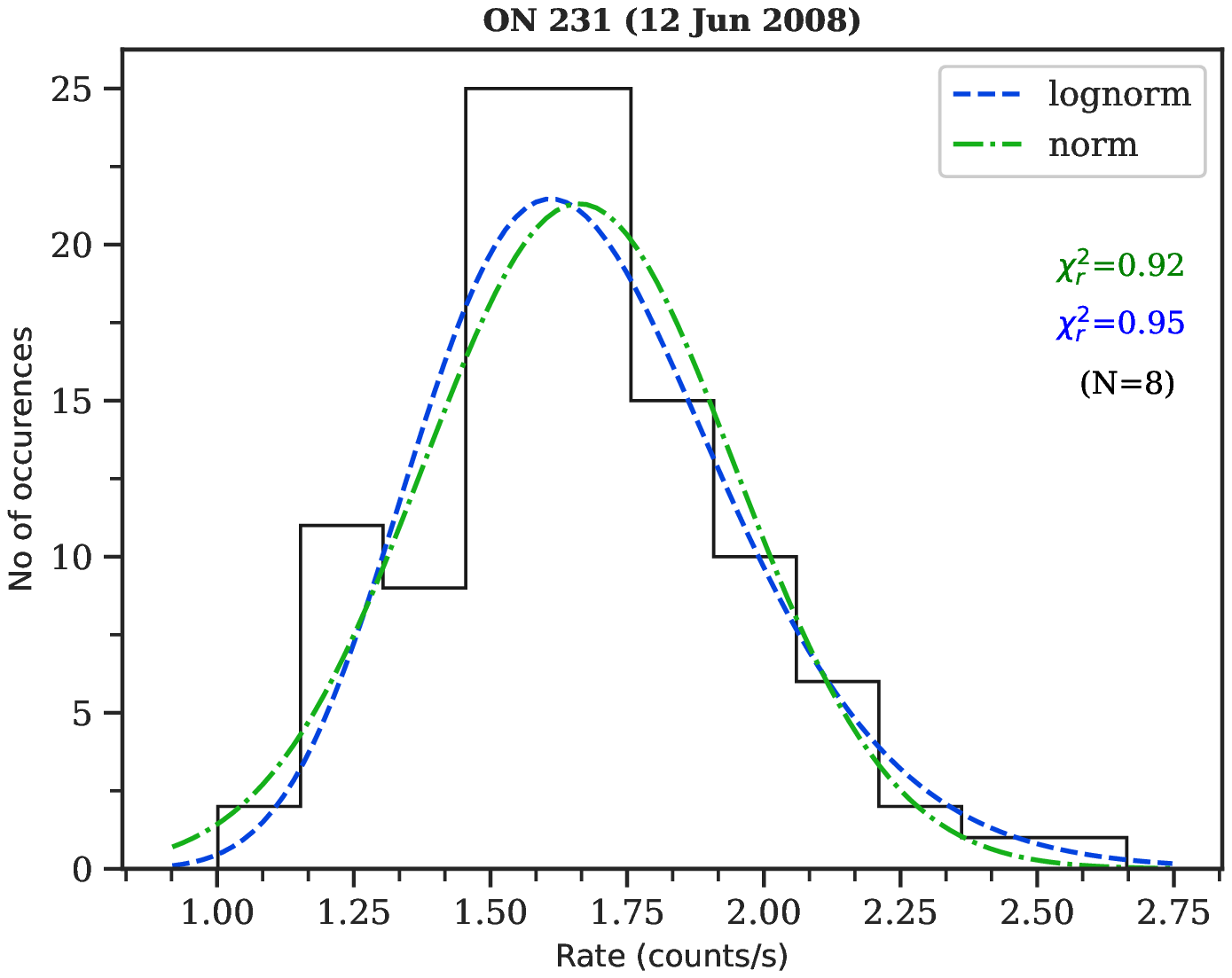}
    \includegraphics[width=6cm]{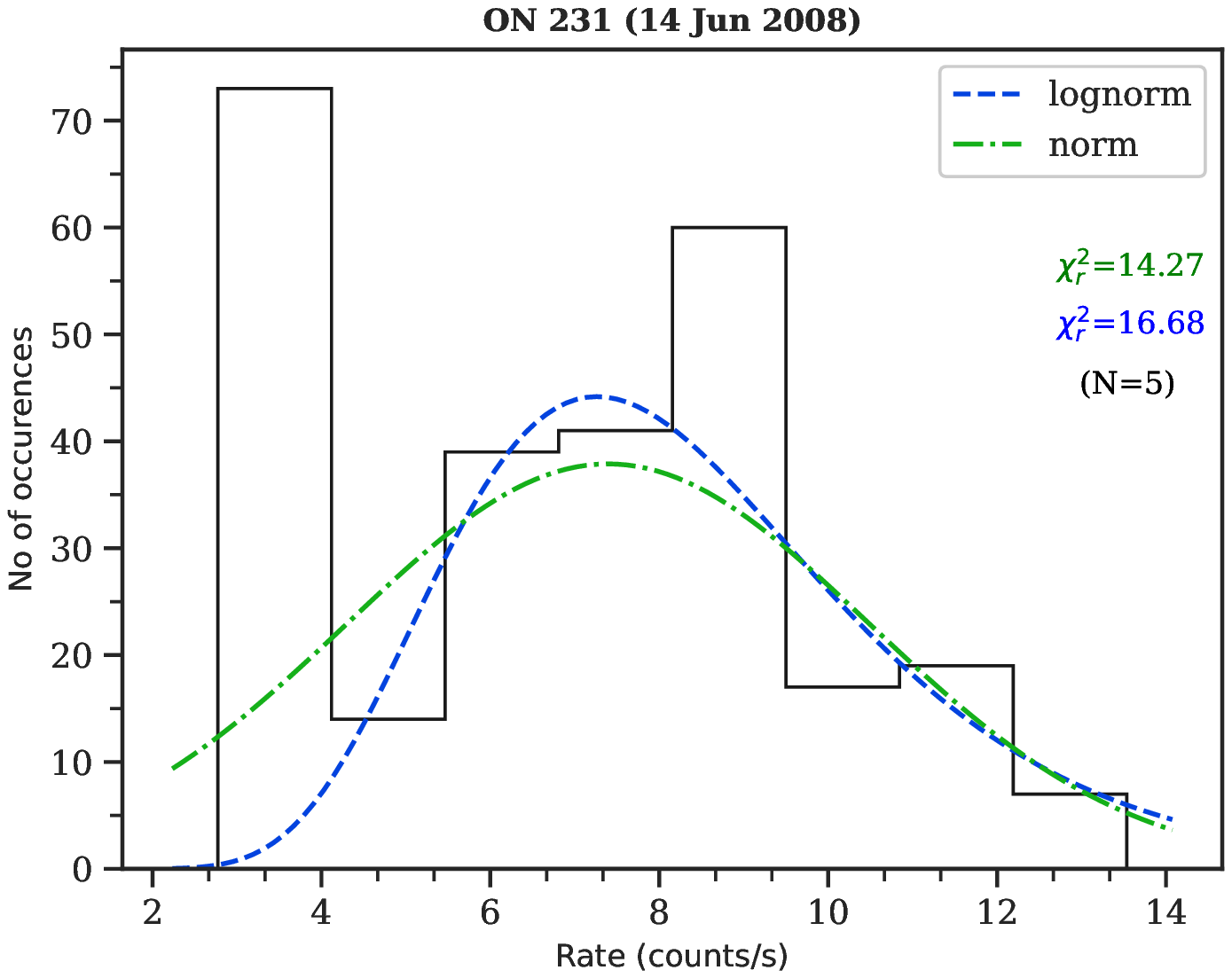}
    \includegraphics[width=6cm]{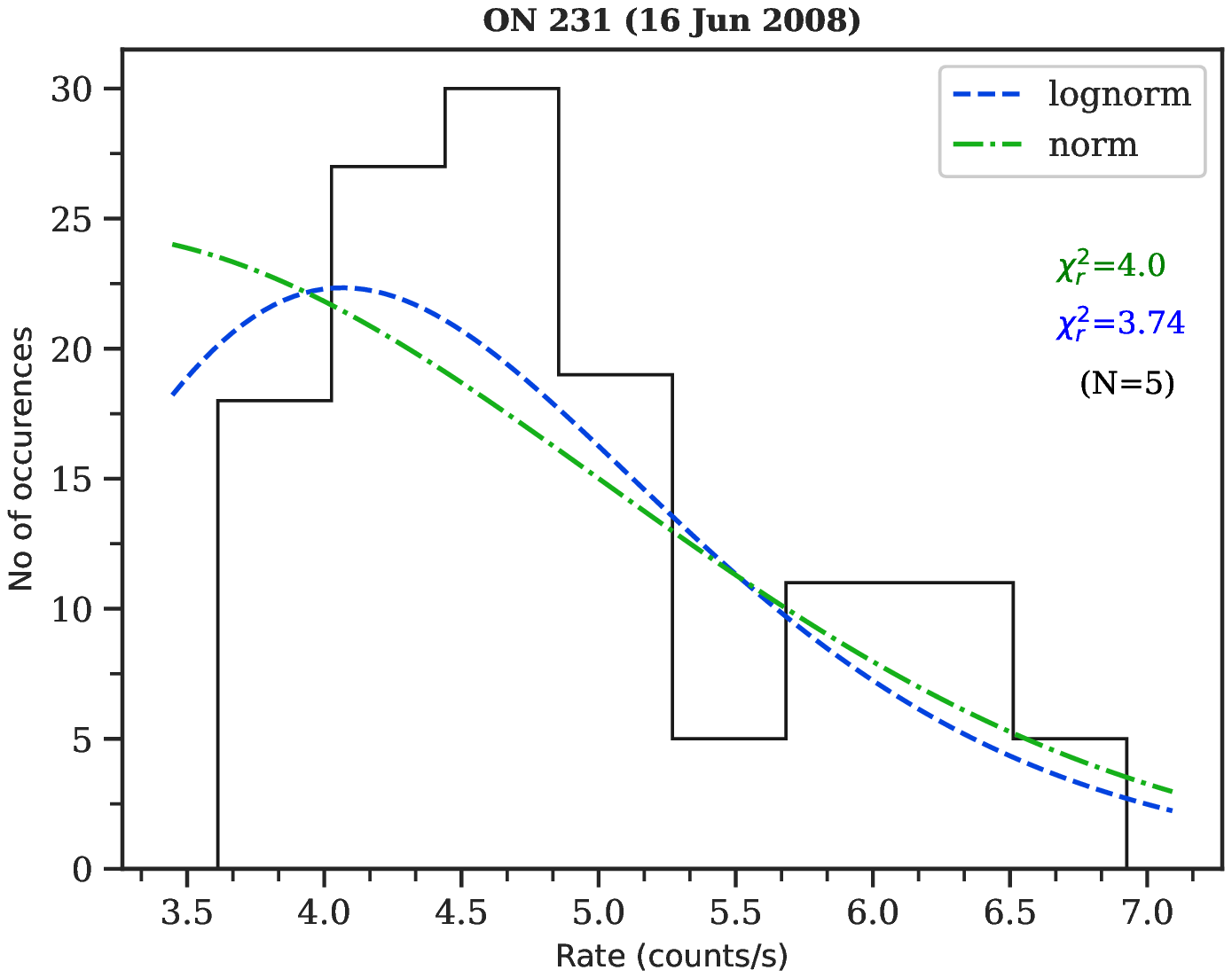}
    \includegraphics[width=6cm]{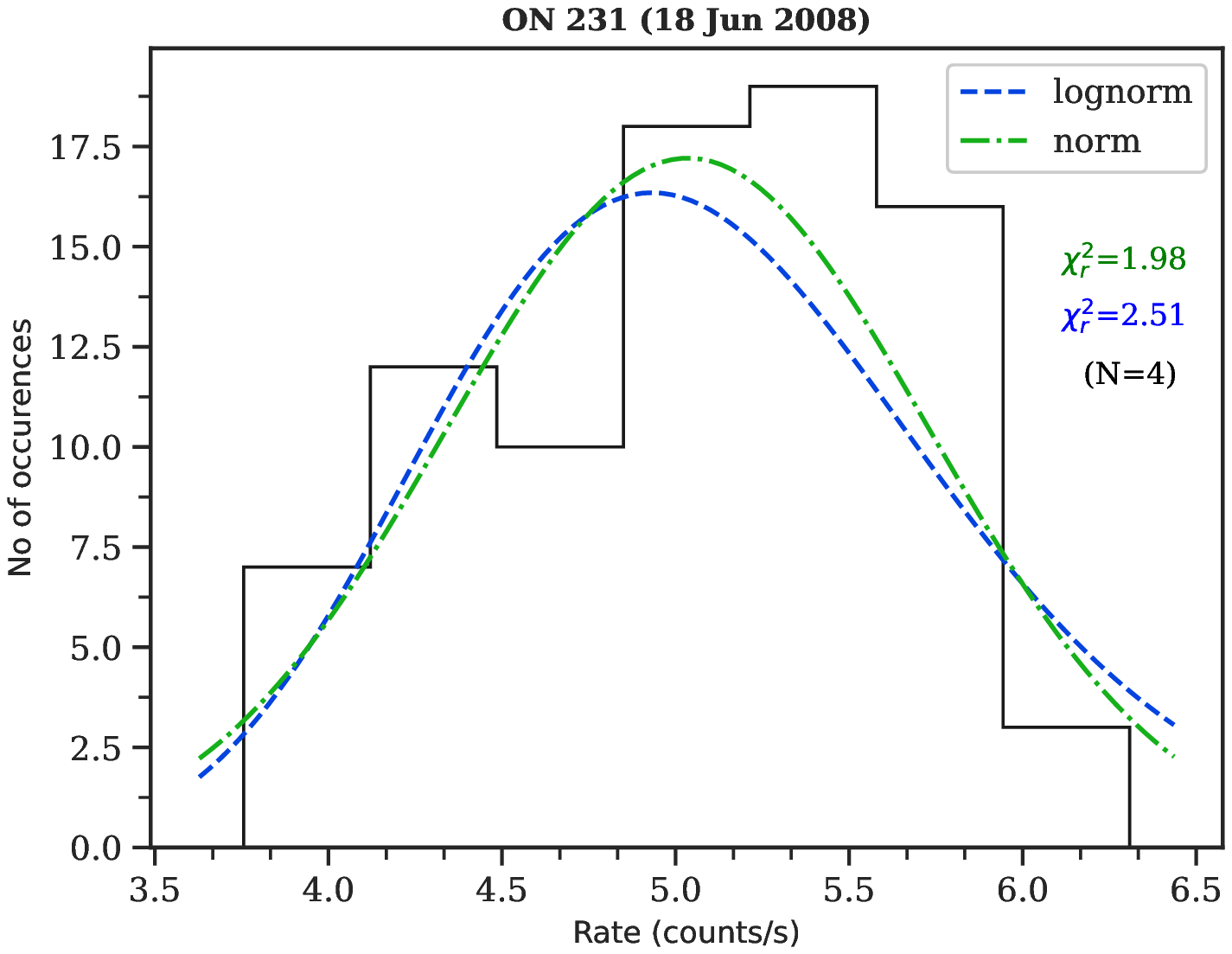}

    \includegraphics[width=6cm]{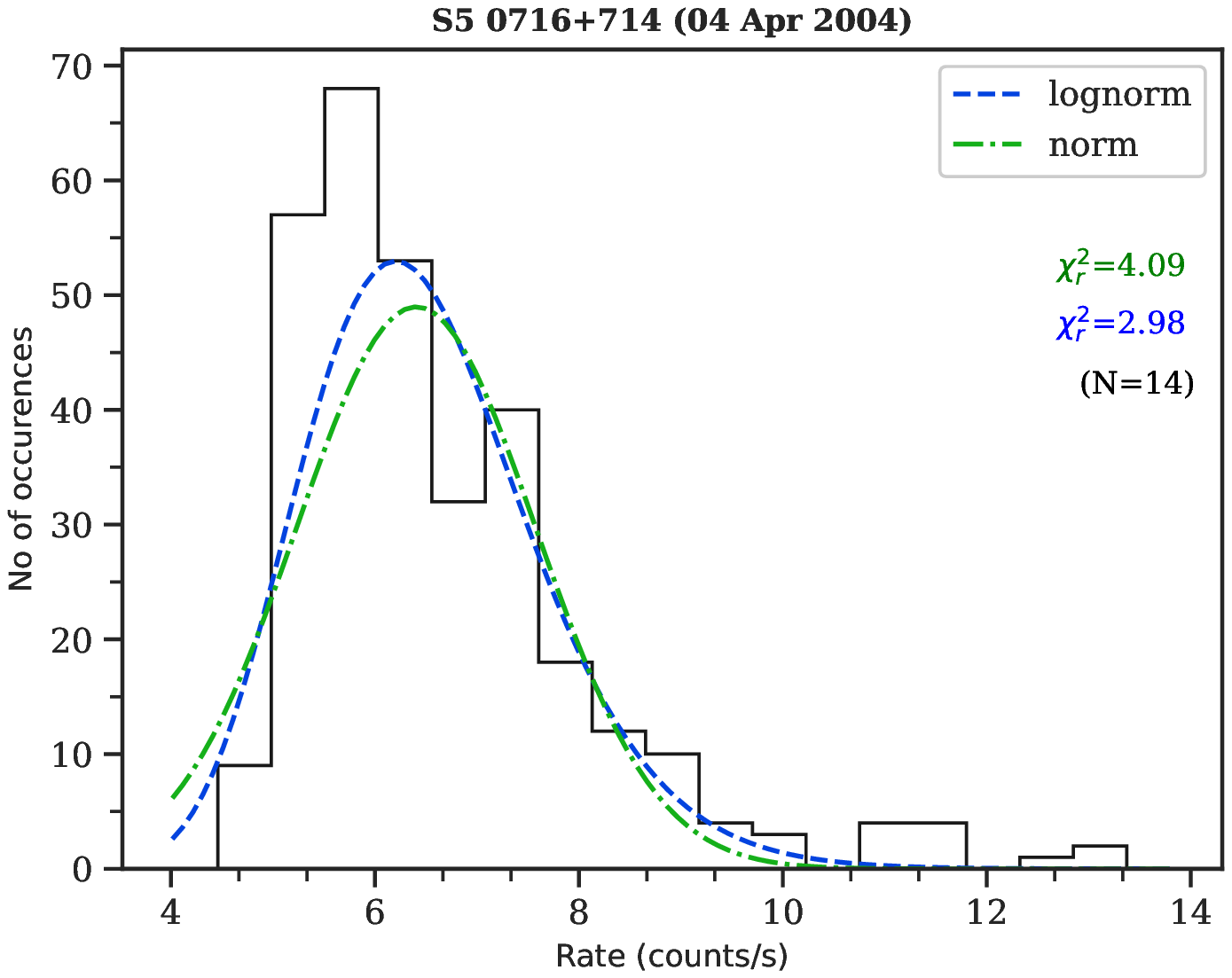}
    \includegraphics[width=6cm]{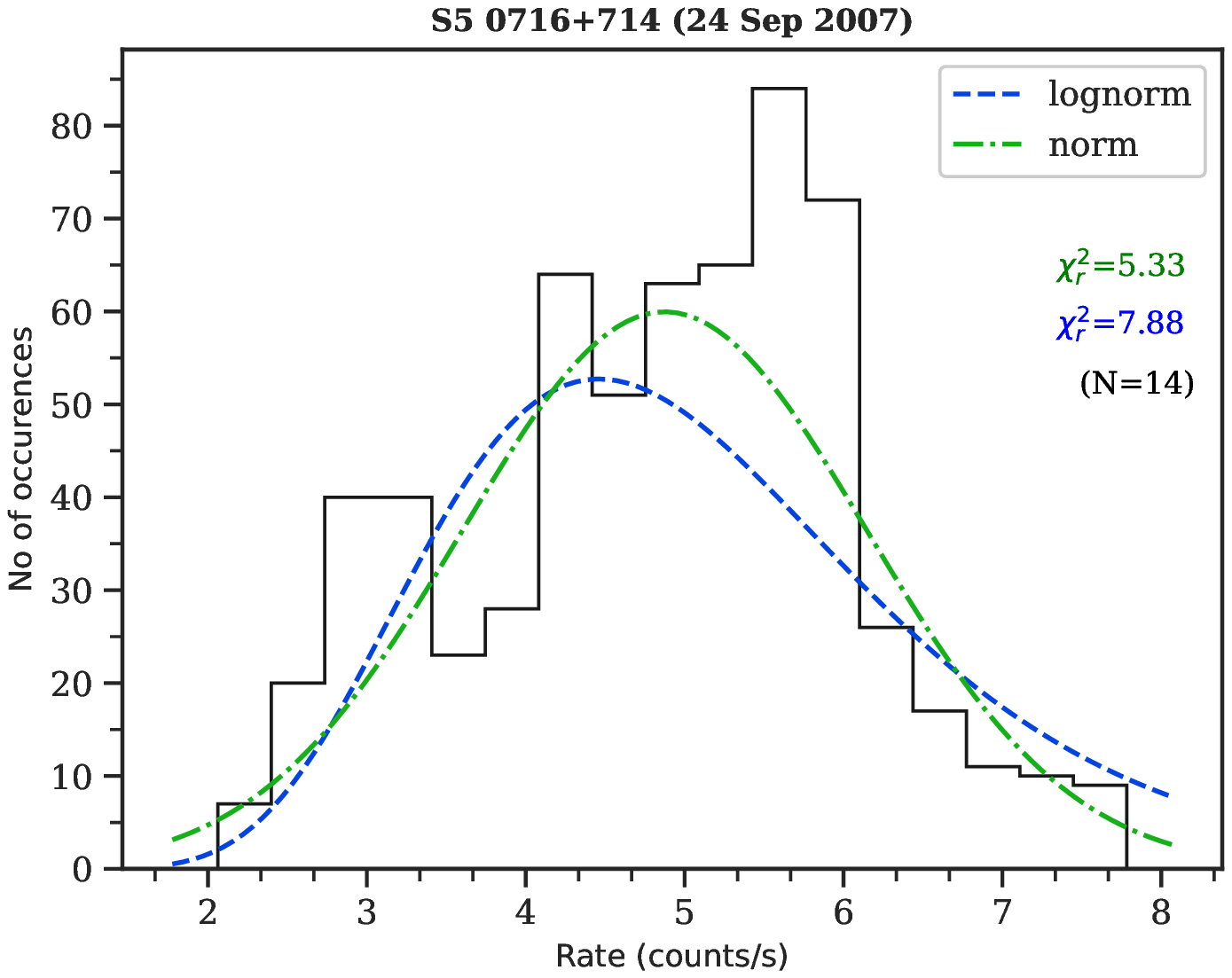}
    \includegraphics[width=6cm]{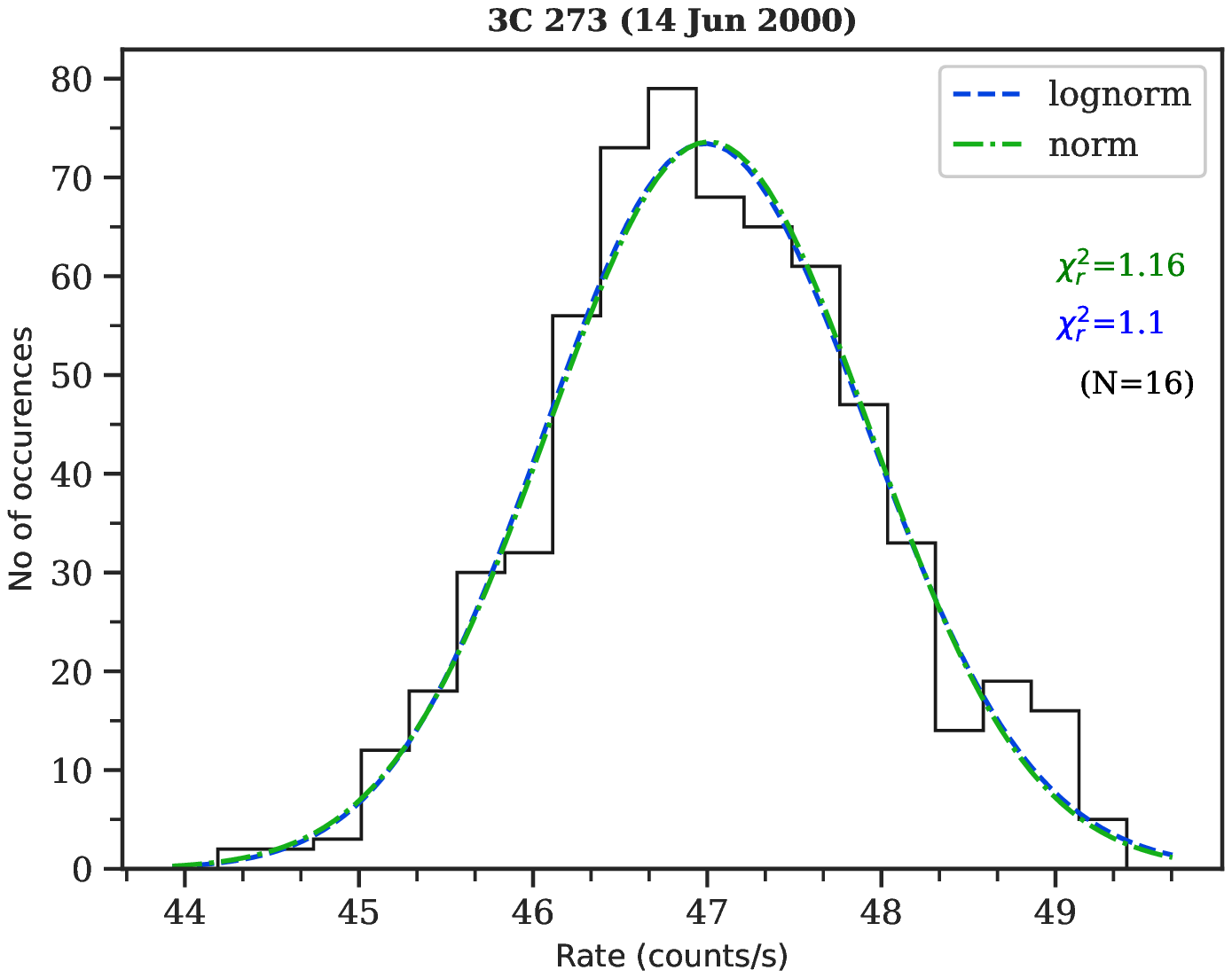}
    
    \includegraphics[width=6cm]{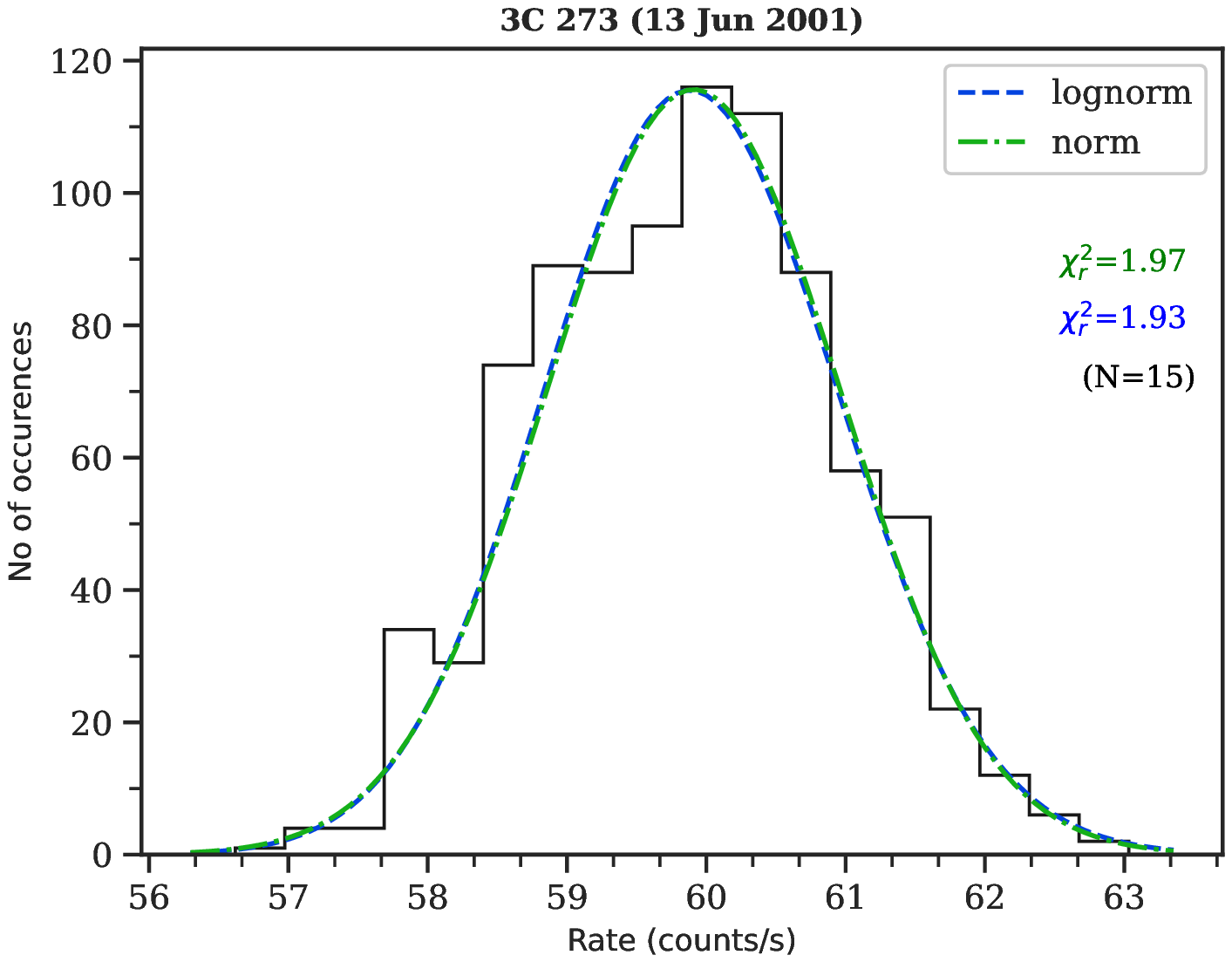}
    \includegraphics[width=6cm]{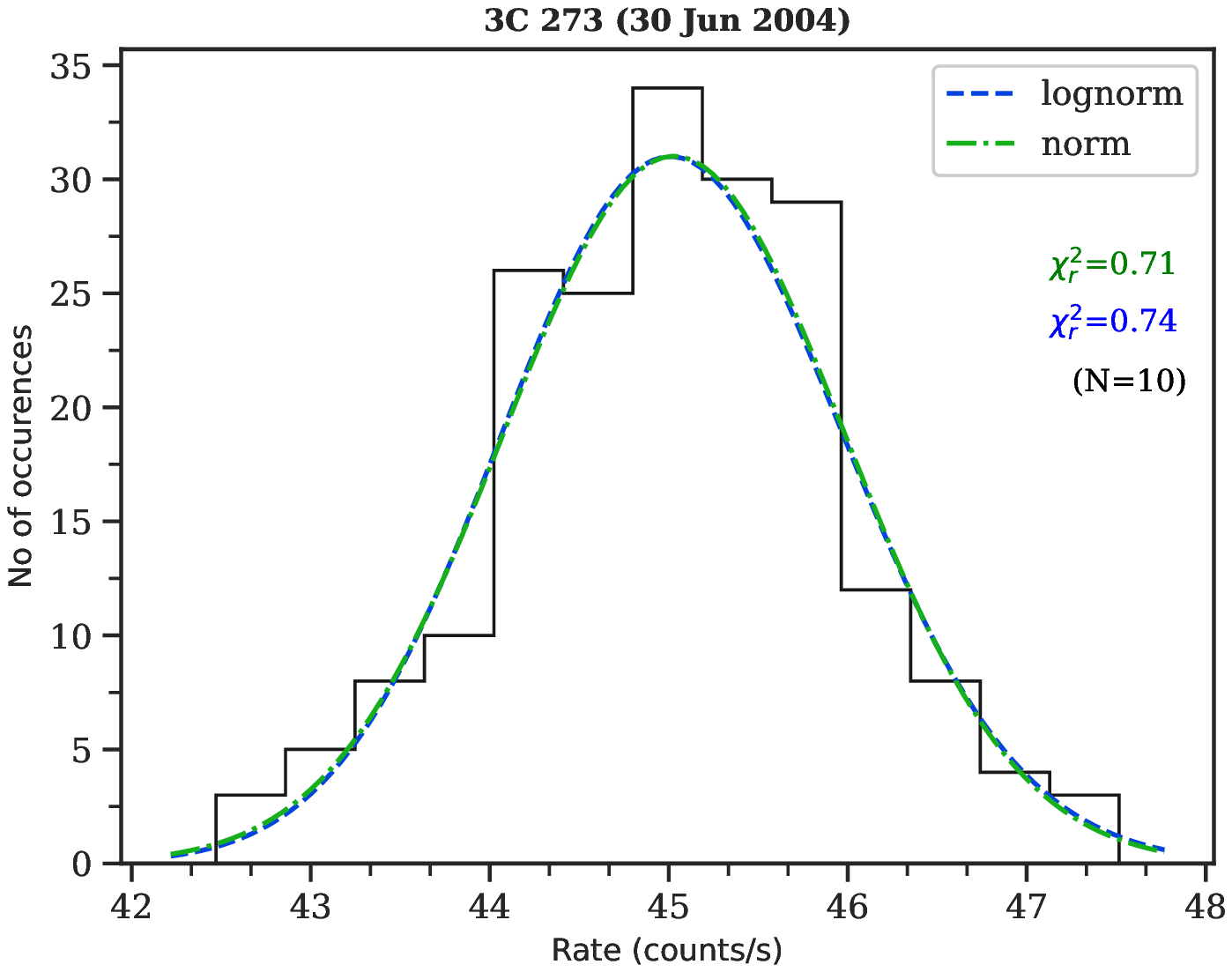}
    \includegraphics[width=6cm]{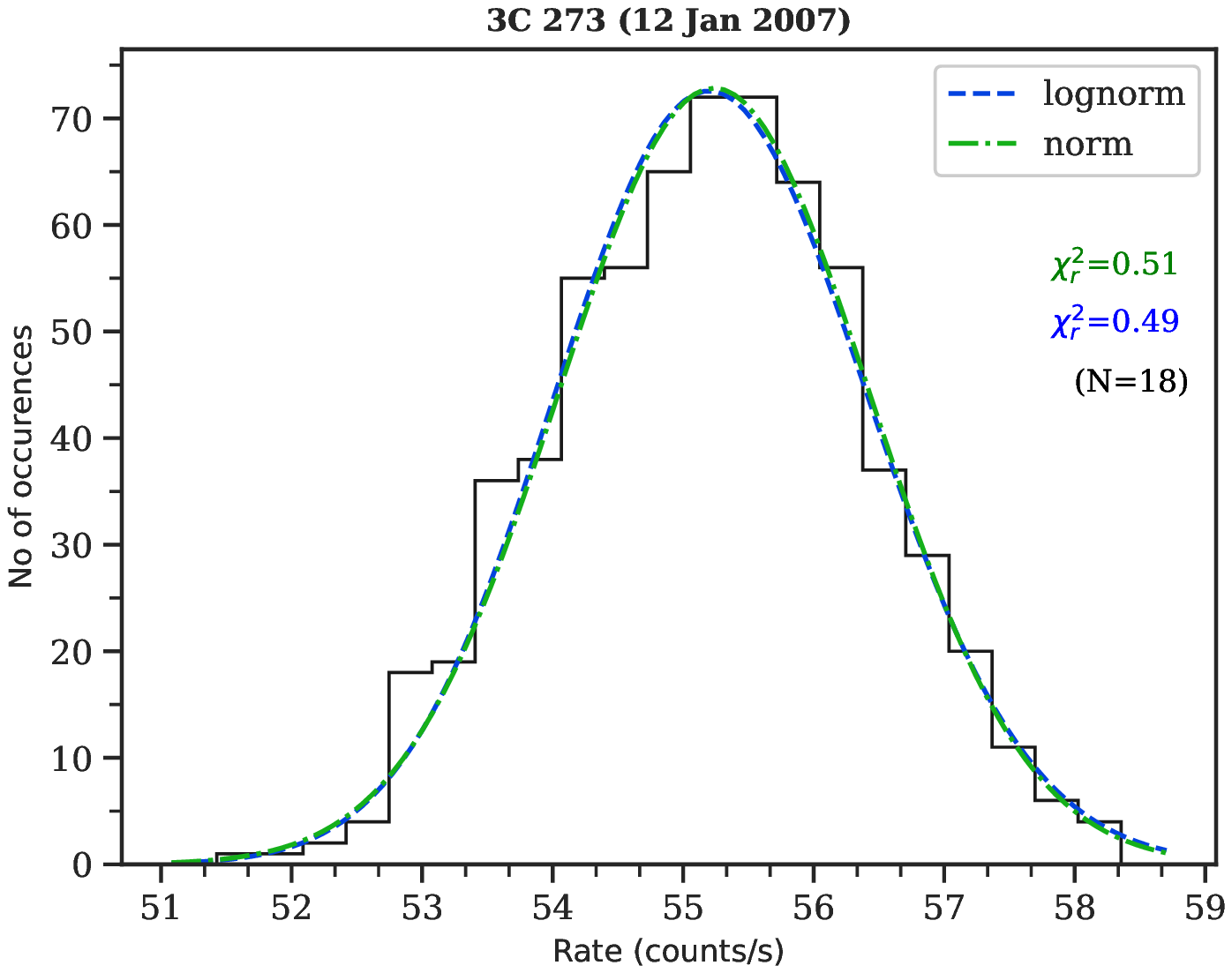}
        \end{adjustwidth}
\caption{\textit{Cont.}}
\end{figure}
\begin{figure}[H]\ContinuedFloat
\begin{adjustwidth}{-\extralength}{0cm}\centering
    \includegraphics[width=6cm]{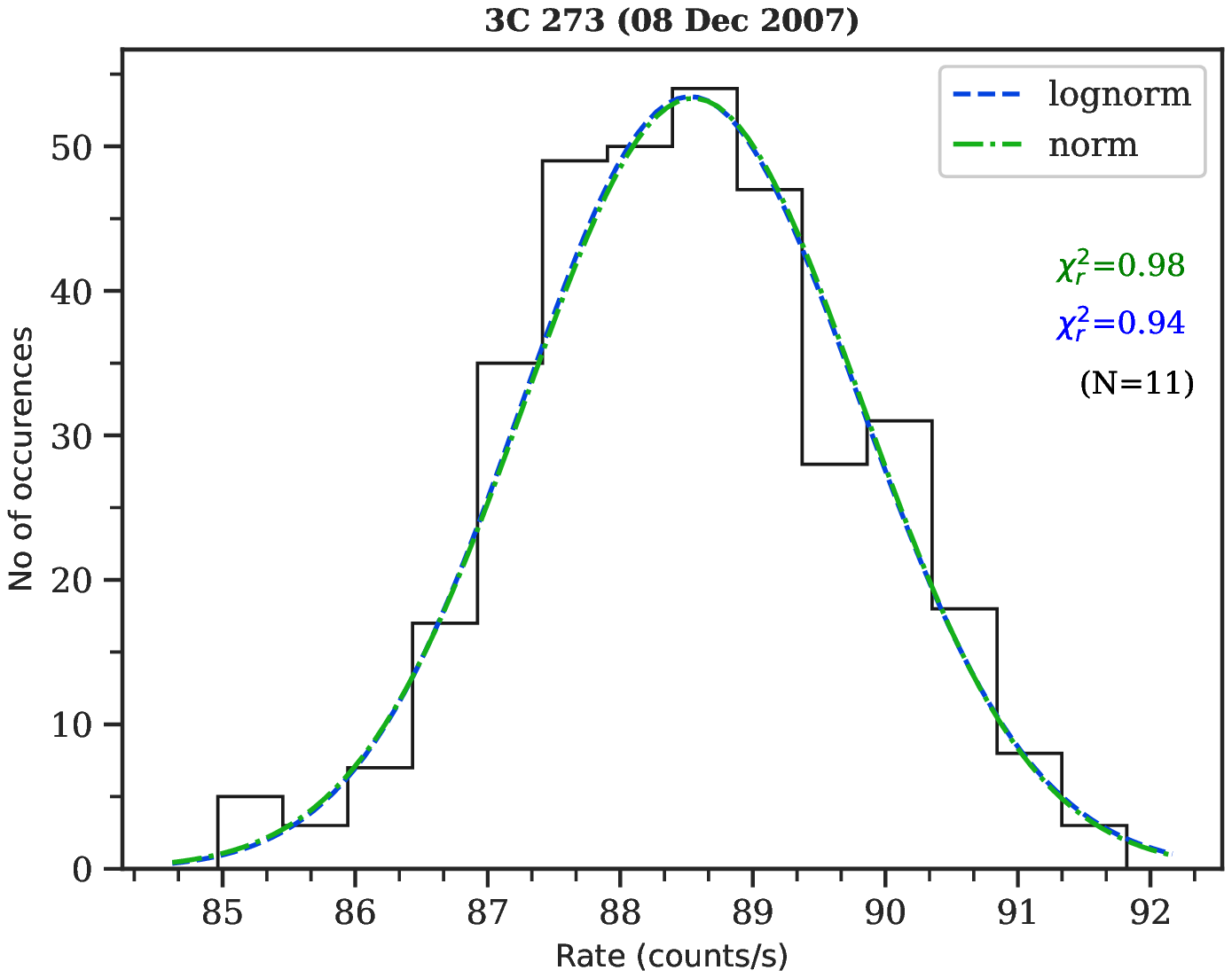}
    \includegraphics[width=6cm]{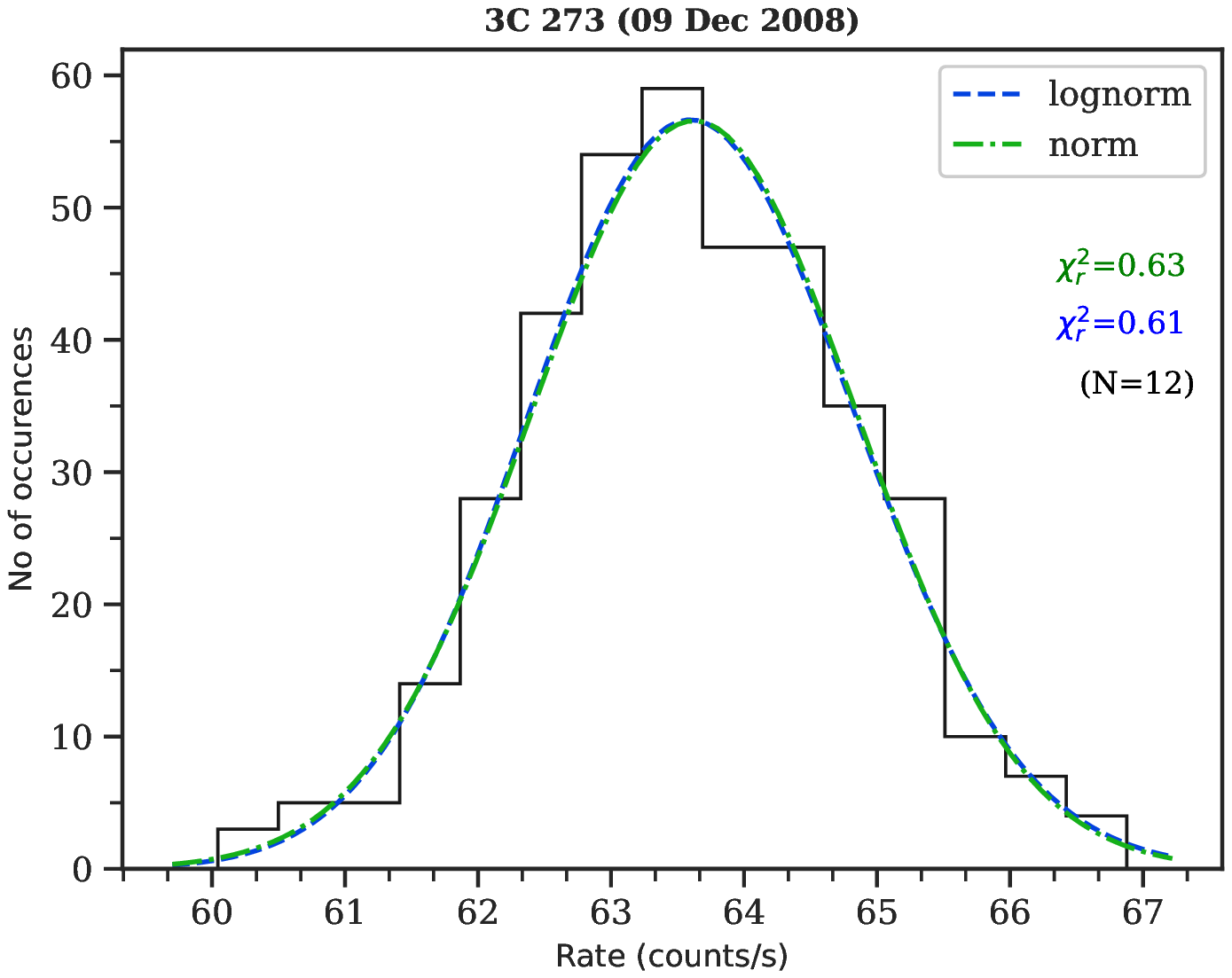}
    \includegraphics[width=6cm]{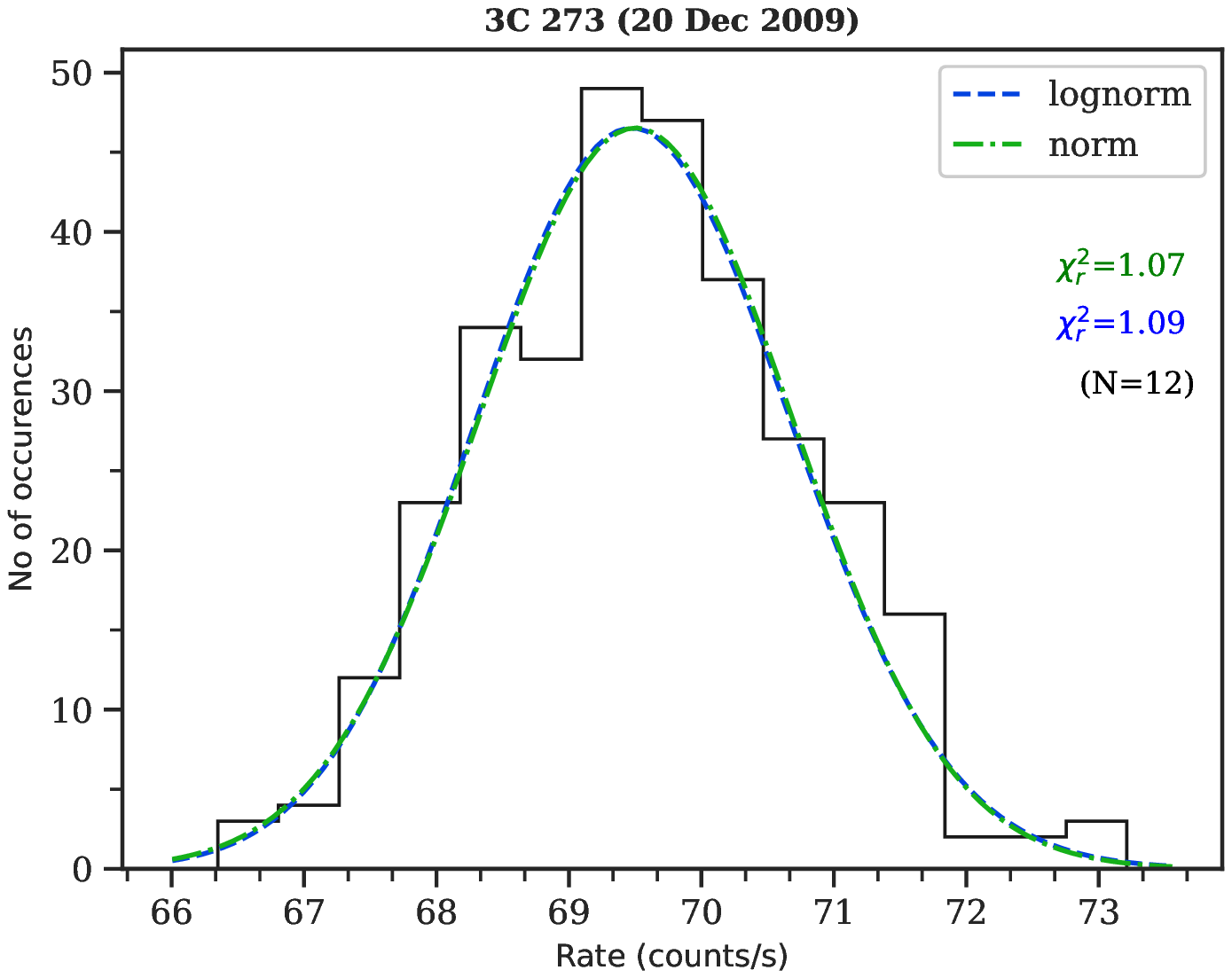}
    
    \includegraphics[width=6cm]{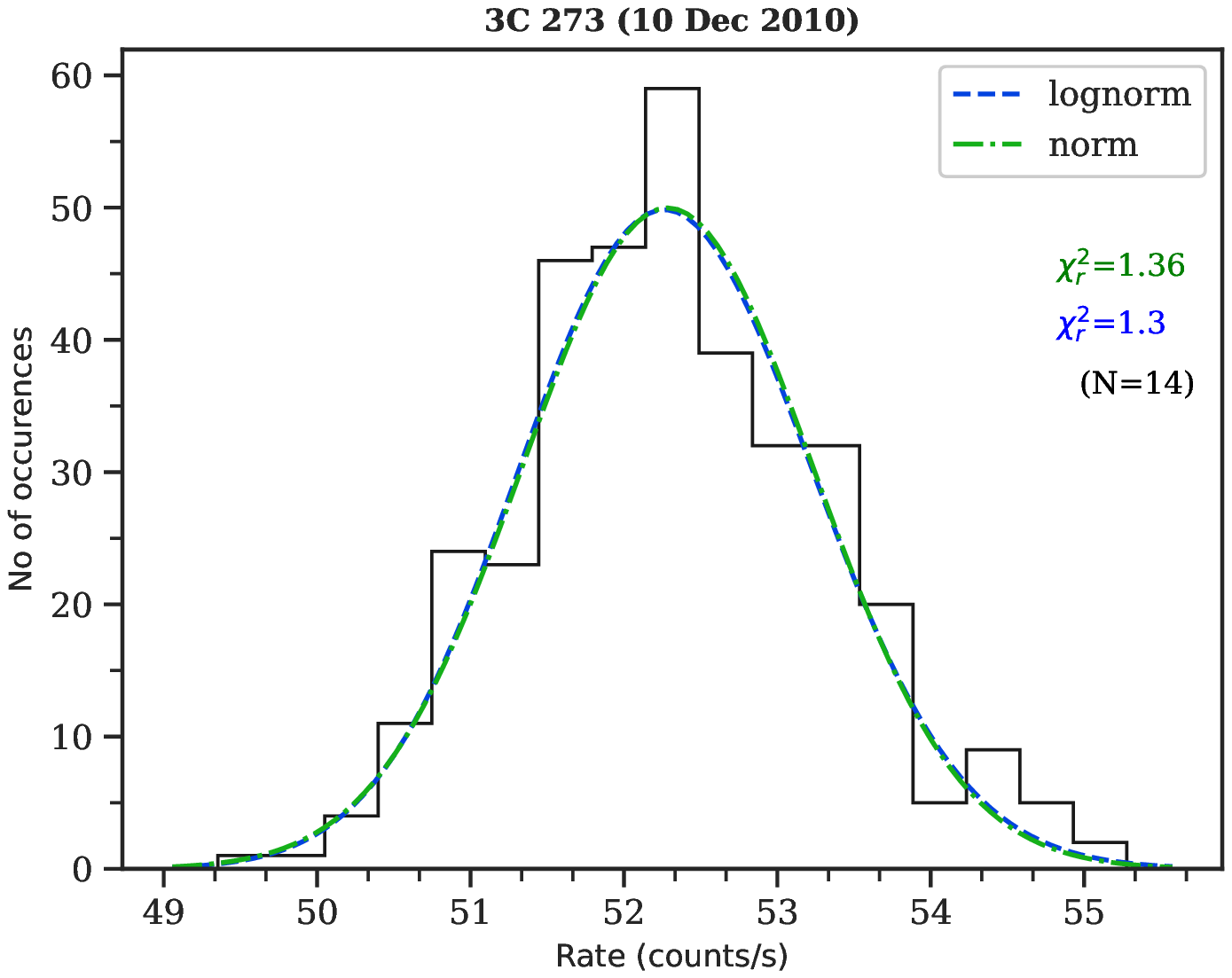}
    \includegraphics[width=6cm]{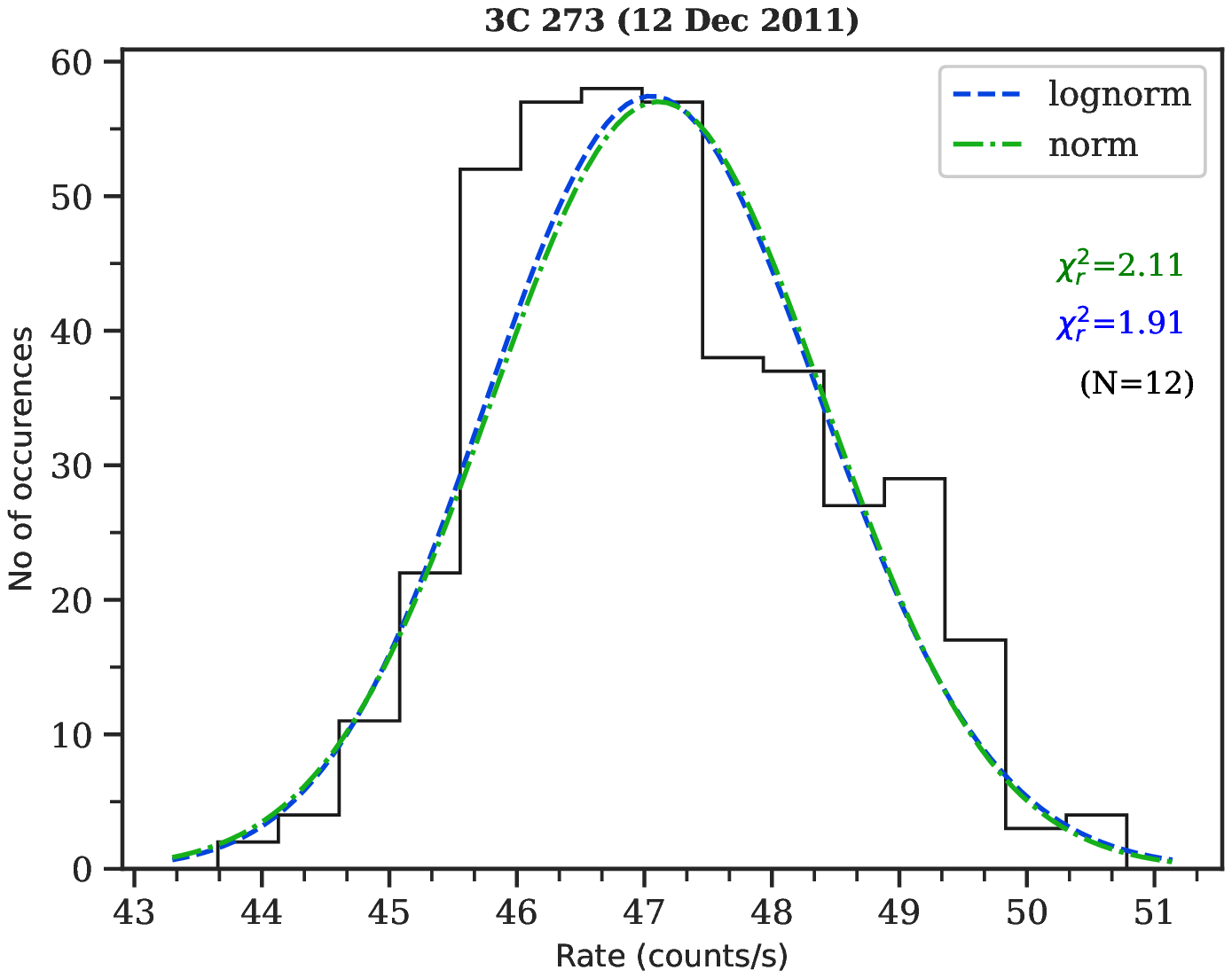}
    \includegraphics[width=6cm]{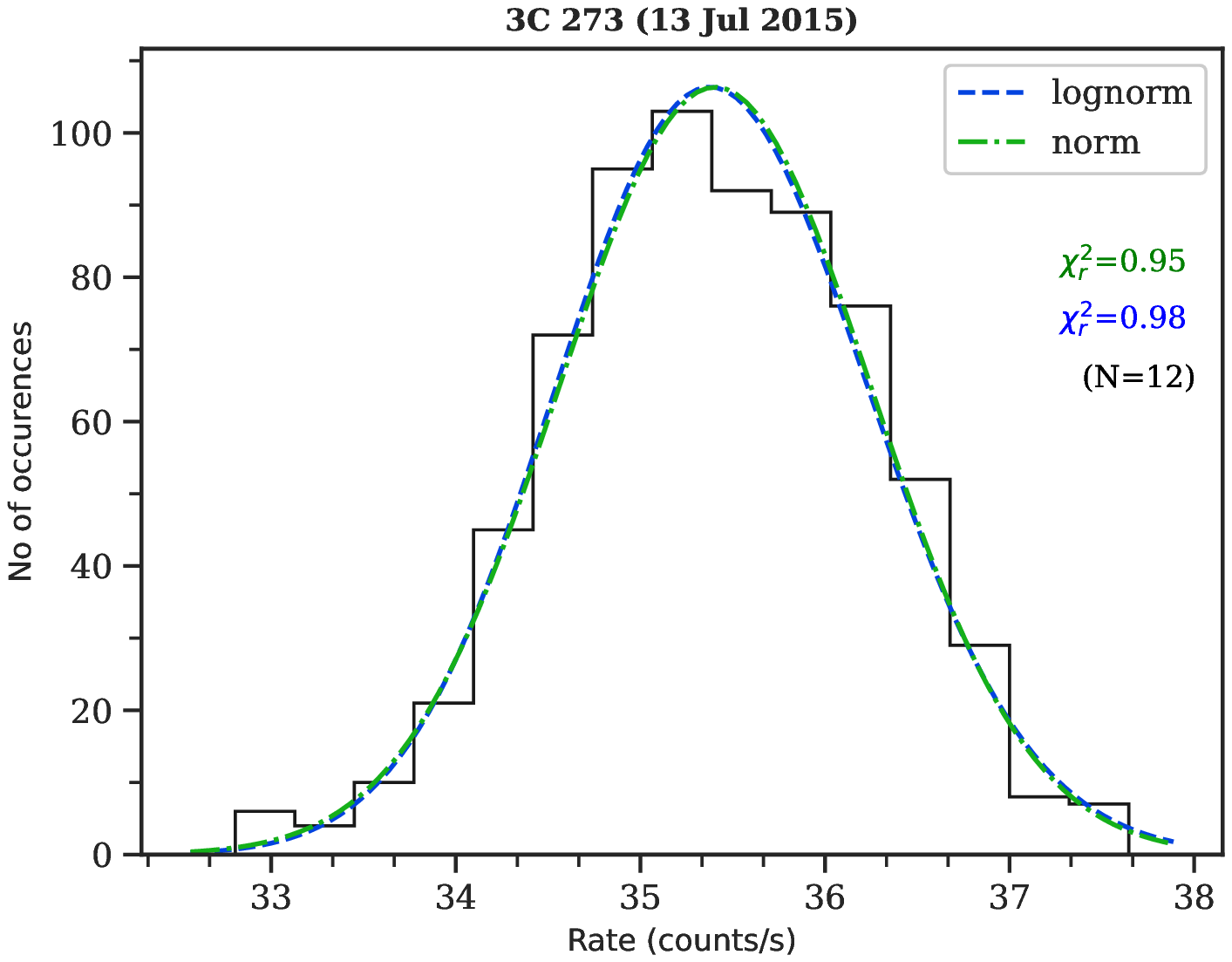}

    \includegraphics[width=6cm]{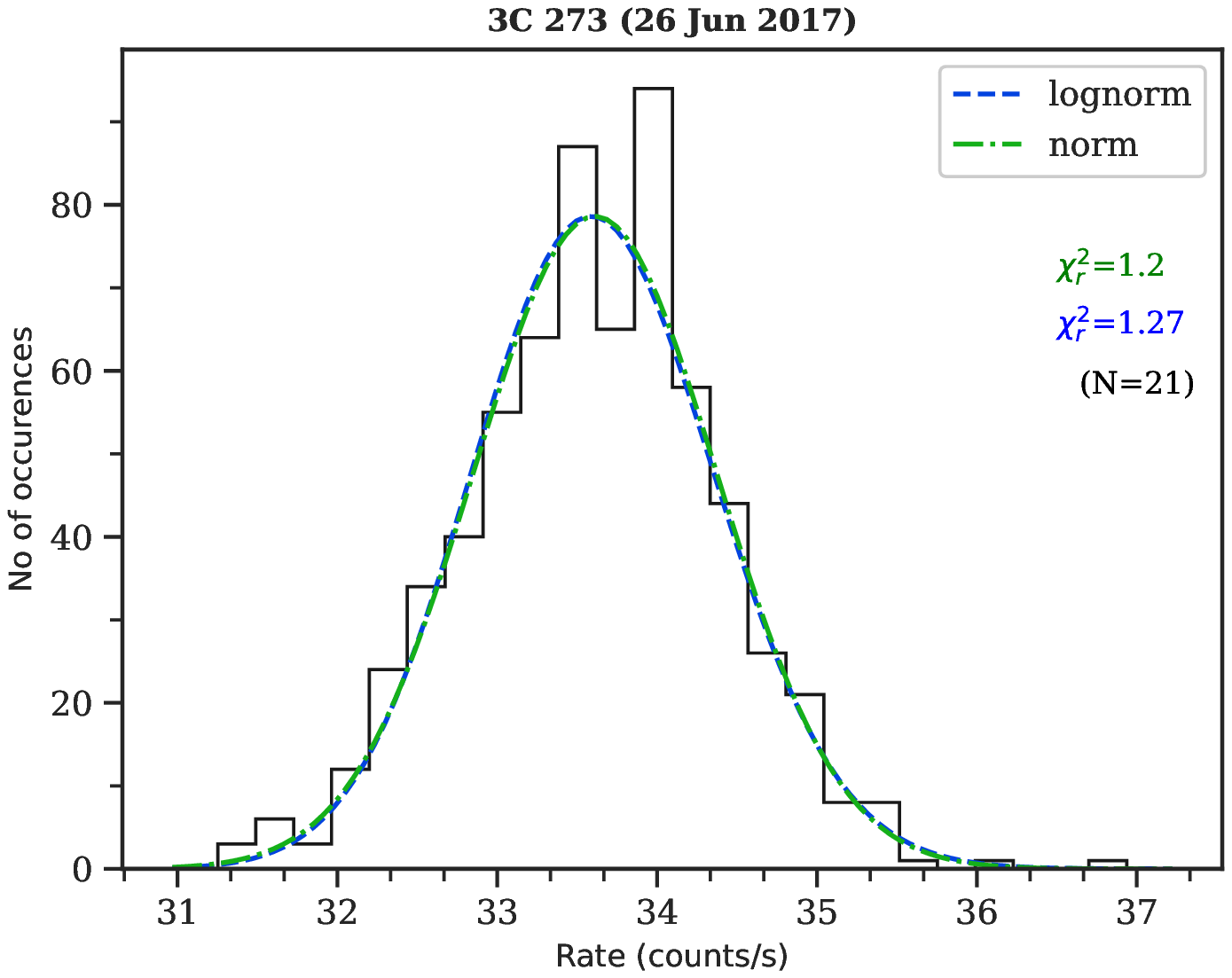}
    \includegraphics[width=6cm]{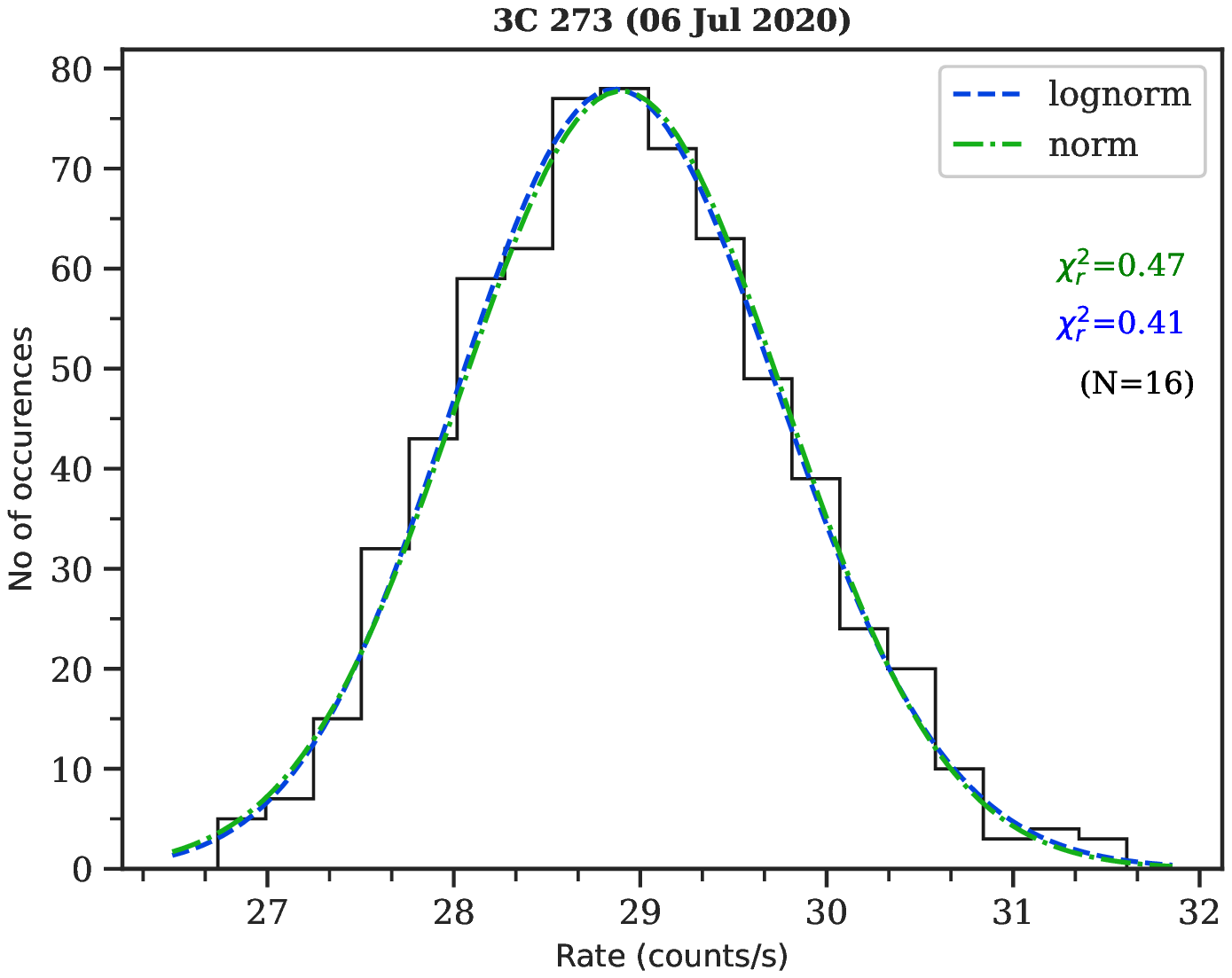}
    \includegraphics[width=6cm]{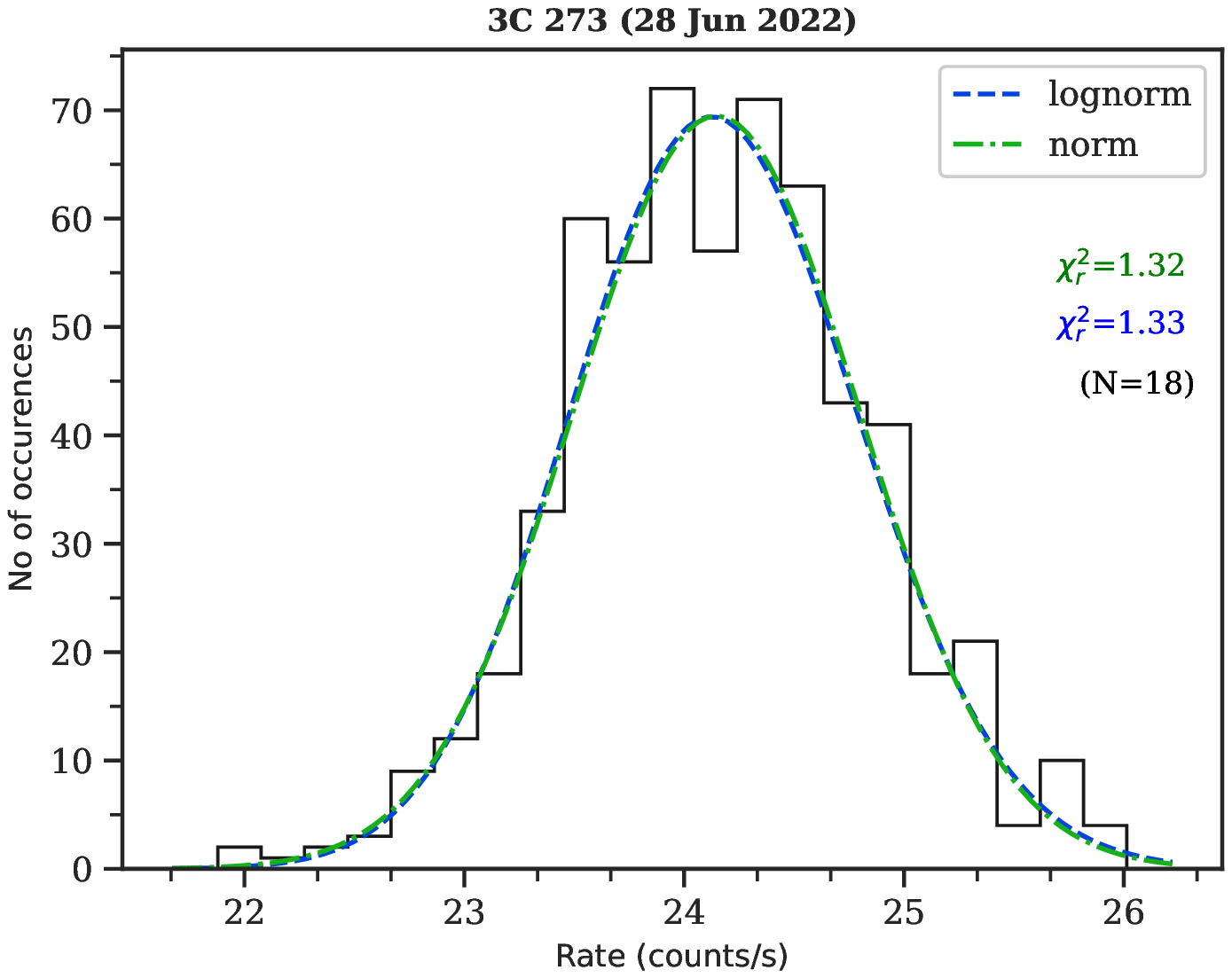}
    \includegraphics[width=6cm]{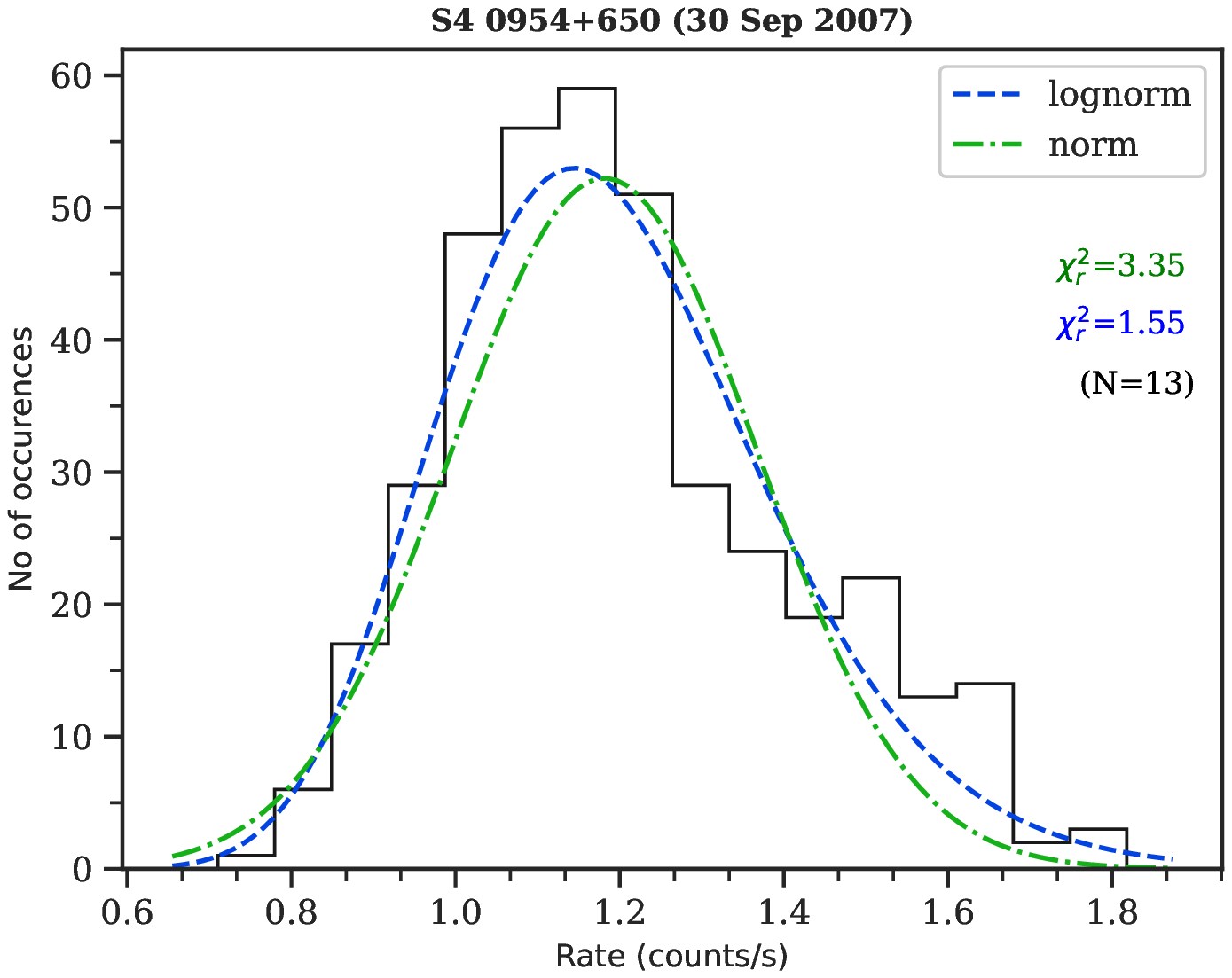}
    \includegraphics[width=6cm]{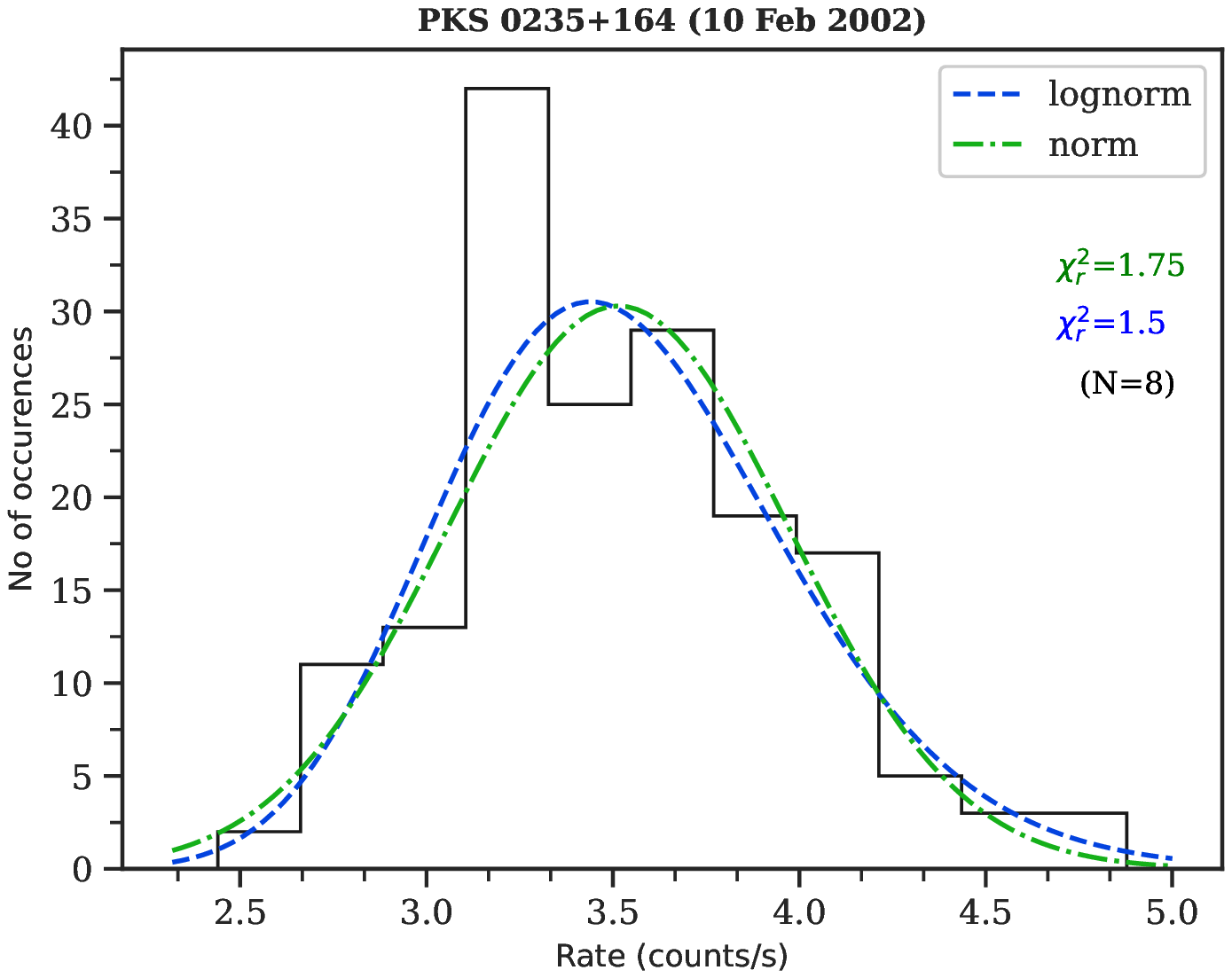}
    \includegraphics[width=6cm]{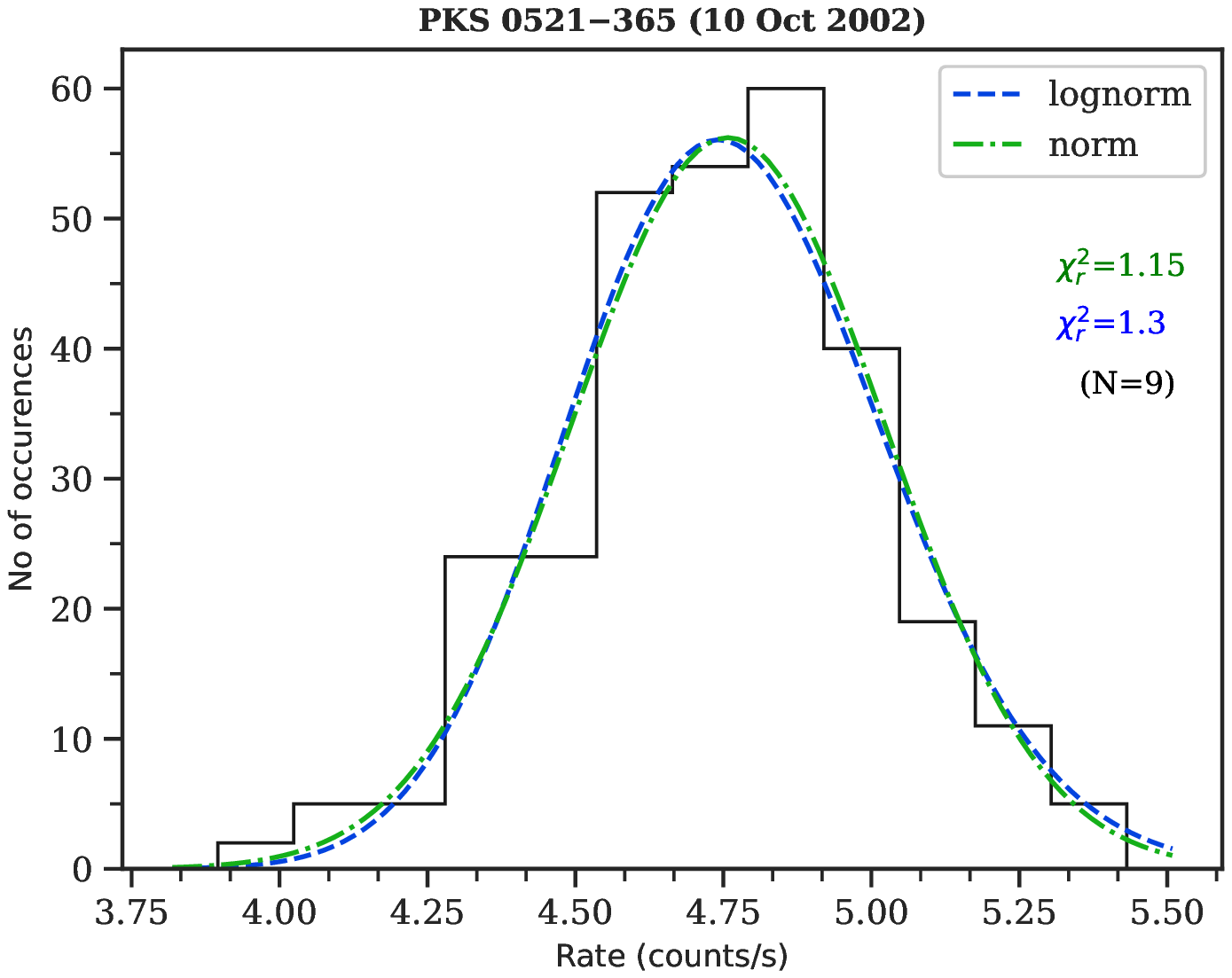}
     
\end{adjustwidth}
\caption{{Histogram} of count rate of light curves of \mbox{XMM-Newton} observations of blazars. Blue dashed lines represent lognormal and green dash-dotted lines represent normal   
PDF fitting results. N represents the degree of freedom of observational data.}
  
\label{fig:histogram}
\end{figure}
\section{Discussion and Conclusions}
\label{Discussion}
In this work, we studied the statistical properties of the X-ray variability of a sample of \mbox{15 blazars}, consisting of ISPs and LSPs observed using the \mbox{XMM-Newton} satellite. We analysed a total of 57
observation epochs for our sample to search for significant variability (at the 3$\sigma$ level) by estimating the fractional variability amplitude $F_{var}$ and found 24 light curves to be variable
for a total of 7 sources. In order to study the cause of variability, we model histograms of the count rate of all of these variable light curves using normal and lognormal distributions. The lognormality of flux
distributions in blazars, if found, is very important as they imply that the X-ray emissions originating from jets have imprints of the accretion disk or its environment, as found in the literature
for many X-ray binaries and Seyfert galaxies, where disk contribution is more compared to  jets \cite{Uttley_McHardy_2005,McHardy_2006,Nakagawa_Mori_2013,Sobolewska_2014}.
Such behaviour was inferred for the famous, nearby HSPs Mrk 421 and Mrk 501 at a Very-High-Energy (VHE) band by \citep{Tluczykont_2010}. The Third LAT AGN Catalog (3LAC; \cite{Ackermann_2015}) 
also showed a lognormal distribution for blazars. {{Kushwaha et al. (2016)}} \cite{Kushwaha_2016} studied an FSRQ PKS 1510--089 using multi-wavelength observations and found that the flux distribution 
of optical and $\gamma$ rays follow two distinctive lognormal profiles, whereas the X-ray flux distribution follows a single lognormal distribution. {Sinha et al. (2016)} \cite{Sinha_2016} also studied 
HSP blazar Mrk 421 and found that the radio to VHE frequencies follows a lognormal distribution. {Shah et~al. (2018)} \cite{Shah_2018} also  analysed $\gamma$-ray data of 50 blazars and found that their flux
distribution follows a lognormal distribution. However, all of the above studies are performed on short- or long-term observations. 

Minute or intra-day variability, as seen in most of the blazars, are difficult to explain using the disc component \cite{Narayan_Piran_2012}, and strongly favour an origin from the 
relativistic jets. In the statistical studies of IDV light curves of blazars, if lognormality is found to be a better model for a normal distribution, it could arise from a few additive 
processes under specific scenarios. Many models have been proposed to explain the IDV of blazars, which involves the shock-in-a-jet model \cite{Marscher_Gear_1985, Bottcher_Dermer_2010}, 
needle-in-a-jet \cite{Ghisellini_Tavecchio_2008}, jets-in-a-jet model \cite{Giannios_2009,Giannios_2010} and turbulent extreme multi-zone (TEMZ) model (e.g., \cite{Marscher_2014}). 
{Biteau and Giebels (2012)} \cite{Biteau_Giebels_2012} studied the properties of blazar light curves using a mini jets-in-a-jet model and found that flux from 
a single region produces a power law histogram and a lognormal distribution in the case of many randomly oriented mini jets. It has been found that the histograms having skewed/lognormal 
profiles followed a linear rms--flux relationship. {Sinha et al. (2018)}~\cite{Sinha_2018} also provided an alternative explanation for the lognormal behaviour of blazar
light curves and proposed that the non-Gaussian signatures in blazar variability could arise from linear fluctuations in the underlying particle acceleration and/or the diffusive 
escape rate of the emitting electrons. Such small Gaussian perturbations propagating in blazar jets could produce non-linear flux distributions and can explain the lognormal behaviour.
{Kushwaha and Pal (2020)} \cite{Kushwaha_2020} studied blazars during high variability phases on IDV timescales and found the flux distributions show a normal profile compared to lognormal ones.

In our studies, we found the lognormal distribution to be a preferred model over the normal one for sources BL Lacertae and S4 0954+650 in two observational epochs. For these observations, 
 we found a linear rms--flux relationship with a Pearson correlation coefficient of 0.61 and a $p$-value of 0.2 for BL Lacertae and 0.63 with a $p$-value of 0.56 for S4 0954+650.
In  11 other observations from 4 sources, namely, ON 231, 3C 273, PKS 0235+164 and PKS 0521-365, we found that the flux distributions  equally fit with both 
distributions, and we can not prefer any model. In the rest of the observations, we are unable to model them  with normal or lognormal distributions. 

If the variability is triggered due to fluctuations in the accretion disk \textls[-15]{and the jet is powered by the central source channelling the energy into non-thermal particles
, then
the observed flux distributions may be an imprint of the accretion disk onto the jet (e.g., \cite{Giebels_2009}).}
All of the above scenarios might be plausible on the long-term timescale variability of blazars. The lack of correlation between the mass of the Supermassive Black Hole (SMBH) and 
the variability timescale in blazars \cite{Kushwaha_2016} suggests that the jet can modify the imprint of accretion disc fluctuation, at least on IDV timescales. 
Based on observations and simulations, blazar jets are found to be highly magnetised systems and, therefore, magnetic reconnection is a potential mechanism for X-ray variability. 
Turbulent magnetic reconnection in the coronal region around the accretion disk can explain the emission from compact accreting sources but fails for blazars where the jet is 
closely pointing to our line of sight. {A magnetic reconnection-based model, mini jets-in-a-jet, can not be favoured for these blazars as we did not find a significant linear relationship between rms and flux. Therefore, the lognormality found in our observations could be attributed to either due to non-Gaussian fluctuations in blazar jets 
or through linear fluctuations of the underlying particle acceleration \cite{Khatoon_2020} or the diffusive escape rate of the emitting electrons \cite{Sinha_2018}. They showed that the perturbation in the acceleration timescale produces a Gaussian distribution in the index that results in a lognormal distribution of the flux. However, the perturbation in the particle cooling rate produces neither a Gaussian nor a lognormal flux distribution, which is also found in many of our observations. Similarly, in the shock acceleration scenario, a small perturbation in the acceleration timescale can lead to the variability of the particle number density, which is a linear combination of Gaussian and lognormal processes. The relative weight of these processes determines the dominant shape of the flux distribution. If the variability arises due to variations in the number density of the accelerated particles, then the flux distribution could appear as both Gaussian and lognormal, which is also seen in our observations.} However, we need more observations
and a bigger sample to characterise the statistical properties of the flux distribution of blazars. 

\vspace{6pt} 



\authorcontributions{K.W. analysed the X-ray data, carried out analysis and wrote the initial manuscript. H.G. conceived the idea and finalised the manuscript. 
All authors have read and agreed to the published version of the manuscript.}

\funding{This research is based on observations obtained with \mbox{XMM-Newton}, an ESA science mission with
instruments and contributions directly funded by ESA
Member States and NASA. K.W. and H.G. acknowledge
the financial support from the Department of Science
and Technology, India, through the INSPIRE Faculty award
IFA17-PH197 at ARIES, Nainital.}

\conflictsofinterest{The authors declare no conflict of interest.} 


\printendnotes[custom]

\begin{adjustwidth}{-\extralength}{0cm}

\reftitle{References}

\end{adjustwidth}

\begin{thebibliography}{999}

\bibitem[{Blandford} and {K{\"o}nigl}(1979)]{Blandford_Konigl1979}
{{Blandford}, R.D.; {K{\"o}nigl}, A.
\newblock {Relativistic jets as compact radio sources.}
\newblock {\em Astrophys. J.} {\bf 1979}, {\em 232},~34--48.
\newblock  } [\href{http://doi.org/10.1086/157262}{CrossRef}]

\bibitem[{Urry} and {Padovani}(1995)]{Urry_padovani_1995}
{Urry}, C.M.; {Padovani}, P.
\newblock {Unified Schemes for Radio-Loud Active Galactic Nuclei}.
\newblock {\em Publ. Astron. Soc. Pac.} {\bf 1995}, {\em 107},~803.
\newblock   [\href{http://dx.doi.org/10.1086/133630}{CrossRef}]

\bibitem[Fan \em{et~al.}(2016)Fan, Yang, Liu, Luo, Lin, Yuan, Xiao, Zhou, Hua,
and Pei]{Fan_2016}
Fan, J.H.; Yang, J.H.; Liu, Y.; Luo, G.Y.; Lin, C.; Yuan, Y.H.; Xiao, H.B.;
Zhou, A.Y.; Hua, T.X.; Pei, Z.Y.
\newblock The spectral energy distributions of Fermi blazars.
\newblock {\em  Astrophys. J. Suppl. Ser.} {\bf 2016}, {\em
226},~20.
\newblock  [\href{http://dx.doi.org/10.3847/0067-0049/226/2/20}{CrossRef}]

\bibitem[{Abdo} \em{et~al.}(2010){Abdo}, {Ackermann}, {Agudo}, {Ajello},
{Aller}, {Aller}, {Angelakis}, {Arkharov}, {Axelsson}, {Bach}, {Baldini},
{Ballet}, {Barbiellini}, {Bastieri}, {Baughman}, {Bechtol}, {Bellazzini},
{Benitez}, {Berdyugin}, {Berenji}, {Blandford}, {Bloom}, {Boettcher},
{Bonamente}, {Borgland}, {Bregeon}, {Brez}, {Brigida}, {Bruel}, {Burnett},
{Burrows}, {Buson}, {Caliandro}, {Calzoletti}, {Cameron}, {Capalbi},
{Caraveo}, {Carosati}, {Casandjian}, {Cavazzuti}, {Cecchi}, {{\c{C}}elik},
{Charles}, {Chaty}, {Chekhtman}, {Chen}, {Chiang}, {Chincarini}, {Ciprini},
{Claus}, {Cohen-Tanugi}, {Colafrancesco}, {Cominsky}, {Conrad}, {Costamante},
{Cutini}, {D'ammando}, {Deitrick}, {D'Elia}, {Dermer}, {de Angelis}, {de
Palma}, {Digel}, {Donnarumma}, {Silva}, {Drell}, {Dubois}, {Dultzin},
{Dumora}, {Falcone}, {Farnier}, {Favuzzi}, {Fegan}, {Focke}, {Forn{\'e}},
{Fortin}, {Frailis}, {Fuhrmann}, {Fukazawa}, {Funk}, {Fusco}, {G{\'o}mez},
{Gargano}, {Gasparrini}, {Gehrels}, {Germani}, {Giebels}, {Giglietto},
{Giommi}, {Giordano}, {Giuliani}, {Glanzman}, {Godfrey}, {Grenier},
{Gronwall}, {Grove}, {Guillemot}, {Guiriec}, {Gurwell}, {Hadasch},
{Hanabata}, {Harding}, {Hayashida}, {Hays}, {Healey}, {Heidt}, {Hiriart},
{Horan}, {Hoversten}, {Hughes}, {Itoh}, {Jackson}, {J{\'o}hannesson},
{Johnson}, {Johnson}, {Jorstad}, {Kadler}, {Kamae}, {Katagiri}, {Kataoka},
{Kawai}, {Kennea}, {Kerr}, {Kimeridze}, {Kn{\"o}dlseder}, {Kocian},
{Kopatskaya}, {Koptelova}, {Konstantinova}, {Kovalev}, {Kovalev},
{Kurtanidze}, {Kuss}, {Lande}, {Larionov}, {Latronico}, {Leto}, {Lindfors},
{Longo}, {Loparco}, {Lott}, {Lovellette}, {Lubrano}, {Madejski}, {Makeev},
{Marchegiani}, {Marscher}, {Marshall}, {Max-Moerbeck}, {Mazziotta},
{McConville}, {McEnery}, {Meurer}, {Michelson}, {Mitthumsiri}, {Mizuno},
{Moiseev}, {Monte}, {Monzani}, {Morselli}, {Moskalenko}, {Murgia},
{Nestoras}, {Nilsson}, {Nizhelsky}, {Nolan}, {Norris}, {Nuss}, {Ohsugi},
{Ojha}, {Omodei}, {Orlando}, {Ormes}, {Osborne}, {Ozaki}, {Pacciani},
{Padovani}, {Pagani}, {Page}, {Paneque}, {Panetta}, {Parent}, {Pasanen},
{Pavlidou}, {Pelassa}, {Pepe}, {Perri}, {Pesce-Rollins}, {Piranomonte},
{Piron}, {Pittori}, {Porter}, {Puccetti}, {Rahoui}, {Rain{\`o}}, {Raiteri},
{Rando}, {Razzano}, {Reimer}, {Reimer}, {Reposeur}, {Richards}, {Ritz},
{Rochester}, {Rodriguez}, {Romani}, {Ros}, {Roth}, {Roustazadeh}, {Ryde},
{Sadrozinski}, {Sadun}, {Sanchez}, {Sander}, {Saz Parkinson}, {Scargle},
{Sellerholm}, {Sgr{\`o}}, {Shaw}, {Sigua}, {Siskind}, {Smith}, {Smith},
{Spandre}, {Spinelli}, {Starck}, {Stevenson}, {Stratta}, {Strickman},
{Suson}, {Tajima}, {Takahashi}, {Takahashi}, {Takalo}, {Tanaka}, {Thayer},
{Thayer}, {Thompson}, {Tibaldo}, {Torres}, {Tosti}, {Tramacere}, {Uchiyama},
{Usher}, {Vasileiou}, {Verrecchia}, {Vilchez}, {Villata}, {Vitale}, {Waite},
{Wang}, {Winer}, {Wood}, {Ylinen}, {Zensus}, {Zhekanis}, and
{Ziegler}]{Abdo_2010}
{Abdo}, A.A.; {Ackermann}, M.; {Agudo}, I.; {Ajello}, M.; {Aller}, H.D.;
{Aller}, M.F.; {Angelakis}, E.; {Arkharov}, A.A.; {Axelsson}, M.; {Bach}, U.;
et~al.
\newblock {The Spectral Energy Distribution of Fermi Bright Blazars}.
\newblock {\em  Astrophys. J.} {\bf 2010}, {\em 716},~30--70.
\newblock   [\href{http://dx.doi.org/10.1088/0004-637X/716/1/30}{CrossRef}]

\bibitem[{Yang} \em{et~al.}(2022){Yang}, {Xiao}, {Wang}, {Yang}, {Pei}, {Wu},
{Yuan}, and {Fan}]{Yang_2022}
{Yang}, W.X.; {Xiao}, H.B.; {Wang}, H.G.; {Yang}, J.H.; {Pei}, Z.Y.; {Wu},
D.X.; {Yuan}, Y.H.; {Fan}, J.H.
\newblock {Correlation between Brightness Variability and Spectral Index
Variability for Fermi Blazars}.
\newblock {\em Res. Astron. Astrophys.} {\bf 2022}, {\em
22},~085002.
\newblock   [\href{http://dx.doi.org/10.1088/1674-4527/ac712c}{CrossRef}]

\bibitem[{Romero} \em{et~al.}(2017){Romero}, {Boettcher}, {Markoff}, and
{Tavecchio}]{Romero_2017}
{Romero}, G.E.; {Boettcher}, M.; {Markoff}, S.; {Tavecchio}, F.
\newblock {Relativistic Jets in Active Galactic Nuclei and Microquasars}.
\newblock {\em Space Sci. Rev.} {\bf 2017}, {\em 207},~5--61.
\newblock   [\href{http://dx.doi.org/10.1007/s11214-016-0328-2}{CrossRef}]

\bibitem[{Dermer} \em{et~al.}(1992){Dermer}, {Schlickeiser}, and
{Mastichiadis}]{Dermer_1992}
{Dermer}, C.D.; {Schlickeiser}, R.; {Mastichiadis}, A.
\newblock {High-energy gamma radiation from extragalactic radio sources.}
\newblock {\em Astron. Astrophys.} {\bf 1992}, {\em 256},~L27--L30.

\bibitem[{Sikora} \em{et~al.}(1994){Sikora}, {Begelman}, and
{Rees}]{Sikora_1994}
{Sikora}, M.; {Begelman}, M.C.; {Rees}, M.J.
\newblock {Comptonization of Diffuse Ambient Radiation by a Relativistic Jet:
The Source of Gamma Rays from Blazars?}
\newblock {\em Astrophys. J.} {\bf 1994}, {\em 421},~153.
\newblock   [\href{http://dx.doi.org/10.1086/173633}{CrossRef}]

\bibitem[{Ghisellini} and {Tavecchio}(2009)]{Ghisellini_Tavecchio_2009}
{{Ghisellini}, G.; {Tavecchio}, F.
{Canonical high-power blazars}.
{\em Mon. Not. R. Astron. Soc.} {\bf 2009}, {\em 397},~985--1002.
 } [\href{http://dx.doi.org/10.1111/j.1365-2966.2009.15007.x}{CrossRef}]

\bibitem[{B{\"o}ttcher} \em{et~al.}(2013){B{\"o}ttcher}, {Reimer}, {Sweeney},
and {Prakash}]{Bottcher_2013}
{B{\"o}ttcher}, M.; {Reimer}, A.; {Sweeney}, K.; {Prakash}, A.
\newblock {Leptonic and Hadronic Modeling of Fermi-detected Blazars}.
\newblock {\em Astrophys. J.} {\bf 2013}, {\em 768},~54.
\newblock   [\href{http://dx.doi.org/10.1088/0004-637X/768/1/54}{CrossRef}]

\bibitem[{Atoyan} and {Dermer}(2003)]{Atoyan_Dermer_2003}
{Atoyan}, A.M.; {Dermer}, C.D.
\newblock {Neutral Beams from Blazar Jets}.
\newblock {\em Astrophys. J.} {\bf 2003}, {\em 586},~79--96.
\newblock   [\href{http://dx.doi.org/10.1086/346261}{CrossRef}]

\bibitem[{Sembay} \em{et~al.}(1993){Sembay}, {Warwick}, {Urry}, {Sokoloski},
{George}, {Makino}, {Ohashi}, and {Tashiro}]{Sembay_1993}
{Sembay}, S.; {Warwick}, R.S.; {Urry}, C.M.; {Sokoloski}, J.; {George}, I.M.;
{Makino}, F.; {Ohashi}, T.; {Tashiro}, M.
\newblock {The X-ray Spectral Variability of the BL Lacertae Type Object PKS
2155-304}.
\newblock {\em  Astrophys. J.} {\bf 1993}, {\em 404},~112.
\newblock  [\href{http://dx.doi.org/10.1086/172263}{CrossRef}]

\bibitem[{Wagner} and {Witzel}(1995)]{Wagner_1995}
{Wagner}, S.J.; {Witzel}, A.
\newblock {Intraday Variability In Quasars and BL Lac Objects}.
\newblock {\em Annu. Rev. Astron. Astrophys.} {\bf 1995}, {\em 33},~163--198.
\newblock  [\href{http://dx.doi.org/10.1146/annurev.aa.33.090195.001115}{CrossRef}]

\bibitem[Ulrich \em{et~al.}(1997)Ulrich, Maraschi, and Urry]{Ulrich_1997}
Ulrich, M.H.; Maraschi, L.; Urry, C.M.
\newblock Variability of active galactic nuclei.
\newblock {\em Annu. Rev. Astron. Astrophys.} {\bf 1997}, {\em
35},~445--502.
\newblock   [\href{http://dx.doi.org/10.1146/annurev.astro.35.1.445}{CrossRef}]

\bibitem[{Gaur} \em{et~al.}(2019){Gaur}, {Gupta}, {Bachev}, {Strigachev},
{Semkov}, {Wiita}, {Kurtanidze}, {Darriba}, {Damljanovic}, {Chanishvili},
{Ibryamov}, {Kurtanidze}, {Nikolashvili}, {Sigua}, and {Vince}]{Gaur_2019}
{Gaur}, H.; {Gupta}, A.C.; {Bachev}, R.; {Strigachev}, A.; {Semkov}, E.;
{Wiita}, P.J.; {Kurtanidze}, O.M.; {Darriba}, A.; {Damljanovic}, G.;
{Chanishvili}, R.G.;  et~al.
\newblock {Optical variability of TeV blazars on long time-scales}.
\newblock {\em Mon. Not. R. Astron. Soc.} {\bf 2019}, {\em 484},~5633--5644.
\newblock  [\href{http://dx.doi.org/10.1093/mnras/stz322}{CrossRef}]

\bibitem[{Gaskell}(2004)]{Gaskell_2004}
{Gaskell}, C.M.
\newblock {Lognormal X-ray Flux Variations in an Extreme Narrow-Line Seyfert 1
Galaxy}.
\newblock {\em  Astrophys. J.} {\bf 2004}, {\em 612},~L21--L24.
\newblock   [\href{http://dx.doi.org/10.1086/424565}{CrossRef}]

\bibitem[{Uttley} \em{et~al.}(2005){Uttley}, {McHardy}, and
{Vaughan}]{Uttley_McHardy_2005}
{Uttley}, P.; {McHardy}, I.M.; {Vaughan}, S.
\newblock {Non-linear X-ray variability in X-ray binaries and active galaxies}.
\newblock {\em Mon. Not. R. Astron. Soc.} {\bf 2005}, {\em 359},~345--362.
\newblock   [\href{http://dx.doi.org/10.1111/j.1365-2966.2005.08886.x}{CrossRef}]

\bibitem[{McHardy}(2010)]{McHardy_2010}
{McHardy}, I.M.
\newblock {\em Lecture Notes in Physics};
\newblock {Belloni, T., Ed.}; {Springer:
Berlin, Germany}, 2010; Volume {{794}, p. 203}. 

\bibitem[{Giebels} and {Degrange}(2009)]{Giebels_2009}
{{Giebels}, B.; {Degrange}, B.
\newblock {Lognormal variability in BL Lacertae}.
\newblock {\em Astron. Astrophys.} {\bf 2009}, {\em 503},~797--799.
\newblock  } [\href{http://dx.doi.org/10.1051/0004-6361/200912303}{CrossRef}]

\bibitem[{Chevalier} \em{et~al.}(2015){Chevalier}, {Kastendieck}, {Rieger},
{Maurin}, {Lenain}, and {Lamanna}]{Chevalier_2015}
{Chevalier}, J.; {Kastendieck}, M.A.; {Rieger}, F.M.; {Maurin}, G.; {Lenain},
J.P.; {Lamanna}, G.
\newblock {Long term variability of the blazar PKS 2155-304}. \emph{arXiv}  \textbf{2015}, arXiv:1509.03104.

\bibitem[{Sinha} \em{et~al.}(2016){Sinha}, {Shukla}, {Saha}, {Acharya},
{Anupama}, {Bhattacharjee}, {Britto}, {Chitnis}, {Prabhu}, {Singh}, and
{Vishwanath}]{Sinha_2016}
{Sinha}, A.; {Shukla}, A.; {Saha}, L.; {Acharya}, B.S.; {Anupama}, G.C.;
{Bhattacharjee}, P.; {Britto}, R.J.; {Chitnis}, V.R.; {Prabhu}, T.P.;
{Singh}, B.B.;  et~al.
\newblock {Long-term study of Mkn 421 with the HAGAR Array of Telescopes}.
\newblock {\em Astron. Astrophys.} {\bf 2016}, {\em 591},~A83.
\newblock   [\href{http://dx.doi.org/10.1051/0004-6361/201628152}{CrossRef}]

\bibitem[Sinha \em{et~al.}(2017)Sinha, Sahayanathan, Acharya, Anupama, Chitnis,
and Singh]{Sinha_2017}
Sinha, A.; Sahayanathan, S.; Acharya, B.S.; Anupama, G.C.; Chitnis, V.R.;
Singh, B.B.
\newblock On the Spectral Curvature of {VHE} Blazar 1ES 1011+496:
Effect of Spatial Particle Diffusion.
\newblock {\em  Astrophys. J.} {\bf 2017}, {\em 836},~83.
\newblock   [\href{http://dx.doi.org/10.3847/1538-4357/836/1/83}{CrossRef}]

\bibitem[{Sinha} \em{et~al.}(2018){Sinha}, {Khatoon}, {Misra}, {Sahayanathan},
{Mandal}, {Gogoi}, and {Bhatt}]{Sinha_2018}
{Sinha}, A.; {Khatoon}, R.; {Misra}, R.; {Sahayanathan}, S.; {Mandal}, S.;
{Gogoi}, R.; {Bhatt}, N.
\newblock {The flux distribution of individual blazars as a key to understand
the dynamics of particle acceleration}.
\newblock {\em Mon. Not. R. Astron. Soc.} {\bf 2018}, {\em 480},~L116--L120.
\newblock   [\href{http://dx.doi.org/10.1093/mnrasl/sly136}{CrossRef}]

\bibitem[{Kushwaha} \em{et~al.}(2016){Kushwaha}, {Chandra}, {Misra},
{Sahayanathan}, {Singh}, and {Baliyan}]{Kushwaha_2016}
{Kushwaha}, P.; {Chandra}, S.; {Misra}, R.; {Sahayanathan}, S.; {Singh}, K.P.;
{Baliyan}, K.S.
\newblock {Evidence for Two Lognormal States in Multi-wavelength Flux Variation
of FSRQ PKS 1510-089}.
\newblock {\em Astrophys. J.} {\bf 2016}, {\em 822},~L13.
\newblock   [\href{http://dx.doi.org/10.3847/2041-8205/822/1/L13}{CrossRef}]

\bibitem[{Kushwaha} and {Pal}(2020)]{Kushwaha_2020}
{Kushwaha}, P.; {Pal}, M.
\newblock {Short-Term X-ray Variability during Different Activity Phases of
Blazars S5 0716+714 and PKS 2155-304}.
\newblock {\em Galaxies} {\bf 2020}, {\em 8},~66.
\newblock   [\href{http://dx.doi.org/10.3390/galaxies8030066}{CrossRef}]

\bibitem[{Shah} \em{et~al.}(2018){Shah}, {Mankuzhiyil}, {Sinha}, {Misra},
{Sahayanathan}, and {Iqbal}]{Shah_2018}
{Shah}, Z.; {Mankuzhiyil}, N.; {Sinha}, A.; {Misra}, R.; {Sahayanathan}, S.;
{Iqbal}, N.
\newblock {Log-normal flux distribution of bright Fermi blazars}.
\newblock {\em Res. Astron. Astrophys.} {\bf 2018}, {\em
18},~141.
\newblock   [\href{http://dx.doi.org/10.1088/1674-4527/18/11/141}{CrossRef}]

\bibitem[{Biteau} and {Giebels}(2012)]{Biteau_Giebels_2012}
{Biteau}, J.; {Giebels}, B.
\newblock {The minijets-in-a-jet statistical model and the rms-flux
correlation}.
\newblock {\em Astron. Astrophys.} {\bf 2012}, {\em 548},~A123.
\newblock   [\href{http://dx.doi.org/10.1051/0004-6361/201220056}{CrossRef}]

\bibitem[{Giannios} \em{et~al.}(2009){Giannios}, {Uzdensky}, and
{Begelman}]{Giannios_2009}
{Giannios}, D.; {Uzdensky}, D.A.; {Begelman}, M.C.
\newblock {Fast TeV variability in blazars: Jets in a jet}.
\newblock {\em Mon. Not. R. Astron. Soc.} {\bf 2009}, {\em 395},~L29--L33.
\newblock   [\href{http://dx.doi.org/10.1111/j.1745-3933.2009.00635.x}{CrossRef}]

\bibitem[{Gaur} \em{et~al.}(2018){Gaur}, {Mohan}, {Wierzcholska}, and
{Gu}]{Gaur_2018}
{Gaur}, H.; {Mohan}, P.; {Wierzcholska}, A.; {Gu}, M.
\newblock {Signature of inverse Compton emission from blazars}.
\newblock {\em Mon. Not. R. Astron. Soc.} {\bf 2018}, {\em 473},~3638--3660.
\newblock   [\href{http://dx.doi.org/10.1093/mnras/stx2553}{CrossRef}]

\bibitem[{Pavana Gowtami} \em{et~al.}(2022){Pavana Gowtami}, {Gaur}, {Gupta},
{Wiita}, {Liao}, and {Ward}]{Gowtami_2022}
{Pavana Gowtami}, G.S.; {Gaur}, H.; {Gupta}, A.C.; {Wiita}, P.J.; {Liao}, M.;
{Ward}, M.
\newblock {X-ray intraday variability and power spectral density profiles of
the blazar 3C 273 with XMM-Newton during 2000--2021}.
\newblock {\em Mon. Not. R. Astron. Soc.} {\bf 2022}, {\em 511},~3101--3112.
\newblock   [\href{http://dx.doi.org/10.1093/mnras/stac286}{CrossRef}]

\bibitem[{Str{\"u}der} \em{et~al.}(2001){Str{\"u}der}, {Briel}, {Dennerl},
{Hartmann}, {Kendziorra}, {Meidinger}, {Pfeffermann}, {Reppin}, {Aschenbach},
{Bornemann}, {Br{\"a}uninger}, {Burkert}, {Elender}, {Freyberg}, {Haberl},
{Hartner}, {Heuschmann}, {Hippmann}, {Kastelic}, {Kemmer}, {Kettenring},
{Kink}, {Krause}, {M{\"u}ller}, {Oppitz}, {Pietsch}, {Popp}, {Predehl},
{Read}, {Stephan}, {St{\"o}tter}, {Tr{\"u}mper}, {Holl}, {Kemmer}, {Soltau},
{St{\"o}tter}, {Weber}, {Weichert}, {von Zanthier}, {Carathanassis}, {Lutz},
{Richter}, {Solc}, {B{\"o}ttcher}, {Kuster}, {Staubert}, {Abbey}, {Holland},
{Turner}, {Balasini}, {Bignami}, {La Palombara}, {Villa}, {Buttler},
{Gianini}, {Lain{\'e}}, {Lumb}, and {Dhez}]{Struder_2001}
{Str{\"u}der}, L.; {Briel}, U.; {Dennerl}, K.; {Hartmann}, R.; {Kendziorra},
E.; {Meidinger}, N.; {Pfeffermann}, E.; {Reppin}, C.; {Aschenbach}, B.;
{Bornemann}, W.;  et~al.
\newblock {The European Photon Imaging Camera on XMM-Newton: The pn-CCD
camera}.
\newblock {\em Astron. Astrophys.} {\bf 2001}, {\em 365},~L18--L26. [\href{http://dx.doi.org/10.1051/0004-6361:20000066}{CrossRef}]

\bibitem[{Vaughan} \em{et~al.}(2003){Vaughan}, {Edelson}, {Warwick}, and
{Uttley}]{Vaughan(2003a)}
{Vaughan}, S.; {Edelson}, R.; {Warwick}, R.S.; {Uttley}, P.
\newblock {On characterizing the variability properties of X-ray light curves
from active galaxies}.
\newblock {\em Mon. Not. R. Astron. Soc.} {\bf 2003}, {\em 345},~1271--1284. [\href{http://dx.doi.org/10.1046/j.1365-2966.2003.07042.x}{CrossRef}]

\bibitem[{Edelson} \em{et~al.}(1990){Edelson}, {Krolik}, and
{Pike}]{EdelsonKrolikPike(1990)}
{Edelson}, R.A.; {Krolik}, J.H.; {Pike}, G.F.
\newblock {Broad-Band Properties of the CfA Seyfert Galaxies. III. Ultraviolet
Variability}.
\newblock {\em Astrophys. J.} {\bf 1990}, {\em 359},~86. [\href{http://dx.doi.org/10.1086/169036}{CrossRef}]

\bibitem[{Rodr{\'\i}guez-Pascual} \em{et~al.}(1997){Rodr{\'\i}guez-Pascual},
{Alloin}, {Clavel}, {Crenshaw}, {Horne}, {Kriss}, {Krolik}, {Malkan},
{Netzer}, {O'Brien}, {Peterson}, {Reichert}, {Wamsteker}, {Alexander},
{Barr}, {Blandford}, {Bregman}, {Carone}, {Clements}, {Courvoisier}, {De
Robertis}, {Dietrich}, {Dottori}, {Edelson}, {Filippenko}, {Gaskell},
{Huchra}, {Hutchings}, {Kollatschny}, {Koratkar}, {Korista}, {Laor},
{MacAlpine}, {Martin}, {Maoz}, {McCollum}, {Morris}, {Perola}, {Pogge},
{Ptak}, {Recondo-Gonz{\'a}lez}, {Rodr{\'\i}guez-Espinoza}, {Rokaki},
{Santos-Lle{\'o}}, {Sekiguchi}, {Shull}, {Snijders}, {Sparke}, {Stirpe},
{Stoner}, {Sun}, {Wagner}, {Wanders}, {Wilkes}, {Winge}, and
{Zheng}]{Rodriguez-Pascual(1997)}
{Rodr{\'\i}guez-Pascual}, P.M.; {Alloin}, D.; {Clavel}, J.; {Crenshaw}, D.M.;
{Horne}, K.; {Kriss}, G.A.; {Krolik}, J.H.; {Malkan}, M.A.; {Netzer}, H.;
{O'Brien}, P.T.;  et~al.
\newblock {Steps toward Determination of the Size and Structure of the
Broad-Line Region in Active Galactic Nuclei. IX. Ultraviolet Observations of
Fairall 9}.
\newblock {\em  Astrophys. J. Suppl. Ser.} {\bf 1997}, {\em 110},~9--20. [\href{http://dx.doi.org/10.1086/312996}{CrossRef}]

\bibitem[Knuth(2019)]{KNUTH_2019}
Knuth, K.H.
\newblock Optimal data-based binning for histograms and histogram-based
probability density models.
\newblock {\em Digit. Signal Process.} {\bf 2019}, {\em 95},~102581.
\newblock   [\href{http://dx.doi.org/10.1016/j.dsp.2019.102581}{CrossRef}]

\bibitem[Anderson and Darling(1952)]{AD_1952}
Anderson, T.W.; Darling, D.A.
\newblock {Asymptotic Theory of Certain ``Goodness of Fit'' Criteria Based on
Stochastic Processes}.
\newblock {\em  Ann. Math. Stat.} {\bf 1952}, {\em 23},~193--212. [\href{http://dx.doi.org/10.1214/aoms/1177729437}{CrossRef}]

\bibitem[Stephens(1977)]{Stephens_1977}
Stephens, M.A.
\newblock {Goodness of fit for the extreme value distribution}.
\newblock {\em Biometrika} {\bf 1977}, {\em 64},~583--588. [\href{http://dx.doi.org/10.1093/biomet/64.3.583}{CrossRef}]

\bibitem[{Mohorian} \em{et~al.}(2022){Mohorian}, {Bhatta}, {Adhikari},
{Dhital}, {P{\'a}nis}, {Dinesh}, {Chaudhary}, {Bachchan}, and
{Stuchl{\'\i}k}]{Mohorian_2022}
{Mohorian}, M.; {Bhatta}, G.; {Adhikari}, T.P.; {Dhital}, N.; {P{\'a}nis}, R.;
{Dinesh}, A.; {Chaudhary}, S.C.; {Bachchan}, R.K.; {Stuchl{\'\i}k}, Z.
\newblock {X-ray timing and spectral variability properties of blazars S5 0716+714, OJ 287, Mrk 501, and RBS 2070}.
\newblock {\em Mon. Not. R. Astron. Soc.} {\bf 2022}, {\em 510},~5280--5301.
\newblock   [\href{http://dx.doi.org/10.1093/mnras/stab3738}{CrossRef}]

\bibitem[{Khatoon} \em{et~al.}(2020){Khatoon}, {Shah}, {Misra}, and
{Gogoi}]{Khatoon_2020}
{Khatoon}, R.; {Shah}, Z.; {Misra}, R.; {Gogoi}, R.
\newblock {Study of long-term flux and photon index distributions of blazars
using RXTE observations}.
\newblock {\em Mon. Not. R. Astron. Soc.} {\bf 2020}, {\em 491},~1934--1940.
\newblock   [\href{http://dx.doi.org/10.1093/mnras/stz3108}{CrossRef}]

\bibitem[{Bhatta}(2021)]{Bhatta_2021}
{Bhatta}, G.
\newblock {Characterizing Long-term Optical Variability Properties of
{\ensuremath{\gamma}}-Ray-bright Blazars}.
\newblock {\em  Astrophys. J.} {\bf 2021}, {\em 923},~7.
\newblock   [\href{http://dx.doi.org/10.3847/1538-4357/ac2819}{CrossRef}]

\bibitem[{Bhatta} and {Dhital}(2020)]{Bhatta_Dhital_2020}
{Bhatta}, G.; {Dhital}, N.
\newblock {The Nature of {\ensuremath{\gamma}}-ray Variability in Blazars}.
\newblock {\em  Astrophys. J.} {\bf 2020}, {\em 891},~120.
\newblock   [\href{http://dx.doi.org/10.3847/1538-4357/ab7455}{CrossRef}]

\bibitem[{Duda} and {Bhatta}(2021)]{Duda_Bhatta_2021}
{Duda}, J.; {Bhatta}, G.
\newblock {Gamma-ray blazar variability: New statistical methods of time-flux
distributions}.
\newblock {\em Mon. Not. R. Astron. Soc.} {\bf 2021}, {\em 508},~1446--1458.
\newblock  [\href{http://dx.doi.org/10.1093/mnras/stab2574}{CrossRef}]

\bibitem[{McHardy} \em{et~al.}(2006){McHardy}, {Koerding}, {Knigge}, {Uttley},
and {Fender}]{McHardy_2006}
{McHardy}, I.M.; {Koerding}, E.; {Knigge}, C.; {Uttley}, P.; {Fender}, R.P.
\newblock {Active galactic nuclei as scaled-up Galactic black holes}.
\newblock {\em Nature} {\bf 2006}, {\em 444},~730--732.
\newblock  [\href{http://dx.doi.org/10.1038/nature05389}{CrossRef}] [\href{http://www.ncbi.nlm.nih.gov/pubmed/17151661}{PubMed}]

\bibitem[{Nakagawa} and {Mori}(2013)]{Nakagawa_Mori_2013}
{Nakagawa}, K.; {Mori}, M.
\newblock {Time Series Analysis of Gamma-Ray Blazars and Implications for the
Central Black-hole Mass}.
\newblock {\em Astrophys. J.} {\bf 2013}, {\em 773},~177.
\newblock   [\href{http://dx.doi.org/10.1088/0004-637X/773/2/177}{CrossRef}]

\bibitem[{Sobolewska} \em{et~al.}(2014){Sobolewska}, {Siemiginowska}, {Kelly},
and {Nalewajko}]{Sobolewska_2014}
{Sobolewska}, M.A.; {Siemiginowska}, A.; {Kelly}, B.C.; {Nalewajko}, K.
\newblock {Stochastic Modeling of the Fermi/LAT {\ensuremath{\gamma}}-Ray
Blazar Variability}.
\newblock {\em Astrophys. J.} {\bf 2014}, {\em 786},~143.
\newblock   [\href{http://dx.doi.org/10.1088/0004-637X/786/2/143}{CrossRef}]

\bibitem[{Tluczykont} \em{et~al.}(2010){Tluczykont}, {Bernardini}, {Satalecka},
{Clavero}, {Shayduk}, and {Kalekin}]{Tluczykont_2010}
{Tluczykont}, M.; {Bernardini}, E.; {Satalecka}, K.; {Clavero}, R.; {Shayduk},
M.; {Kalekin}, O.
\newblock {Long-term lightcurves from combined unified very high energy
{\ensuremath{\gamma}}-ray data}.
\newblock {\em Astron. Astrophys.} {\bf 2010}, {\em 524},~A48.
\newblock   [\href{http://dx.doi.org/10.1051/0004-6361/201015193}{CrossRef}]

\bibitem[{Ackermann} \em{et~al.}(2015){Ackermann}, {Ajello}, {Atwood},
{Baldini}, {Ballet}, {Barbiellini}, {Bastieri}, {Becerra Gonzalez},
{Bellazzini}, {Bissaldi}, {Blandford}, {Bloom}, {Bonino}, {Bottacini},
{Brandt}, {Bregeon}, {Britto}, {Bruel}, {Buehler}, {Buson}, {Caliandro},
{Cameron}, {Caragiulo}, {Caraveo}, {Carpenter}, {Casandjian}, {Cavazzuti},
{Cecchi}, {Charles}, {Chekhtman}, {Cheung}, {Chiang}, {Chiaro}, {Ciprini},
{Claus}, {Cohen-Tanugi}, {Cominsky}, {Conrad}, {Cutini}, {D'Abrusco},
{D'Ammando}, {de Angelis}, {Desiante}, {Digel}, {Di Venere}, {Drell},
{Favuzzi}, {Fegan}, {Ferrara}, {Finke}, {Focke}, {Franckowiak}, {Fuhrmann},
{Fukazawa}, {Furniss}, {Fusco}, {Gargano}, {Gasparrini}, {Giglietto},
{Giommi}, {Giordano}, {Giroletti}, {Glanzman}, {Godfrey}, {Grenier}, {Grove},
{Guiriec}, {Hewitt}, {Hill}, {Horan}, {Itoh}, {J{\'o}hannesson}, {Johnson},
{Johnson}, {Kataoka}, {Kawano}, {Krauss}, {Kuss}, {La Mura}, {Larsson},
{Latronico}, {Leto}, {Li}, {Li}, {Longo}, {Loparco}, {Lott}, {Lovellette},
{Lubrano}, {Madejski}, {Mayer}, {Mazziotta}, {McEnery}, {Michelson},
{Mizuno}, {Moiseev}, {Monzani}, {Morselli}, {Moskalenko}, {Murgia}, {Nuss},
{Ohno}, {Ohsugi}, {Ojha}, {Omodei}, {Orienti}, {Orlando}, {Paggi}, {Paneque},
{Perkins}, {Pesce-Rollins}, {Piron}, {Pivato}, {Porter}, {Rain{\`o}},
{Rando}, {Razzano}, {Razzaque}, {Reimer}, {Reimer}, {Romani}, {Salvetti},
{Schaal}, {Schinzel}, {Schulz}, {Sgr{\`o}}, {Siskind}, {Sokolovsky}, {Spada},
{Spandre}, {Spinelli}, {Stawarz}, {Suson}, {Takahashi}, {Takahashi},
{Tanaka}, {Thayer}, {Thayer}, {Tibaldo}, {Torres}, {Torresi}, {Tosti},
{Troja}, {Uchiyama}, {Vianello}, {Winer}, {Wood}, and
{Zimmer}]{Ackermann_2015}
{Ackermann}, M.; {Ajello}, M.; {Atwood}, W.B.; {Baldini}, L.; {Ballet}, J.;
{Barbiellini}, G.; {Bastieri}, D.; {Becerra Gonzalez}, J.; {Bellazzini}, R.;
{Bissaldi}, E.;  et~al.
\newblock {The Third Catalog of Active Galactic Nuclei Detected by the Fermi
Large Area Telescope}.
\newblock {\em Astrophys. J.} {\bf 2015}, {\em 810},~14.
\newblock   [\href{http://dx.doi.org/10.1088/0004-637X/810/1/14}{CrossRef}]

\bibitem[{Narayan} and {Piran}(2012)]{Narayan_Piran_2012}
{Narayan}, R.; {Piran}, T.
\newblock {Variability in blazars: Clues from PKS 2155-304}.
\newblock {\em Mon. Not. R. Astron. Soc.} {\bf 2012}, {\em 420},~604--612.
\newblock   [\href{http://dx.doi.org/10.1111/j.1365-2966.2011.20069.x}{CrossRef}]

\bibitem[{Marscher} and {Gear}(1985)]{Marscher_Gear_1985}
{Marscher}, A.P.; {Gear}, W.K.
\newblock {Models for high-frequency radio outbursts in extragalactic sources,
with application to the early 1983 millimeter-to-infrared flare of 3C 273.}
\newblock {\em Astrophys. J.} {\bf 1985}, {\em 298},~114--127.
\newblock   [\href{http://dx.doi.org/10.1086/163592}{CrossRef}]

\bibitem[{B{\"o}ttcher} and {Dermer}(2010)]{Bottcher_Dermer_2010}
{B{\"o}ttcher}, M.; {Dermer}, C.D.
\newblock {Timing Signatures of the Internal-Shock Model for Blazars}.
\newblock {\em Astrophys. J.} {\bf 2010}, {\em 711},~445--460.
\newblock   [\href{http://dx.doi.org/10.1088/0004-637X/711/1/445}{CrossRef}]

\bibitem[{Ghisellini} and {Tavecchio}(2008)]{Ghisellini_Tavecchio_2008}
{Ghisellini}, G.; {Tavecchio}, F.
\newblock {Rapid variability in TeV blazars: The case of PKS2155-304}.
\newblock {\em Mon. Not. R. Astron. Soc.} {\bf 2008}, {\em 386},~L28--L32.
\newblock   [\href{http://dx.doi.org/10.1111/j.1745-3933.2008.00454.x}{CrossRef}]


\bibitem[{Giannios} \em{et~al.}(2010){Giannios}, {Uzdensky}, and
{Begelman}]{Giannios_2010}
{Giannios}, D.; {Uzdensky}, D.A.; {Begelman}, M.C.
\newblock {Fast TeV variability from misaligned minijets in the jet of M87}.
\newblock {\em Mon. Not. R. Astron. Soc.} {\bf 2010}, {\em 402},~1649--1656.
\newblock  [\href{http://dx.doi.org/10.1111/j.1365-2966.2009.16045.x}{CrossRef}]

\bibitem[{Marscher}(2014)]{Marscher_2014}
{Marscher}, A.P.
\newblock {Turbulent, Extreme Multi-zone Model for Simulating Flux and
Polarization Variability in Blazars}.
\newblock {\em Astrophys. J.} {\bf 2014}, {\em 780},~87.
\newblock  [\href{http://dx.doi.org/10.1088/0004-637X/780/1/87}{CrossRef}]

\end{thebibliography}
\end{document}